\documentclass{emulateapj}
\usepackage{lscape}
\usepackage{epsfig}
\usepackage{natbib}

\newcommand\msun{\cal{M}_\odot}

\shortauthors{Bochanski et al.}
\shorttitle{SDSS M Dwarf Luminosity \& Mass Functions}
\begin{document}

\title{The Luminosity and Mass Functions of Low--Mass Stars in the Galactic Disk: II.  The Field}

\author{John J. Bochanski\altaffilmark{1,2},
	 Suzanne L. Hawley\altaffilmark{1},
	 Kevin R. Covey\altaffilmark{3}
	 Andrew A. West\altaffilmark{2,4},
	 I. Neill Reid\altaffilmark{5}
	 David A. Golimowski\altaffilmark{5},
	 \v{Z}eljko Ivezi\'{c}\altaffilmark{1}
	 }

\altaffiltext{1}{Astronomy Department, University of Washington,
   Box 351580, Seattle, WA  98195\\
email: jjb@mit.edu}
\altaffiltext{2}{Kavli Institute for Astrophysics and Space Research, Massachusetts Institute of Technology, Building 37, 77 Massachusetts Avenue, Cambridge, MA 02139}
\altaffiltext{3}{Department of Astronomy, Cornell University, Ithaca, NY, 14853}
\altaffiltext{4}{Department of Astronomy, Boston University, 725 Commonwealth Avenue, Boston, MA 02215}
\altaffiltext{5}{Space Telescope Science Institute,
3700 San Martin Drive, Baltimore, MD 21218 }

\begin{abstract}
We report on new measurements of the luminosity function (LF) and mass function (MF) of field low--mass dwarfs derived from Sloan Digital Sky Survey (SDSS) Data Release 6 (DR6) photometry.  The analysis incorporates $\sim$ 15 million low--mass stars (0.1 $\msun <$  $\cal{M}$  $<  0.8 \msun$), spread over 8,400 square degrees.   Stellar distances are estimated using new photometric parallax relations, constructed from $ugriz$ photometry of nearby low--mass stars with trigonometric parallaxes.  We use a technique that simultaneously measures
Galactic structure and the stellar LF from $7 < M_r < 16$.  We compare the LF to previous studies and convert to a MF using the mass--luminosity relations of \cite{2000A&A...364..217D}.  The system MF, measured over  $-1.0 <$ log ${\cal M} / \msun$ $< -0.1$, is well--described by a log--normal distribution with $\cal{M}_{\circ}$ = 0.25 $\msun$.  We stress that our results should not be extrapolated to other mass regimes.  Our work generally agrees with prior low--mass stellar MFs and places strong constraints in future star--formation studies of the Milky Way.
\end{abstract}


\section{Introduction}
Low--mass dwarfs (0.1 $\msun <$ $\cal{M}$  $<  0.8 \msun$) are, by number, the dominant stellar population of the Milky Way.   These long-lived \citep{1997ApJ...482..420L} and ubiquitous objects comprise $\sim70\%$ of all stars, yet their diminutive luminosities ($L \lesssim 0.05 L_{\odot}$) have traditionally prohibited their study in large numbers.  However, in recent years, the development of large--format CCDs has enabled accurate photometric surveys over wide solid angles on the sky, such as the Two--Micron All Sky Survey \citep[2MASS; ][]{2006AJ....131.1163S} and the Sloan Digital Sky Survey \citep[SDSS; ][]{2000AJ....120.1579Y}.  These projects obtained precise ($\sigma \lesssim 5\%$) and deep ($r \sim 22, J \sim 16.5$) photometry of large solid angles ($\gtrsim 10^{4}$ sq$.$ deg$.$).   The resulting photometric datasets contain millions of low--mass stars, enabling statistical investigations of their properties. 
In particular, 2MASS photometry led to the discovery of two new spectral classes: L and T \citep{1999ApJ...519..802K, 2002ApJ...564..421B}, and was used to trace the structure of the Sagittarius dwarf galaxy \citep{2003ApJ...599.1082M} with M giants.  SDSS data led to the discovery the first field methane brown dwarf, which was the coolest substellar object known at the time of its discovery \citep{1999ApJ...522L..61S}.  Other notable SDSS results include the discovery of new stellar streams in the Halo \citep[e.g.][]{2003ApJ...588..824Y,2006ApJ...642L.137B} and new Milky Way companions \citep[e.g.,][]{2005AJ....129.2692W,2007ApJ...654..897B}, as well as unprecedented $in$ $situ$ mapping of the stellar density \citep[][]{2008ApJ...673..864J} and metallicity \citep{Ivezic08} distributions of the Milky Way and confirmation of the dual-halo structure of the Milky Way \citep{2007Natur.450.1020C}.  SDSS has proven to be a valuable resource for statistical investigations of the properties of low--mass stars, including their magnetic activity and chromospheric properties \citep{2004AJ....128..426W, 2008AJ....135..785W}, flare characteristics \citep{2009arXiv0906.2030K} and their use as tracers of Galactic structure and kinematics \citep{2007AJ....134.2418B, 2009AJ....137.4149F}.

Despite the advances made in other cool--star topics, two fundamental properties, the luminosity and mass functions, remain uncertain.  The luminosity function (LF) describes the number density of stars as a function of absolute magnitude ($\Phi(M) = dN / dM$).  The mass function (MF), typically inferred from the LF, is defined as the number density per unit mass ($\psi({\cal M}) = dN / d{\cal M}$).   For low--mass stars, with lifetimes much greater than the Hubble time, the observed present-day mass function (PDMF) in the field should trace the initial mass function (IMF).  Following \cite{1955ApJ...121..161S}, the IMF has usually been characterized by a power--law $\psi({\cal M}) = dN / d{\cal M} \propto {\cal M}^{- \alpha}$, with the exponent $\alpha$ varying over a wide range, from 0.5 to 2.5.  However, some studies have preferred a log--normal distribution.  Previous investigations are summarized in Table \ref{chap6:table:field_mfs} (MF) and Table \ref{chap5:table:existing_studies} (LF), which show the total number of stars included and solid angle surveyed in each study.

Present uncertainties in the LF and MF can be attributed to disparate measurement techniques that are a result of trade--offs in observing strategy.  Previous investigations of the LF and MF have fallen in one of two categories:  nearby, volume--limited studies of trigonometric parallax stars; or pencil--beam surveys of distant stars over a small solid angle.  Tables \ref{chap6:table:field_mfs} \& \ref{chap5:table:existing_studies} detail the techniques used by modern investigations of the field LF and MF.  In both cases, sample sizes were limited to a few thousand stars, prohibiting detailed statistical measurements.  The solid angle shown in each table distinguishes the two types of surveys. There is considerable disagreement between the nearby, volume--limited investigations \citep{1997AJ....113.2246R} and pencil beam studies of distant stars \citep{1998AJ....116.2513M, 2001ApJ...555..393Z, 2006AA...447..185S}.  It has been suggested that this discrepancy is due to the presence of unresolved binary stars in the pencil beam surveys \citep{1993MNRAS.262..545K, 2003ApJ...586L.133C}, but this has not been shown conclusively \citep{1997AJ....113.2246R}.  We investigate the effect of unresolved binaries in \S \ref{sec:binaries}.  

Using a sample drawn from SDSS, 2MASS and Guide Star Catalog photometry, and supplemented with SDSS spectroscopy, \cite{covey08} performed the largest field low--mass LF and MF investigation to date.  Covering 30 sq.\ deg.\ and containing $\sim30,000$ low--mass stars, their sample measured the LF using absolute magnitudes estimated from photometric colors, and quantified the contamination rate by obtaining spectra of every red point source in a 1 sq.\ deg.\ calibration region.  The \cite{covey08} sample serves as a calibration study for the present work, as it quantified the completeness, bias and contamination rate of the SDSS and 2MASS photometric samples.  While their study focused on a limited sightline, the current investigation expands to the entire SDSS footprint, increasing the solid angle by a factor of $\sim300$.

In \S \ref{sec:observations}, we describe the SDSS photometry used to measure the field LF and MF.  The color--absolute magnitude calibration is discussed in \S \ref{sec:phot_pi}.  In \S \ref{sec:analysis}, we introduce a new technique for measuring the LF of large, deep photometric datasets, and compare to previous analyses.  The resulting ``raw'' LF is corrected for systematic effects such as unresolved binarity, metallicity gradients and changes in Galactic structure in \S \ref{sec:correction}.  The final LF and our MF is presented in \S \ref{sec:LF} and \S \ref{sec:MF}.  Our conclusions follow in \S \ref{sec:conclusions}.

\begin{turnpage}
\begin{deluxetable*}{llllllll}
\tablewidth{0pt}
 
 \tablecaption{Major Low--Mass Field IMF Studies}
 \tabletypesize{\scriptsize}
 
 \tablehead{
 \colhead{Authors} &
 \colhead{$N_{\rm Stars}$} &
 \colhead{$\Omega$ (sq. deg.)} &
 \colhead{Filter(s)} &
 \colhead{Depth} &
 \colhead{Mass Range} &
 \colhead{$\alpha$, ${\cal M}_{\circ}$}&
 \colhead{Notes}
 }
 \startdata
\cite{1955ApJ...121..161S} & \nodata    &  \nodata     &    $V$     &   \nodata      &  $0.3\msun$ - 10 $\msun$   & $ \alpha =$ 2.35   & Compiled LFs from contemporaries\tablenotemark{a} \\
\cite{1979ApJS...41..513M} & \nodata    &  \nodata     &   $V$      &   \nodata      &   $0.1\msun$ - 60 $\msun$  &  ${\cal M}_{\circ} \simeq 0.1 \msun$  & Log-normal fit, Compilation of 3 field LFs\tablenotemark{b}  \\
\cite{1990MNRAS.244...76K} & \nodata    &  \nodata     &   $V$      &   \nodata      &  $0.1\msun$ - 0.9 $\msun$  &     ${\cal M}_{\circ} \simeq 0.23 \msun$, $\alpha = 0.70$ & Adopted LFs of \cite{1986FCPh...11....1S} and \cite{1989MNRAS.238..709S}  \\
\cite{1993MNRAS.262..545K} & \nodata    &   \nodata    &  $V,I$     &  \nodata       &   $0.08\msun$ - 0.5 $\msun$     &   0.70 $< \alpha < 1.85$       & LF from \cite{1983nssl.conf..163W} and \cite{1989MNRAS.238..709S}  \\
\cite{1993ApJ...414..279T} &  3,500   &  280 deg$^2$   &   $I,K$  &  $I \lesssim 17.5$   &  $0.1\msun$ - 0.5 $\msun$   &  \nodata       & Turnover at 0.25 $\msun$  \\
\cite{1997AJ....113.2246R} &  151     &  $\delta~>~-30^\circ$  & $V,I$   & $d < 8 $pc &   $0.08\msun $- 1.2 $\msun$     & $ \alpha =$ 1.2    &  Solar Neighborhood\tablenotemark{c} \\
\cite{1998AJ....116.2513M} & 1,500 & 0.83 deg$^2$  & $V,R$   & $V \lesssim 23.5$   & $0.1 \msun$ - 0.6 $\msun$ & 1.3 &   \\
\cite{2001ApJ...555..393Z}  &  $\sim 1,400$   &  $\sim 0.4$ deg$^2$     &   $V,I$      & $18  \lesssim I \lesssim 24$     &   $0.1 \msun$  - 0.6 $\msun$    & $ \alpha =$ 0.45  & HST observations   \\
\cite{2002Sci...295...82K} &  \nodata   &  \nodata     &  $V,I$       &  \nodata       &   $0.08\msun$ - 0.50 $\msun$     &  $ \alpha =$ 1.3      &  Compiled contemporary LFs\tablenotemark{d}  \\
\cite{2002AJ....124.2721R} &  558   &  3$\pi$ ster.    &   $B,V$  &  $d \lesssim 20$pc &   $0.1 \msun$ - 3.0 $\msun$     &  $\alpha \simeq 1.3$  &  Solar Neighborhood Survey  \\
\cite{2003PASP..115..763C} &  \nodata   &  \nodata      &   $V,K$      &  \nodata    &   $0.1 \msun$ - 1.0 $\msun$     &   ${\cal M}_{\circ} = 0.22 \msun$      & Review of contemporary field LFs\tablenotemark{e} \\
\cite{2006AA...447..185S} & 3,600  &  $\sim3$ deg$^2$     &   $r^{\prime},i^{\prime}$   &  $i^{\prime} \sim 21$       &  ${\cal M} <$ 0.25 $\msun$      &  $ \alpha =$ 2.5    & CFHT MegaCAM observations  \\
\cite{covey08} &  $\sim29 \times 10^3$   &   30 deg$^2$    &   $i,J$      &   $J=16.2$      &    $0.1\msun$ - 0.8 $\msun$    & ${\cal M}_{\circ} = 0.29 \msun$   & Matched SDSS \& 2MASS observations  \\
This Study & $\sim15 \times 10^6$    &  8,400 deg$^2$     &  $r,i,z$       &  $16 < r < 22$       &  $0.1\msun$ - 0.8 $\msun$      &  ${\cal M}_{\circ} = 0.25 \msun$        &  SDSS observations\\
\enddata

\tablenotetext{a}{Salpeter averaged luminosity functions from \cite{1925PGro...38D...1V,1936PGro...47....1V} and \cite{1939POMin...2..121L,1941NYASA..42..201L}.}
\tablenotetext{b}{Their adopted LF was averaged from the LFs of \cite{1966VA......7..141M}, \cite{1968MNRAS.139..221L} and \cite{1974HiA.....3..395W}.}
\tablenotetext{c}{The ``8 parsec'' sample was compiled by \cite{1997AJ....113.2246R}, with later additions from \cite{1999ApJ...521..613R}, \cite{2003AJ....125..354R}, and \cite{2007AJ....133..439C}.}
\tablenotetext{d}{\cite{2002Sci...295...82K} presents a comprehensive summary of MFs derived from the field and clusters over a wide mass range.  For low-mass stars in the field, he refers to\\ \cite{1999ApJ...521..613R}, \cite{1999ApJ...526L..17H}, \cite{2001ApJ...554.1274C} and \cite{2001ApJ...555..393Z}.}
\tablenotetext{e}{\cite{2003PASP..115..763C} compared the  LFs of \cite{1986AJ.....91..621D}, \cite{1990ApJ...350..334H} and \cite{2001ApJ...555..393Z}.}

\label{chap6:table:field_mfs}
\end{deluxetable*}
\end{turnpage}

\begin{turnpage}
\begin{deluxetable*}{lllllll}
 \tablewidth{0pt}
 
 \tablecaption{Major Low-Mass Stellar Field LF Studies}
 \tabletypesize{\scriptsize }
 
 \tablehead{
 \colhead{Authors} &
 \colhead{$N_{\rm Stars}$} &
 \colhead{$\Omega$ (sq. deg.)} &
 \colhead{Filter(s)} &
 \colhead{Depth} &
 \colhead{Distance Method} &
 \colhead{Notes}
 }
 \startdata
 \cite{1989MNRAS.238..709S} & 178    &  18.88                     & $V,I$ & $ I < 16$       & Phot. $\pi$ & $VI$ photometry        \\
 \cite{1994AJ....108.1437H} & 92     &  $\delta > -25^\circ$      & $V$   & $d = 5$ pc      & Trig. $\pi$ &  Spec., CSN3 photometry\\
 \cite{1995AJ....110.1838R} & 520    &  $\delta > -30^\circ$      & $V$   & $d \sim 20$pc   & Trig. $\pi$ &  Spec., CNS3 photometry\tablenotemark{a}\\
 \cite{1997AJ....113.2246R} & 151    &  $\delta > -30^\circ$      & $V$   & $d = 8$ pc      & Trig. $\pi$ &  Spec., CNS3 photometry\tablenotemark{b}\\
 \cite{1998AJ....116.2513M} & 4,005  & 0.83                       & $V$   & $V \sim 23.5$   & Phot. $\pi$ & $UBVRI$ photometry \\ 
 \cite{2001ApJ...555..393Z} & 1,413  & $\sim 0.4$                 & $V$   & $I \sim 26.5$   & Phot. $\pi$ & HST photometry \tablenotemark{c}\\
 \cite{2002AJ....124.2721R} & 558    &  $\delta > -30^\circ$      & $V$   & $d \sim 20$pc   & Trig. $\pi$ & Spec., CNS3 photometry \\ 
 \cite{2007AJ....133..439C} & 99     & 14,823                     & $J$   & $J \sim 17$, $d \sim 20$pc    & Phot. $\pi$, Trig. $\pi$ & Spec., 2MASS photometry   \\
 \cite{covey08}             & $\sim 29 \times 10^3$   & 29        & $J$   & $J = 16.2 $     & Phot. $\pi$ & SDSS \& 2MASS photometry \\
 This Study &               $\sim 15 \times 10^6$     & 8,417     & $r,J$ & $r = 22$        & Phot. $\pi$ & SDSS photometry \\
 \enddata

 \label{chap5:table:existing_studies}
 \tablenotetext{a}{See \cite{1991adc..rept.....G} for details on sources of photometry.}
 \tablenotetext{b}{The 8 pc sample was further extended by \cite{1999ApJ...521..613R} and \cite{2003AJ....125..354R} and presented in $J$ by \cite{2007AJ....133..439C}.}
 \tablenotetext{c}{Some of the HST observations in this study were presented by \cite{1996ApJ...465..759G} and \cite{1997ApJ...482..913G}.}
\end{deluxetable*}
\end{turnpage}

\section{Observations}\label{sec:observations}
\subsection{SDSS Photometry}
The Sloan Digital Sky Survey \citep[SDSS;][]{2000AJ....120.1579Y,2002AJ....123..485S} employed a 2.5m telescope \citep{2006AJ....131.2332G} at Apache Point Observatory (APO) to conduct a photometric survey in the optical $ugriz$ filters \citep[][]{1996AJ....111.1748F,2007AJ....134..973I}.  The sky was imaged using a time-delayed integration technique. Great circles on the sky were scanned along six camera columns, each consisting of five 2048 $\times$ 2048 SITe$/$Tektronix CCDs with an exposure time of $\sim$54 seconds \citep{1998AJ....116.3040G}.  A custom photometric pipeline \citep[Photo;][]{2001ASPC..238..269L} was constructed to analyze each image and perform photometry.  Calibration onto a standard star network \citep{2002AJ....123.2121S} was accomplished using observations from the ``Photometric Telescope'' \citep[PT; ][]{2001AJ....122.2129H,2006AN....327..821T}.  Further discussion of PT calibrations for low--mass stars can be found in \cite{2007AJ....134.2430D}.  Absolute astrometric accuracy is better than 0.1\arcsec~\citep{2003AJ....125.1559P}.  Centered on the Northern Galactic Cap, the imaging data spans $\sim$ 10,000 square degrees, and is 95\% complete to $r \sim 22.2$  \citep{2002AJ....123..485S,2008ApJS..175..297A}.  When the North Galactic Pole was not visible from APO, $\sim$ 300 sq$.$ deg were scanned along the $\delta=0$ region known as ``Stripe 82'' to empirically quantify completeness and photometric precision \citep{2007AJ....134..973I}.  Over 357 million unique photometric objects have been identified in the latest public data release \citep[DR7,][]{2009ApJS..182..543A}.  The photometric precision of SDSS is unrivaled for a survey of this size, with typical errors $\lesssim$ 0.02 mag \citep{2004AN....325..583I,2007AJ....134..973I}. 

\subsection{Sample Selection}\label{chap5:sec:flags}
\begin{figure}[htbp]
\begin{center}$
\begin{array}{c} 
\includegraphics[scale=0.3]{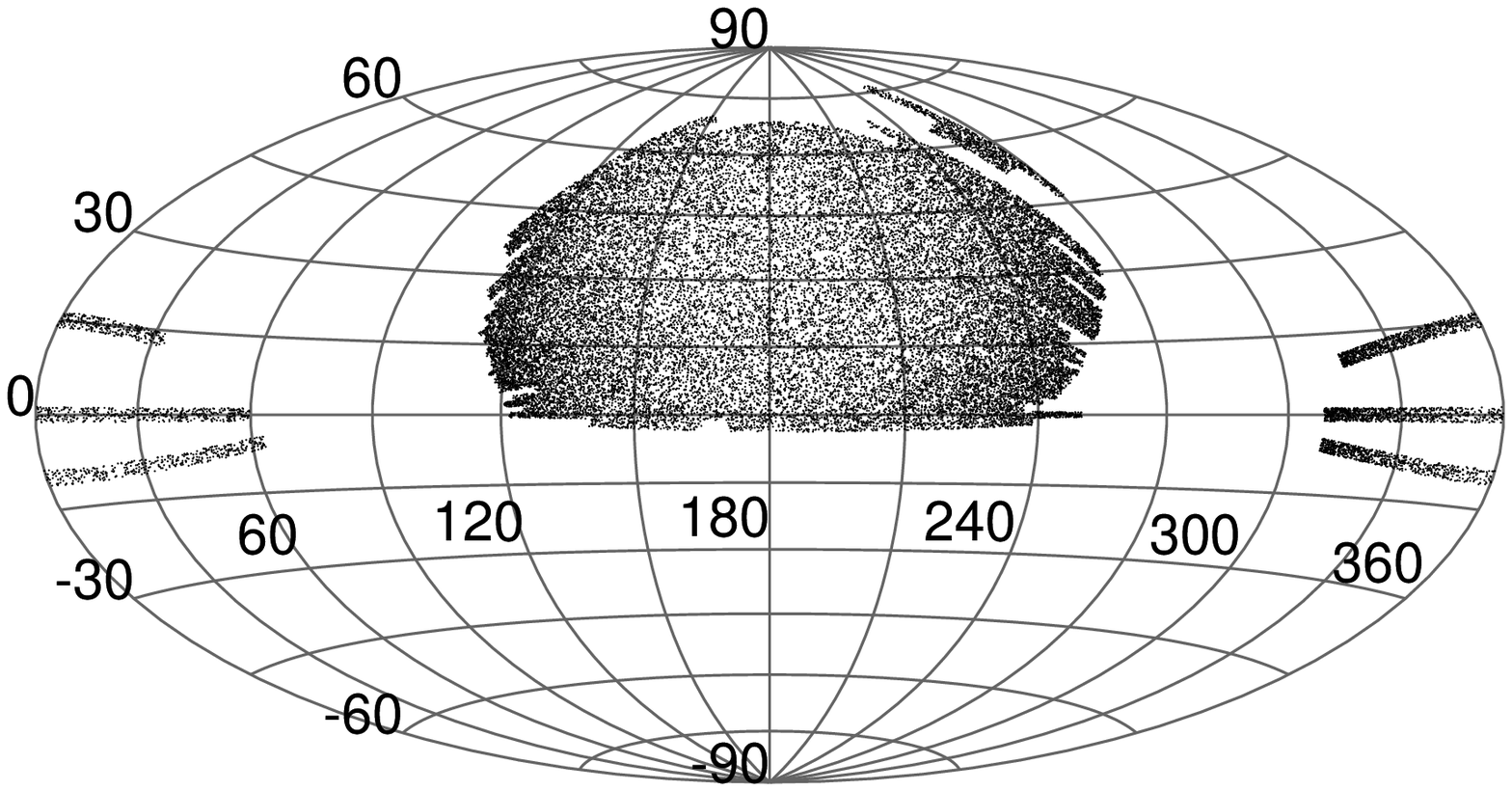} \\
\includegraphics[scale=0.3]{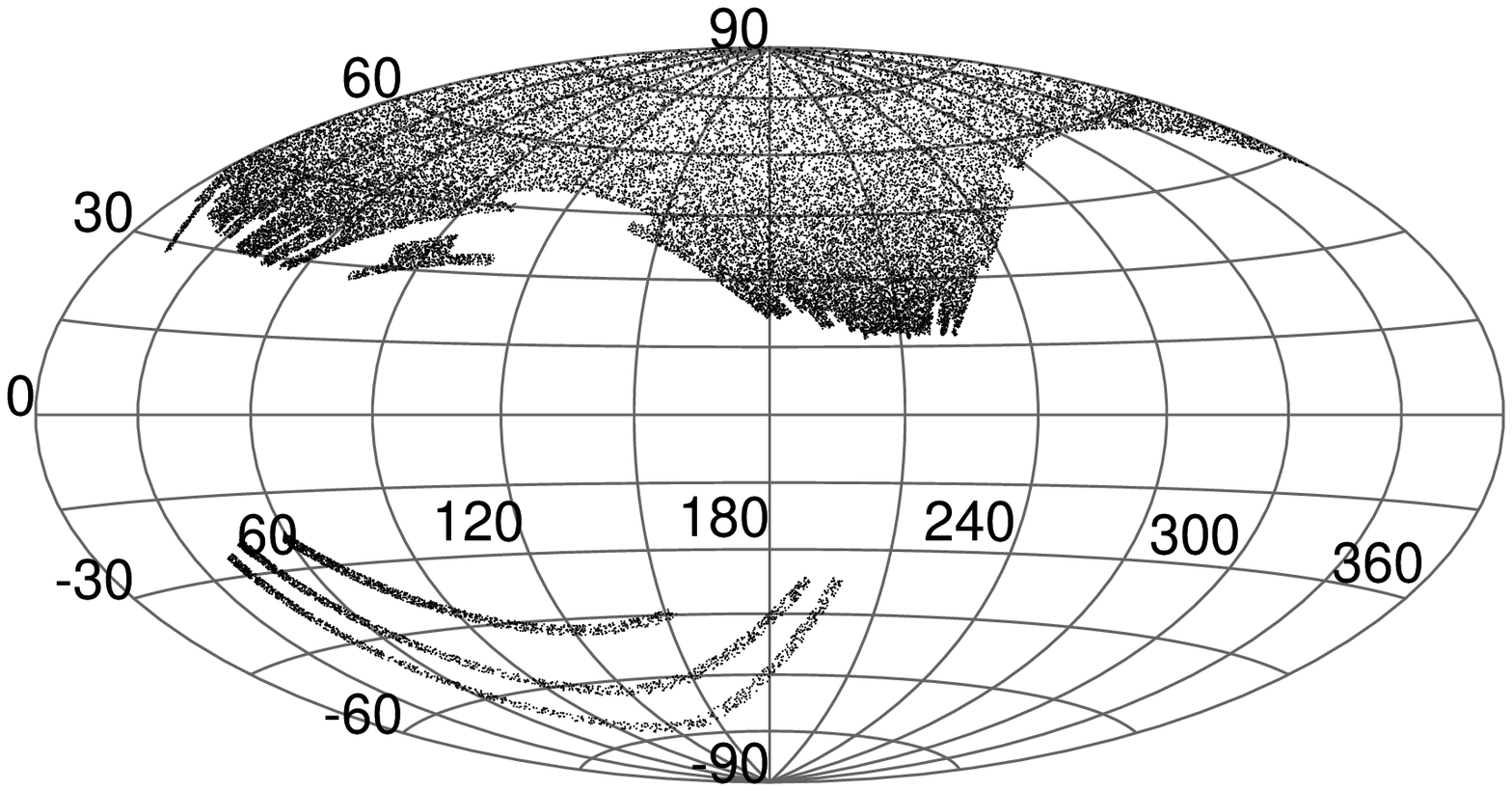}
\end{array}$
\end{center}
\caption{Aitoff projections of the SDSS DR6 Legacy footprint in equatorial (top panel) and Galactic (bottom panel) coordinates.  To aid figure clarity, only 0.2\% of the final sample is shown. }
\label{chap5:fig:map}
\end{figure}

We queried the SDSS catalog archive server (CAS) through the $casjobs$ website \citep{2005cs........2072O}\footnote{http://casjobs.sdss.org/CasJobs/} for point sources with the following criteria:
\begin{itemize}
 \item{The observations fell within the Data Release 6 - Legacy \citep[DR6;][]{2008ApJS..175..297A} footprint.  The equatorial and Galactic coordinate maps of the sample are shown in Figure \ref{chap5:fig:map}.}  
 \item{The photometric objects were flagged as \textsc{PRIMARY}.  This flag serves two purposes.  First, it implies that the \textsc{GOOD} flag has been set, where \textsc{GOOD $\equiv$  !BRIGHT AND (!BLENDED OR NODEBLEND OR N\_CHILD $=$ 0)}.  \textsc{BRIGHT} refers to duplicate detections of bright objects and the other set of flags ensures that stars were not deblended and counted twice. The \textsc{PRIMARY} flag indicates that objects imaged multiple times are only counted once 2\footnote{Note that ! indicates the NOT logical operator.}.}
\item{The object was classified morphologically as a star. \textsc{(TYPE = 6)}.}
\item{The photometric objects fell within the following brightness and color limits:\\
\begin{center}
$ i < 22.0, z < 21.2$\\
$r - i \geq 0.3, i -z \geq 0.2$\\
\end{center}
The first two cuts extend past the 95\% completeness limits of the survey \citep[$i < 21.3$, $z < 20.5$; ][]{2002AJ....123..485S}, but more conservative completeness cuts are enforced below.  The latter two cuts ensure that the stars have red colors typical of M dwarfs \citep{2007AJ....133..531B, 2007AJ....134.2398C, 2008AJ....135..785W}.}
\end{itemize}

This query produced 32,816,619 matches.  To ensure complete photometry, we required $16 < r < 22$.  These cuts conservatively account for the bright end of SDSS photometry, since the detectors saturate near 15$^{\rm th}$ magnitude \citep{2002AJ....123..485S}.  At the faint end, the  $r < 22$ limit is slightly brighter than the formal 95\% completness limits.  23,323,453 stars remain after these brightness cuts.

\begin{figure*}[htbp]
\begin{center}$
\begin{array}{cc}
\includegraphics[scale=0.3]{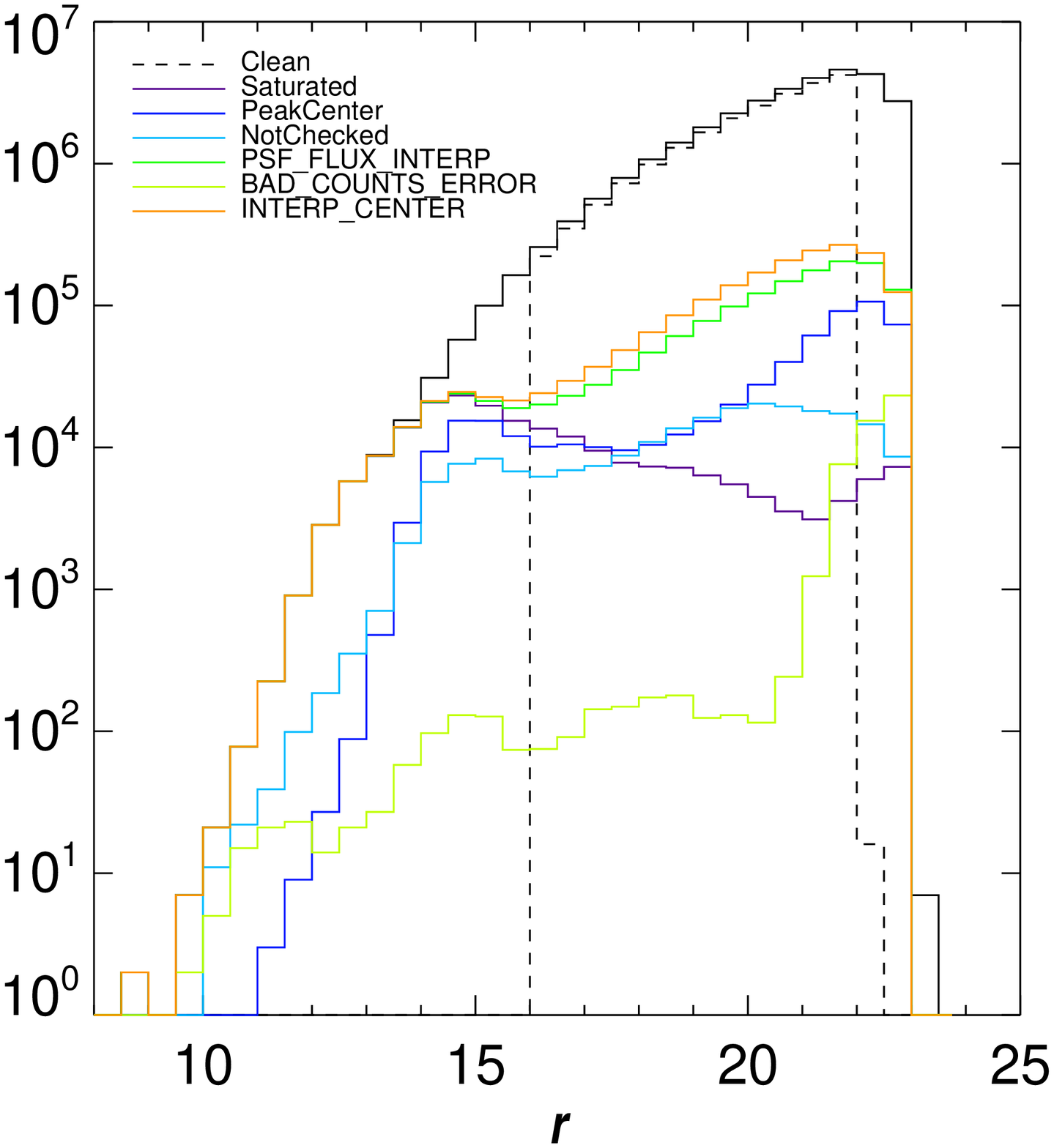} &
\includegraphics[scale=0.3]{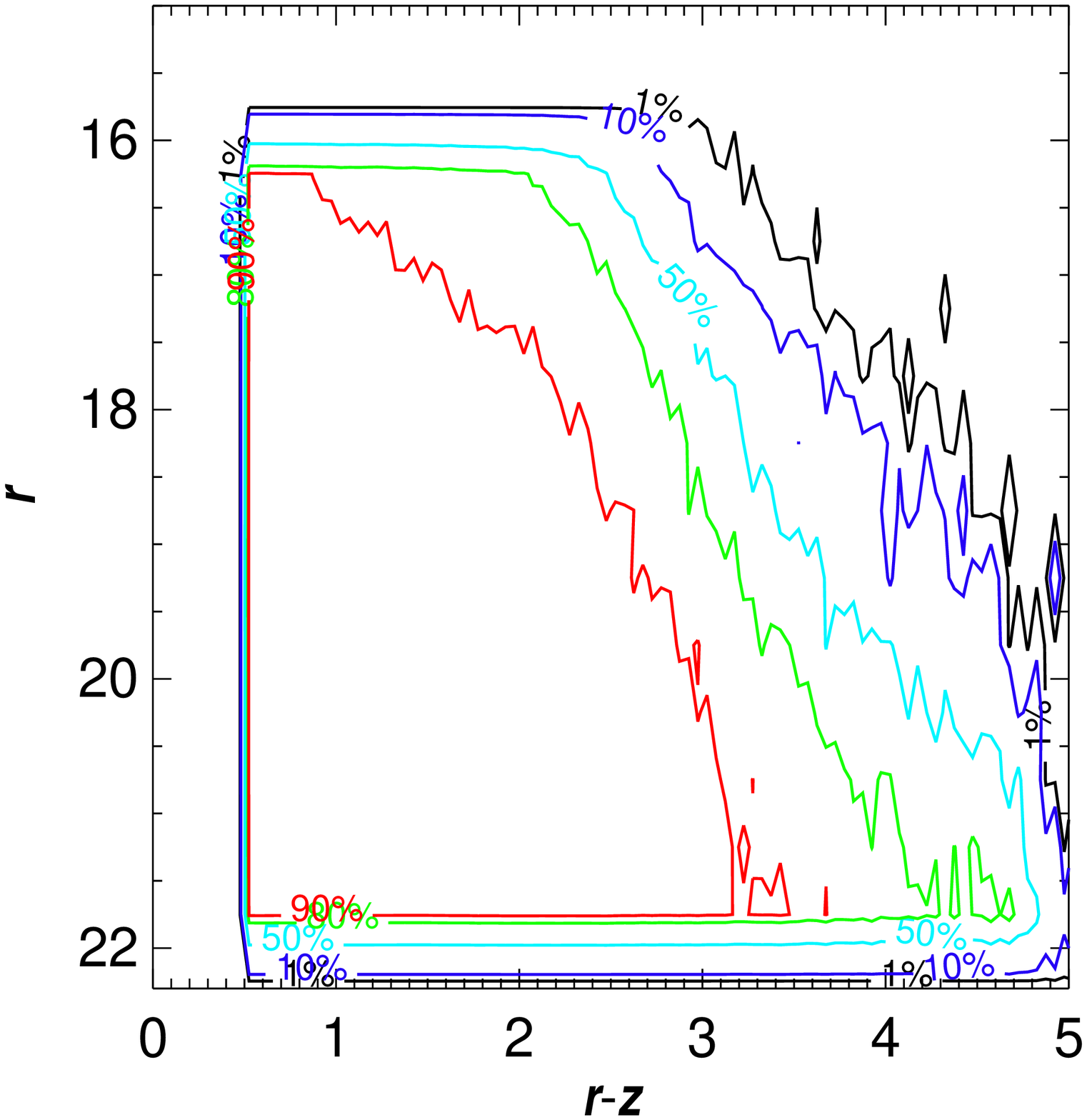} \\
\end{array}$
\end{center}
\caption{Left Panel: Histogram of $r$--band photometry (0.5 mag bins) showing the effects of flag cuts on the sample.  Each flag is labeled with a different color as noted in the legend.  The ``clean'' sample (dashed line) is complete from $16 < r < 22$.  Each flag is plotted separately, but many objects with faulty photometry have multiple flags set.  Right Panel:  Percentage of stars retained after the described flag cuts as a function of color and brightness.  Note that for the majority of the sample, over $90\%$ of the stars are being retained.  Each contour is labelled.}
 \label{chap5:fig:flag_cuts_histo}
\end{figure*}

\begin{figure}[htbp]
\centering
\includegraphics[scale=0.4]{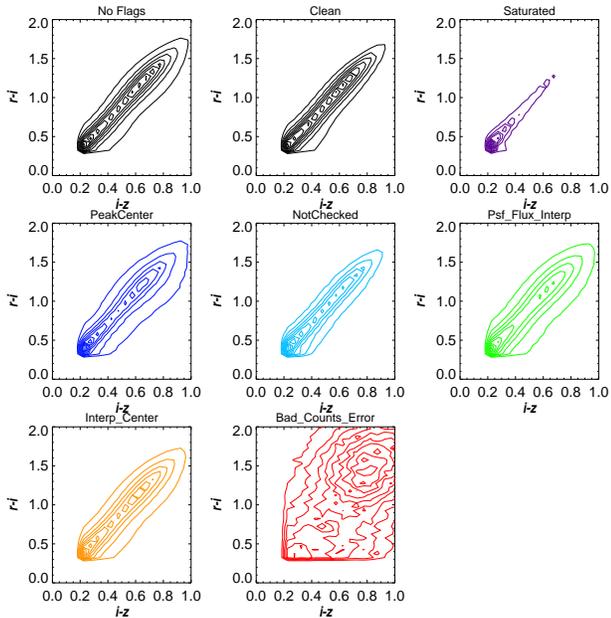}
\caption{The $r-i, i-z$ color-color diagrams for the various flag cuts discussed above.  The contours increase at 10\% intervals.  Of note are the relatively wide loci of the objects with the PSF\_FLUX\_INTERP, PEAK\_CENTER and INTERP\_CENTER flags set.  BAD\_COUNTS\_ERROR objects demonstrate considerable scatter.  SATURATED objects are mostly bluer, indicating that they are probably higher luminosity stars.}
\label{chap5:fig:flags_color_color}
\end{figure}

SDSS provides many photometric flags that assess the quality of each measurement.  These flags are described in detail by \cite{2002AJ....123..485S} and in the SDSS web documentation\footnote{ http://www.astro.princeton.edu/\~{}rhl/flags.html and\\ http://www.sdss.org/dr7/products/catalogs/flags\_detail.html provide excellent documentation of flag properties.}.  With the following series of flag cuts, the $\sim 23$ million photometric objects were cleaned to a complete, accurate sample.  Since only the $r,i$ and $z$ filters were used in this analysis, all of the following flags were only applied to those filters.  The $r$--band distribution of sources is shown in Figure \ref{chap5:fig:flag_cuts_histo}, along with the subset eliminated by each flag cut described below.  The color--color diagrams for each of these subsets are shown in Figure \ref{chap5:fig:flags_color_color}.

Saturated photometry was removed by selecting against objects with the SATURATED flag set.  As seen in Figure \ref{chap5:fig:flag_cuts_histo}, this cut removes mostly objects with $r < 15$.  However, there were some fainter stars within the footprint of bright, saturated stars.   These stars are also marked as SATURATED and not included in our sample. NOTCHECKED was used to further clean saturated stars from the photometry.  This flag marks areas on the sky where Photo did not search for local maxima, such as the cores of saturated stars.  Similarly, we eliminated sources with PEAKCENTER set, where the center of a photometric object is identified by the peak pixel, and not a more sophisticated centroiding algorithm.  As seen in Figure \ref{chap5:fig:flag_cuts_histo}, both of these flags composed a small fraction of the total number of observations and are more common near the bright and faint ends of the photometry.  Saturated objects and very low signal-to-noise observations will fail many of these tests.

The last set of flags examines the structure of the PSF after it has been measured.  
  The PSF\_FLUX\_INTERP flag is set when over 20\% of the star's PSF is interpolated.  While \cite{2002AJ....123..485S} claim that this procedure generally provides trustworthy photometry, they warn of cases where this may not be true.  Visual inspection of the ($r-i, i-z$) color-color diagram in Figure \ref{chap5:fig:flags_color_color} confirmed the latter, showing a wider locus than other flag cuts.  The INTERP\_CENTER flag is set when a pixel within three pixels of the center of a star is interpolated.  The $(r-i, i-z)$ color-color diagram of objects with INTERP\_CENTER set is also wide and the fit to the PSF could be significantly affected by an interpolated pixel near its center \citep{2002AJ....123..485S}.  Thus, stars with these flags set were removed.  Finally, BAD\_COUNTS\_ERROR is set when a significant fraction of the star's PSF is interpolated over and the photometric error estimates should not be trusted.  Table \ref{chap5:table:flag_cuts} lists the number of stars in the sample with each flag set.  For the final ``clean'' sample, we defined the following metaflag:

\begin{center}
clean = (!SATURATED$_{r,i,z}$ AND !PEAKCENTER$_{r,i,z}$ AND !NOTCHECKED$_{r,i,z}$ AND !PSF\_FLUX\_INTERP$_{r,i,z}$ AND !INTERP\_CENTER$_{r,i,z}$ AND !BAD\_COUNTS\_ERROR$_{r,i,z}$ AND ($16 <$ psfmag\_r $< 22$))
\end{center} 

After the flag cuts the stellar sample was composed of 21,418,445 stars.

The final cut applied to the stellar sample was based on distance.  As explained below in \S \ref{sec:density}, stellar densities were calculated within a $4 \times 4 \times 4$ kpc$^{3}$ cube centered on the Sun.  Thus, only stars within this volume were retained, and the final number of stars in the sample is 15,340,771\footnote{The reported number is based on the $(M_r, r-z)$ color--magnitude relation (CMR).  As explained in \S \ref{sec:correction}, changes to the CMR can add or subtract stars from the volume.}.
 
\begin{deluxetable*}{lll}
\tablewidth{0pt}
 
 \tablecaption{Flag Cuts in SDSS DR6 sample}
 \tabletypesize{\small}
 
 \tablehead{
 \colhead{Flag} &
 \colhead{Num. of Stars} &
 \colhead{Description} 
 }
 \startdata
SATURATED & 246,316  &  Pixel(s) saturated within the PSF  \\
PSF\_FLUX\_INTERP & 1,609,439  & $>20 \%$ of the PSF interpolated \\
INTERP\_CENTER & 1,993,063  & Interpolated pixel within 3 pixels of the center\\
BAD\_COUNTS\_ERROR  & 97,697  &  Significant interpolation, underestimated errors  \\
PEAKCENTER  & 598,108   & Center found by peak pixel, centroiding failed  \\
NOTCHECKED  & 230,375  &  Peak of PSF not examined, probably saturated  \\
``CLEAN''   & 21,418,445 passed   &  Stars that passed quality \& completeness cuts  \\
``CUBE''    & 15,340,771 passed & ``Clean'' stars within 4 kpc$^3$ cube \\
\enddata
\label{chap5:table:flag_cuts}
\end{deluxetable*}

\begin{figure*}[htbp]
\begin{center}$
\begin{array}{ccc}
\includegraphics[scale=0.3]{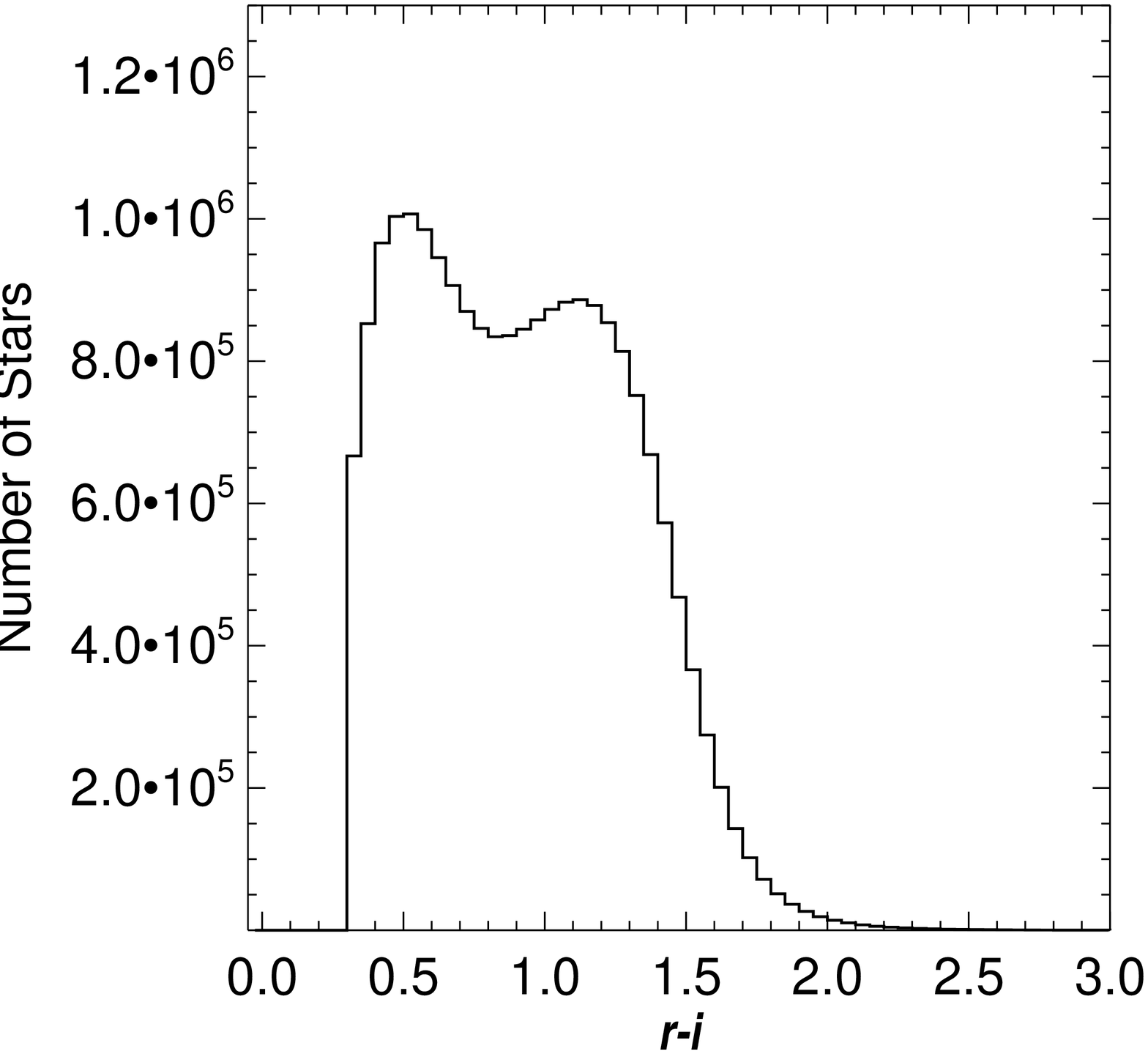}

\includegraphics[scale=0.3]{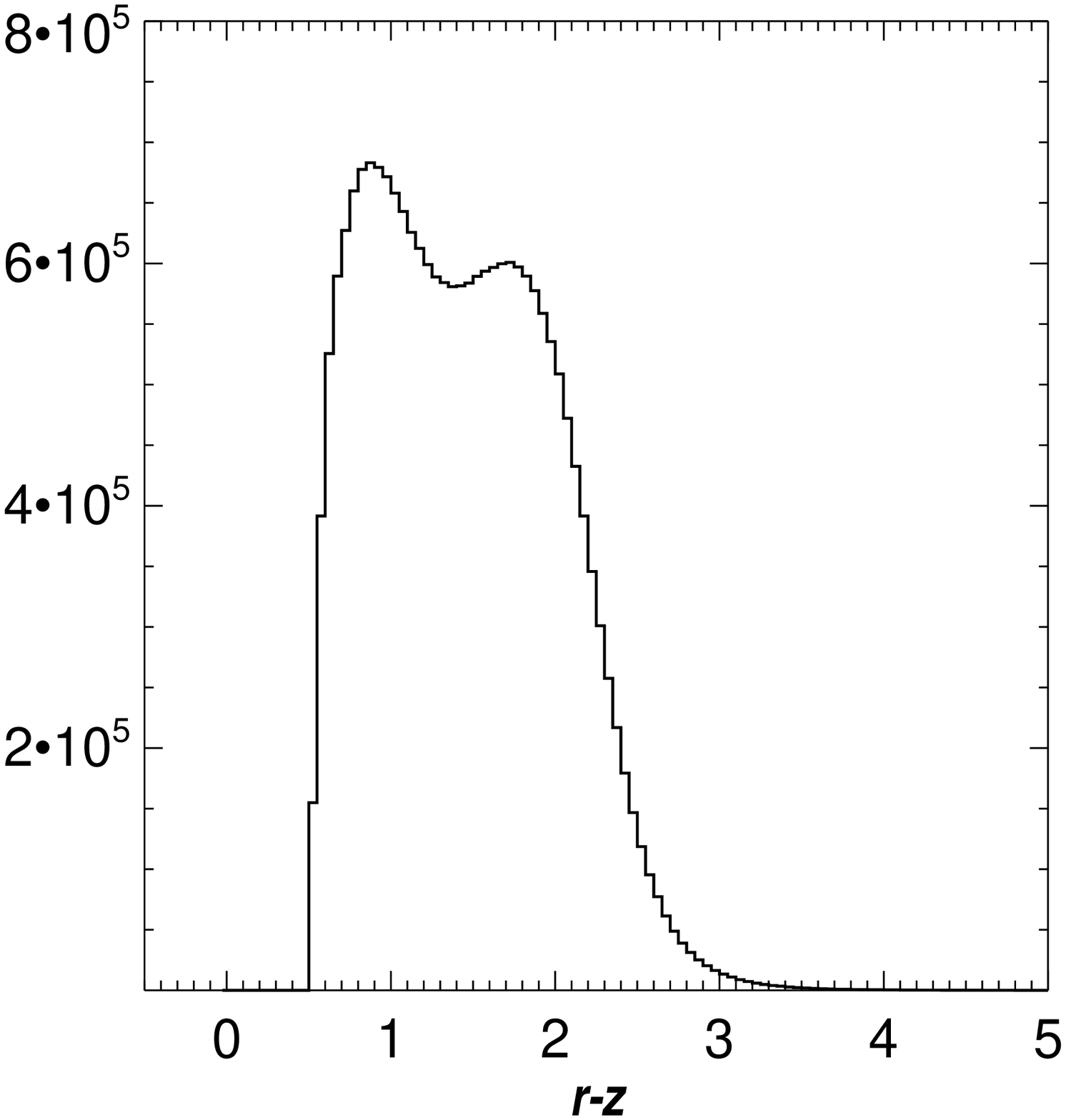}
\end{array}$
\end{center}
\caption{Histograms of color for the final stellar sample in $r-i$ and $r-z$. }
\label{chap5:fig:color_hists}
\end{figure*}

\begin{figure*}[htbp]
\begin{center}$
\begin{array}{cc}
\includegraphics[scale=0.3]{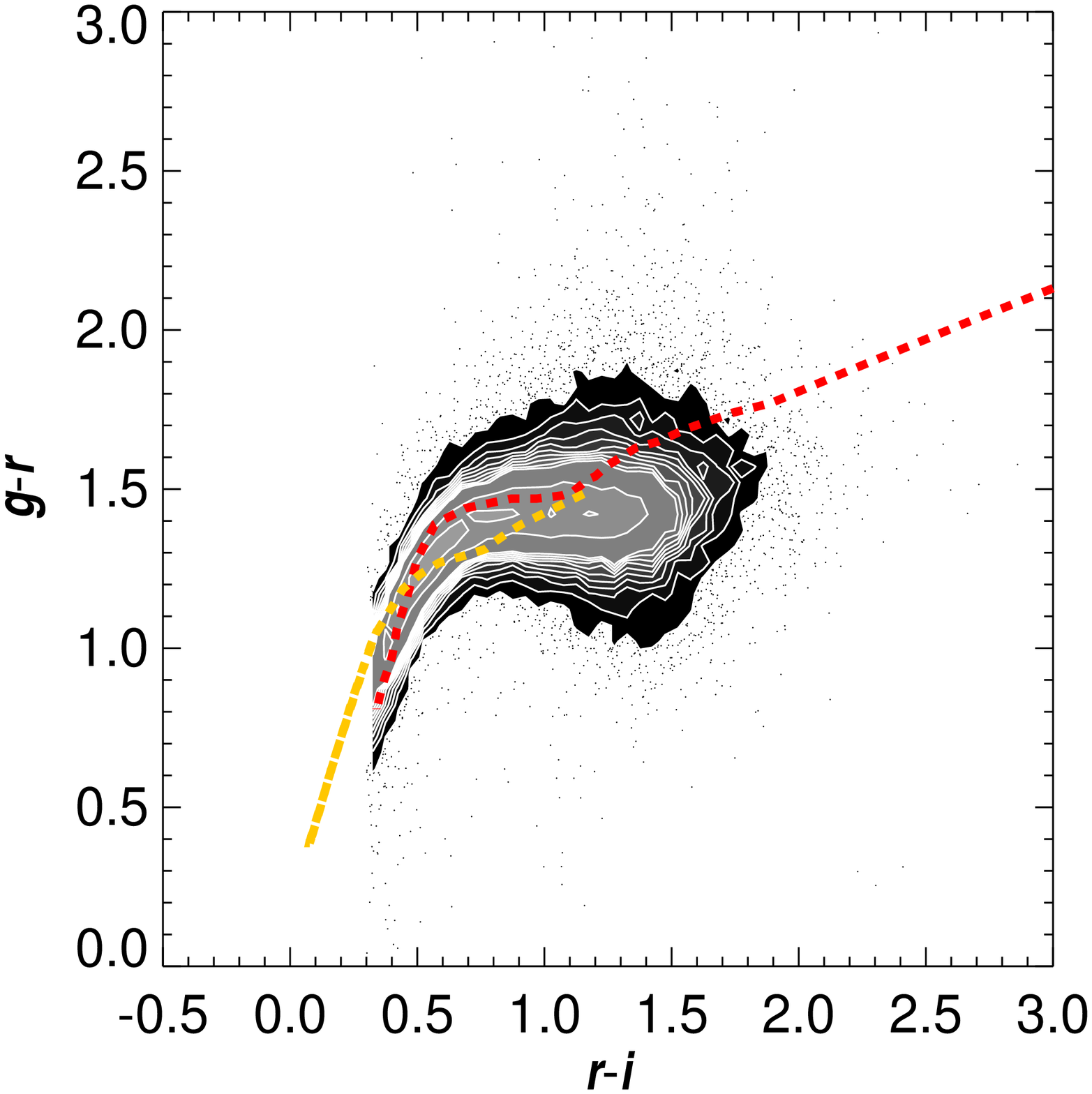} 
\includegraphics[scale=0.3]{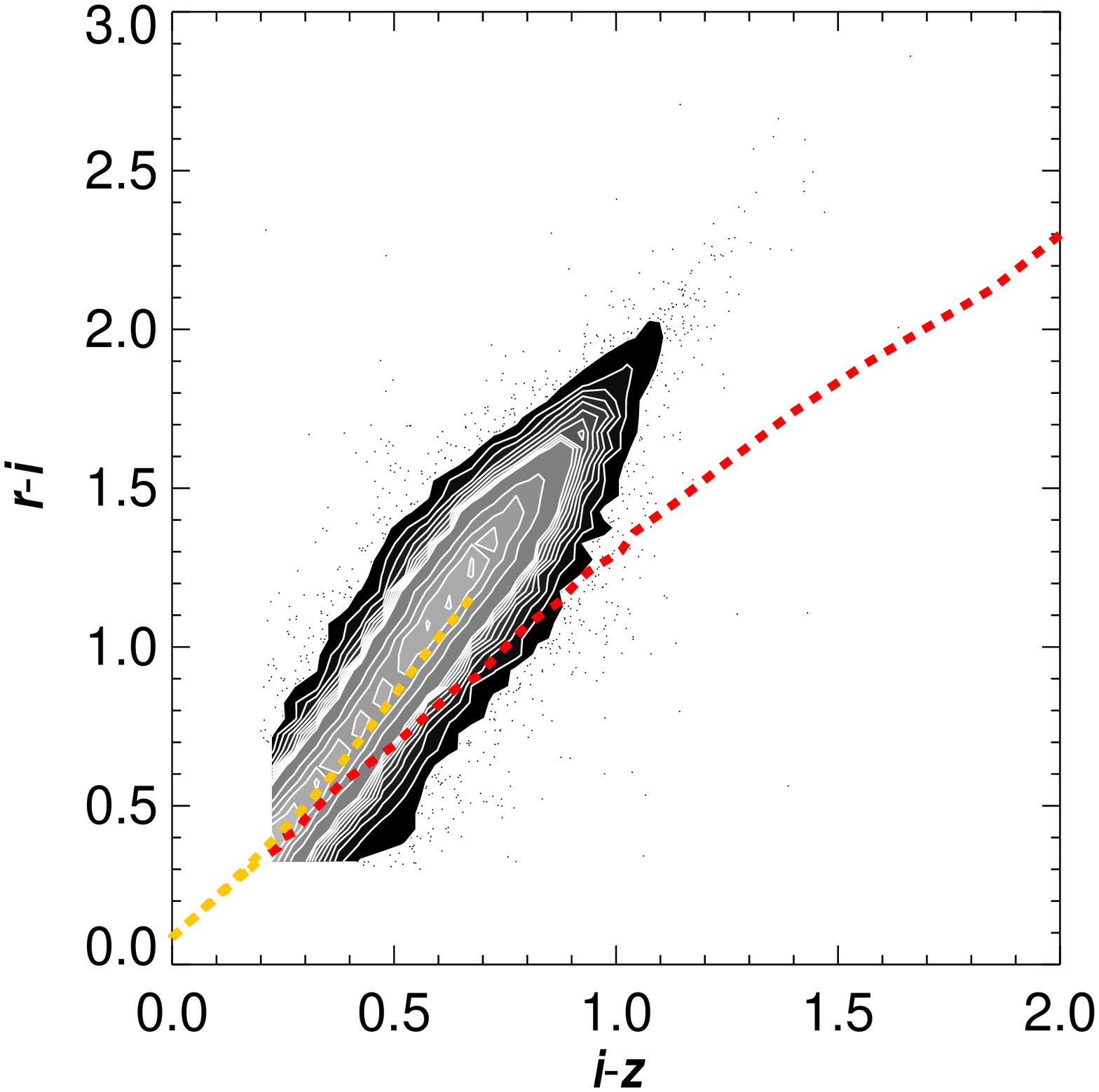}  
\end{array}$
\end{center}
\caption{Color-color diagrams of the final photometric sample with the 5 Gyr isochrones of \cite[][red dashed line]{1998A&A...337..403B}  and \cite[][yellow dashed line]{2004A&A...422..205G} overplotted. The contours represent 0.2\% of our entire sample, with contours increasing every 10 stars per 0.05 color-color bin.  Note that the model predictions fail by nearly one magnitude in some locations of the stellar locus.}
\label{chap5:fig:color_color_plots}
\end{figure*}

In Figure \ref{chap5:fig:color_hists}, histograms of the $r-i$, $i-z$ and $r-z$ colors are shown.  These color histograms map directly to absolute magnitude, since color--magnitude relations (CMRs) are used to estimate absolute magnitude and distance. The structure seen in the color histograms at $r-i\sim 1.5$ and $r-z \sim 2.2$ results from the convolution of the peak of the LF with the Galactic stellar density profile over the volume probed by SDSS.  Removing the density gradients and normalizing by the volume sampled constitutes the majority of the effort needed to convert these color histograms into a LF.  The $(g-r, r-i)$ and $(r-i, i-z)$ color--color diagrams are shown in Figure \ref{chap5:fig:color_color_plots}, along with the model predictions of \cite{1998A&A...337..403B} and \cite{2004A&A...422..205G}.  It is clearly evident that the models fail to reproduce the stellar locus, with discrepancies as large as $\sim 1$ mag.  These models should not be employed as color--absolute-magnitude relations for low--mass stars.

\subsection{Star--Galaxy Separation}\label{sec:star-gal}
\begin{figure}[htbp]
\centering
\includegraphics[angle=90,scale=0.3]{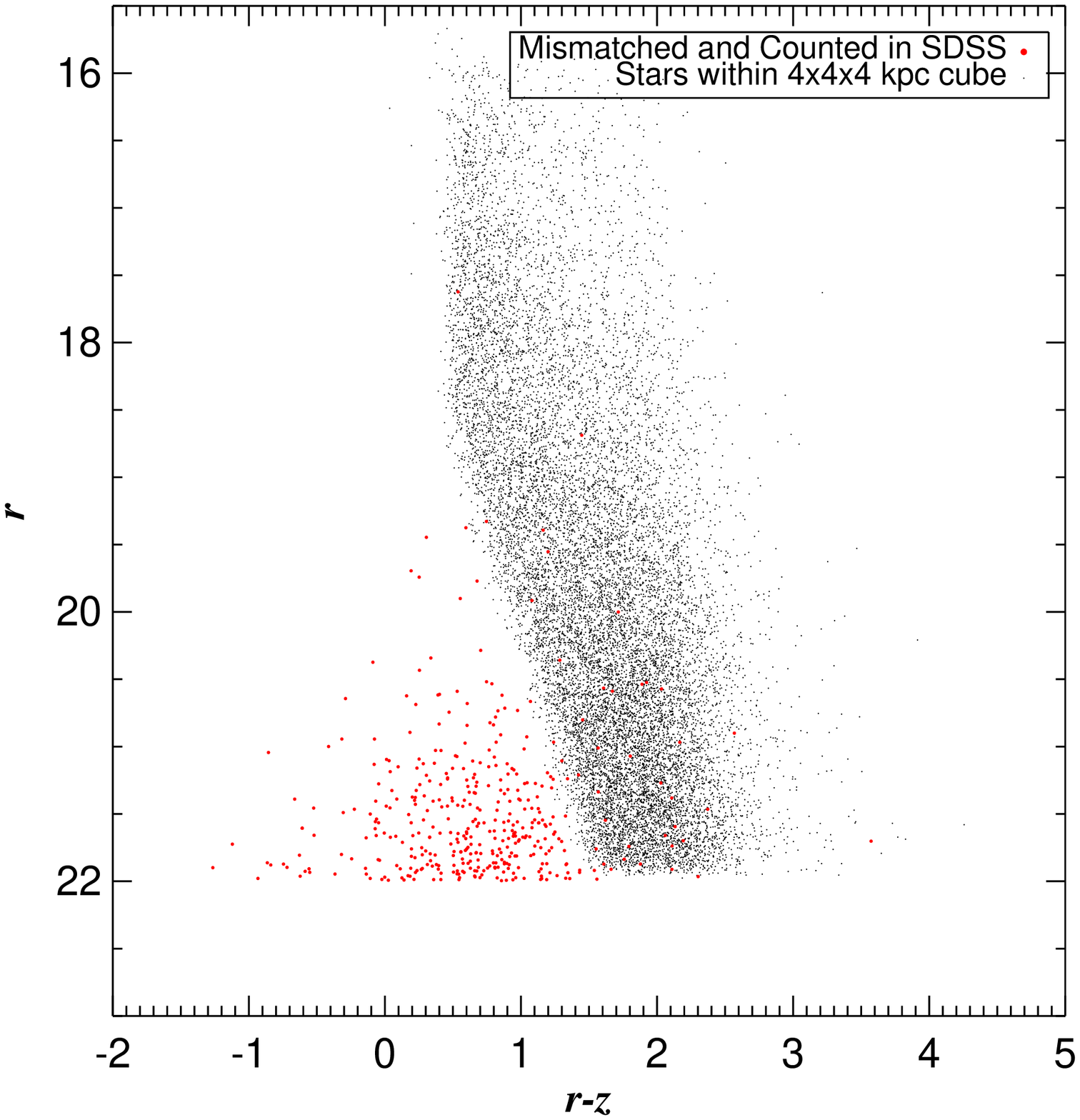} 
\caption{The Hess diagram for objects identified as stars in the SDSS pipeline, but as galaxies with high--resolution ACS imaging in the COSMOS footprint (red filled circles).   The black points show 0.02\% of the final stellar sample used in the present analysis.    Note that galaxy contamination is the most significant at faint, blue colors.  These colors and magnitudes are not probed by our analysis, since these objects lie beyond our $4\times4\times4$ kpc distance cut.}
\label{fig:stargal}
\end{figure}

With any deep photometric survey, accurate star--galaxy separation is a requisite for many astronomical investigations.   At faint magnitudes, galaxies far outnumber stars, especially at the Galactic latitudes covered by SDSS. Star--galaxy identification is done automatically in the SDSS pipeline, based on the brightness and morphology of a given source.  \cite{2001ASPC..238..269L} investigated the fidelity of this process, using overlap between HST observations and early SDSS photometry.  They showed that star--galaxy separation is accurate for more than  $> 95\%$ of objects to a magnitude limit of $r \sim 21.5$.  Since the present sample extends to $r = 22$, we re-investigated the star--galaxy separation efficiency of the SDSS pipeline.  We matched the SDSS pipeline photometry to the Hubble Space Telescope ACS images within the COSMOS \citep{2007ApJS..172....1S} footprint.  The details of this analysis will be published in a later paper (Bochanski et al., 2010, in prep).  In Figure \ref{fig:stargal}, we plot the colors and brightnesses of COSMOS galaxies identified as stars by the SDSS pipeline (red filled circles), along with a representative subsample of 0.02\% of the stars in our sample.  This figure demonstrates that for the majority of the stars in the present analysis, the SDSS morphological indentifications are adequate and contamination by galaxies is not a major systematic.   

\section{Calibration: Photometric Parallax}\label{sec:phot_pi}
Accurate absolute magnitude estimates are necessary to measure the stellar field LF.  Trigonometric parallaxes, such as those measured by Hipparcos \citep{1997yCat.1239....0E,2007A&A...474..653V},  offer the  most direct method for calculating absolute magnitude.  Unfortunately, trigonometric parallaxes are not available for many faint stars, including the overwhelming majority of the low--mass dwarfs observed by SDSS.  Thus, other methods must be employed to estimate a star's absolute magnitude (and distance).  Two common techniques, known as photometric parallax and  spectroscopic parallax, use a star's color or spectral type, respectively.  These methods are calibrated by sources with known absolute magnitudes (nearby trigonometric parallax stars, clusters, etc.) and mathematical relations are fit to their color (or spectral type) -- absolute magnitude locus.  Thus, the color of a star can be used to estimate its absolute magnitude, and in turn, its distance, by the well-known distance modulus ($m - M$):

\begin{large}
\begin{equation}
m_{\lambda,1} - M_{\lambda,1}(\small{m_{\lambda,1} - m_{\lambda,2}}) = 5 {\rm log}~ d - 5
\label{chap4:eqn:distance_mod}
\end{equation}
\end{large}

\noindent where $d$ is the distance, $m_{\lambda,1}$ is the apparent magnitude in one filter, and $m_{\lambda,1} - m_{\lambda,2}$ is the color from two filters, which is used to calculate the absolute magnitude, $M_{\lambda,1}$.   

\begin{figure}[htbp]
  \centering
  \includegraphics[scale=0.45]{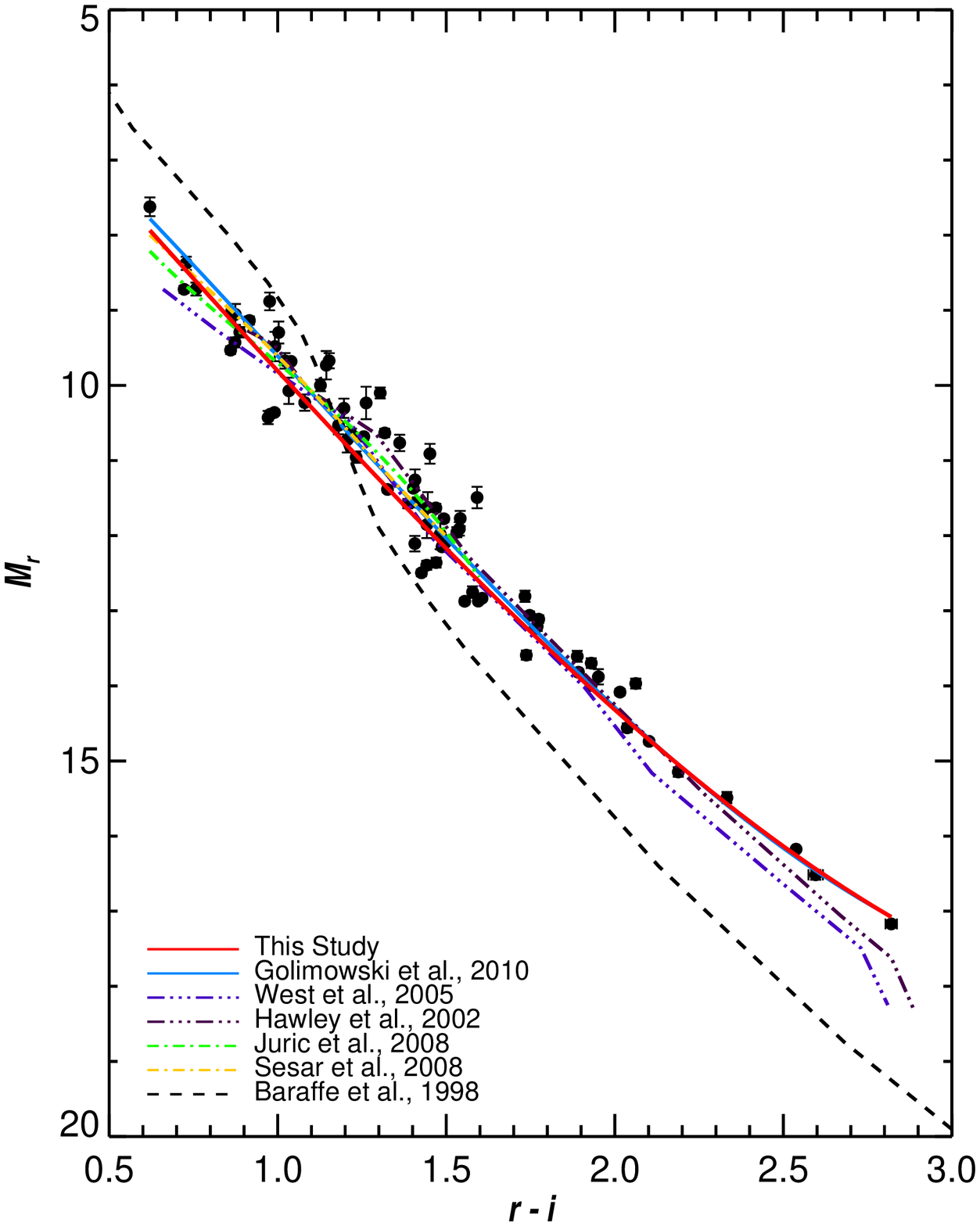}
  \caption{$M_r$ vs. $r - i $ CMD. The parallax stars from the nearby star sample are shown as filled circles and the best fit line from Table \ref{table:cmrs} is the solid red line.  Other existing parallax relations are plotted for comparison: \citet[purple dash-dot line]{2005PASP..117..706W},  \citet[their ``bright'' relation; green dash-dot line]{2008ApJ...673..864J}, \citet[yellow dash-dot line]{Sesar08}, \citet[solid blue line]{Golimowski08}.  The original \cite{2005PASP..117..706W} relations have been transformed using the data from their Table 1.  In addition, the 5 Gyr isochrone from the  \cite{1998A&A...337..403B} models appears as the dashed line.}
  \label{chap4:fig:photpis}
\end{figure}

\begin{figure}[htbp]
  \centering
  \includegraphics[scale=0.45]{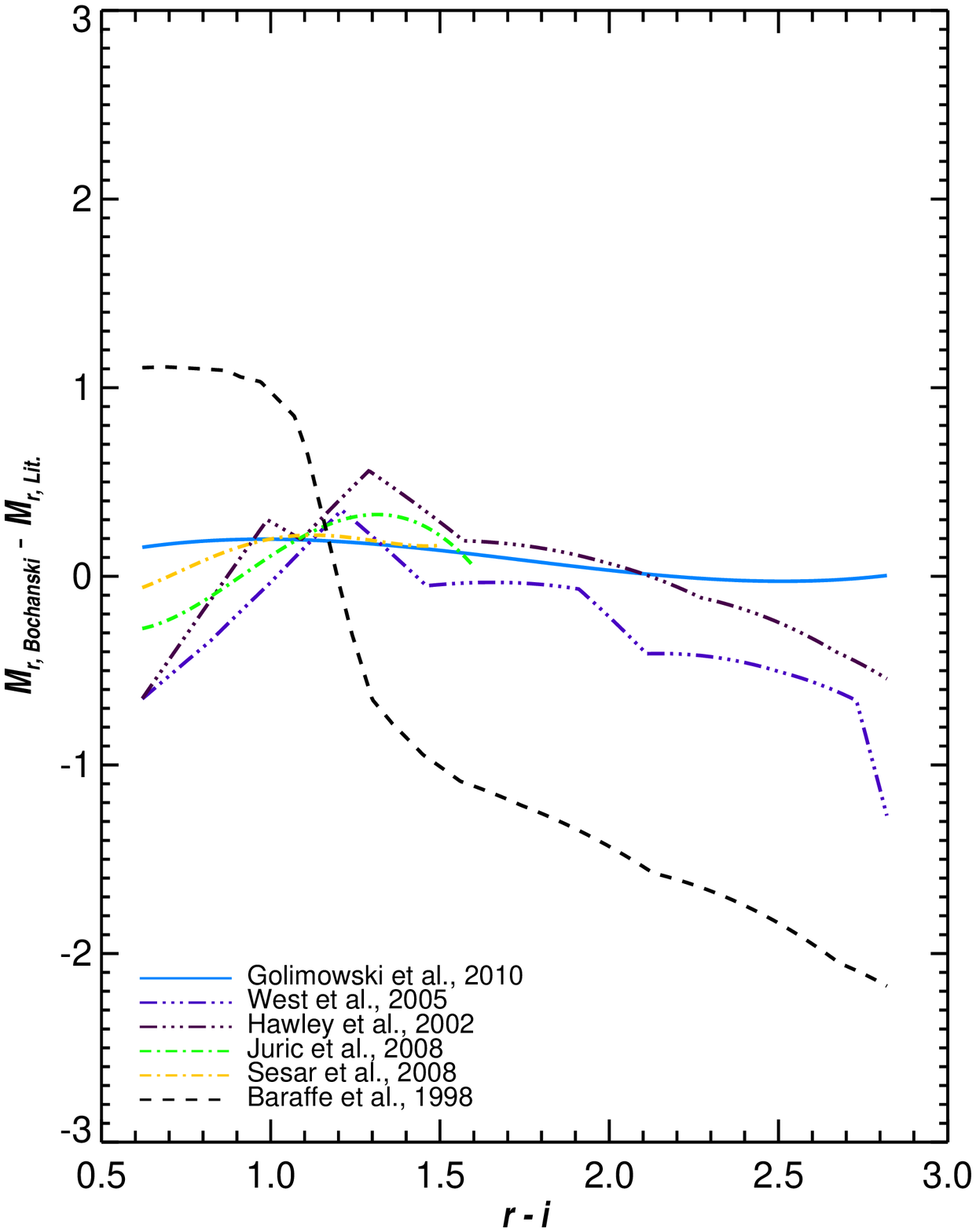}
  \caption{Differences between the $M_r$ vs. $r-i$ relations in Figure \ref{chap4:fig:photpis}.  The line styles are the same as Figure \ref{chap4:fig:photpis}.  Note that the observed photometric parallax relations generally agree to $\sim 0.5$ mags, while the 5 Gyr isochrone of \cite{1998A&A...337..403B} does not agree with the empirical results.}
  \label{chap4:fig:r_vs_ri_comp}
\end{figure}

There have been multiple photometric parallax relations\footnote{Photometric parallax relations are often referred to as ``color-magnitude relations'' (CMRs).  We use both names interchangeably throughout this manuscript.}, as shown in Figure \ref{chap4:fig:photpis}, constructed for low--mass stars observed by SDSS \citep{2002AJ....123.3409H,2002AAS...20111511W,2005PASP..117..706W,2008ApJ...673..864J,Sesar08,Golimowski08}.  There is a spread among the relations, seen in Figure \ref{chap4:fig:r_vs_ri_comp}, which are valid over different color ranges.  Additional photometry in $ugrizJHK_s$ of a large sample of nearby stars with well--measured trigonometric parallaxes is required to provide a reliable relation.  Fortunately, an observing program led by \cite{Golimowski08} acquired such observations and they kindly provided their data prior to publication.  The resulting CMRs are used to estimate the absolute magnitude and distance to all the stars in our sample, as described below.

\subsection{Photometric Telescope  Photometry}
The nearby star survey \citep{Golimowski08} targeted stars with the colors of low--mass dwarfs and precise trigonometric parallaxes.  The majority of targets were drawn from the Research Consortium on Nearby Stars (RECONS) catalog \citep[e.g.,][]{1994AJ....108.1437H,1995AJ....109..797K,2004AJ....128.2460H}.  Most of the stars selected from the RECONS sample are within 10 pc, with good parallactic precision ($\sigma_\pi / \pi \lesssim 0.1$).  In addition to RECONS targets, the nearby sample included K dwarfs from the \cite{1979lccs.book.....L} and \cite{1971lpms.book.....G} proper motion surveys.  Parallax measurements for these additional stars were obtained from the Hipparcos \citep{1997yCat.1239....0E} or General Catalogue of Trigonometric Stellar Parallaxes \citep[the ``Yale'' catalog; ][]{1995gcts.book.....V} surveys.

Near-infrared $JHK_s$ photometry was obtained from the 2MASS Point Source Catalog \citep{2003yCat.2246....0C}.  Acquiring $ugriz$ photometry proved more problematic.  Since typical SDSS photometry saturates near $r \sim$ 15, most of the nearby stars were too bright to be directly imaged with the 2.5m telescope. Instead, the 0.5m Photometric Telescope (PT) was used to obtain $(ugriz)^\prime$ photometry\footnote{$(ugriz)^\prime$ refers to $u^{\prime}g^{\prime}r^{\prime}i^{\prime}z^{\prime}$ photometry, which is defined by the standard stars of \cite{2002AJ....123.2121S} observed by the USNO 1m telescope.} of these stars.  The PT was active every night the 2.5m telescope was used in imaging mode during the SDSS, observing patches of the nightly footprint to determine the photometric solution for the night and to calibrate the zero-point of the 2.5m observations \citep{2002AJ....123.2121S,2006AN....327..821T}.  \cite{Golimowski08} obtained $(ugriz)^\prime$ photometry of the parallax sample over 20 nights for 268 low--mass stars. The transformations of \cite{2006AN....327..821T} and the \cite{2007AJ....134.2430D} corrections were applied to the nearby star photometry to transform the ``primed'' PT photometry to the native ``unprimed'' 2.5m system (see  \cite{2007AJ....134.2430D} for more details).  

To produce a reliable photometric parallax relation, the following criteria were imposed on the sample.  First, stars with large photometric errors ($\sigma > 0.1$ mags) in the $griz$ bands were removed.  Next, high signal--to--noise 2MASS photometry was selected, by choosing stars with their ph\_qual flag equal to `AAA'.  This flag corresponds to a signal--to--noise ratio $> 10$ and photometric uncertainties $<$ 0.1 mags in the $JHK_s$ bands. Next, a limit on parallactic accuracy of $\sigma_\pi / \pi < 0.10$ was enforced.   It ensured that the bias introduced by a parallax--limited sample, described by \cite{1973PASP...85..573L}, is minimized.  Since many of the stars in the nearby star sample have precise parallaxes ( $\sigma_\pi / \pi < 0.04$), the Lutz-Kelker correction is essentially negligible \citep[$ < -0.05$;][]{1979MNRAS.186..875H}.  Finally, contaminants such as known subdwarfs, known binaries, suspected flares, or white dwarfs were culled from the nearby star sample.

\subsection{Additional Photometry}
To augment the original PT observations, we searched the literature for other low--mass stars with accurate parallaxes and $ugriz$ and $JHK_s$ photometry.  The studies of \cite{2002AJ....124.1170D} and \cite{2004AJ....127.2948V} supplemented the original sample and provided accurate parallaxes ($\sigma_\pi / \pi \lesssim 0.1$) of late M and L dwarfs.  Several of those stars were observed with the SDSS 2.5m telescope, obviating the need for transformations between the primed and unprimed $ugriz$ systems.  Six late M and L dwarfs were added from these catalogs, extending the parallax sample in color from  $r-i \sim 2.5$ to $r-i \sim 3.0$ and in $M_r$ from 16 to 20.   Our final sample is given in \cite{bochanskithesis}.

\subsection{Color Magnitude Relations}
Multiple color--absolute magnitude diagrams (CMDs) in the $ugriz$ and $JHK_s$ bandpasses were constructed using the photometry and parallaxes described above.  The CMDs were individually inspected, fitting the main sequence with linear, 2nd, 3rd, and 4th--order polynomials.  Piecewise functions were also tested, placing discontinuities by eye along the main sequence. There is extensive discussion in the literature of a ``break'' in the main sequence near spectral type M4 \citep[or $V-I \sim 2.8$, see][]{1996AJ....112.2799H, 1997AJ....113.2246R, 2002AJ....123.2806R,2005nlds.book.....R}.   Certain colors, such as $V-I$ show evidence of a break \citep[Fig. 10 of][]{2002AJ....123.2806R}, while other colors, such as $V-K$, do not \citep[Fig. 9 of][]{2002AJ....123.2806R}.  We did not enforce a break in our fits.  Finally, the rms scatter about the fit for each CMD was computed and the relation that produced the smallest scatter for each color-absolute magnitude combination was retained.  Note that the rms scatter was dominated by the intrinsic width of the main sequence, as both our photometry and distances have small ($\sim$ 2\%) uncertainties.

\begin{figure}[htbp]
    \centering
       \includegraphics[scale=0.45]{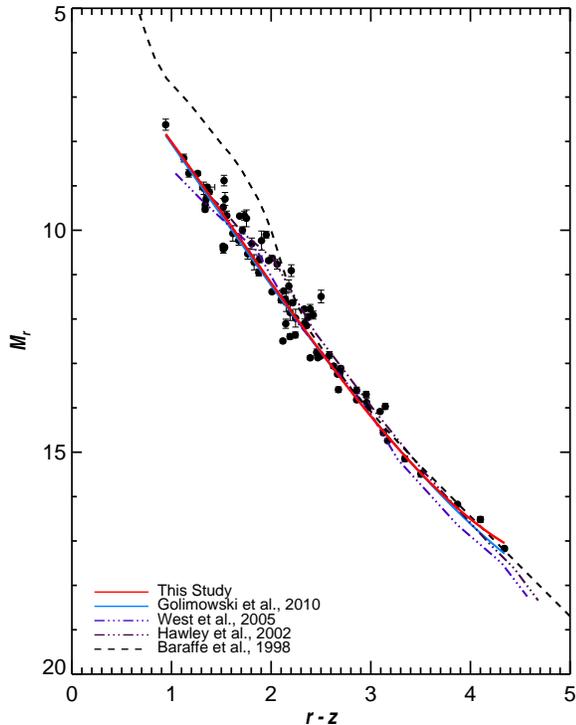}
    \caption{$M_r$ vs. $r - z $ CMD. Symbols and lines are the same as Figure \ref{chap4:fig:photpis}. }
\label{chap4:fig:r_vs_rz}
\end{figure}

\begin{figure}[htbp]
    \centering
       \includegraphics[scale=0.45]{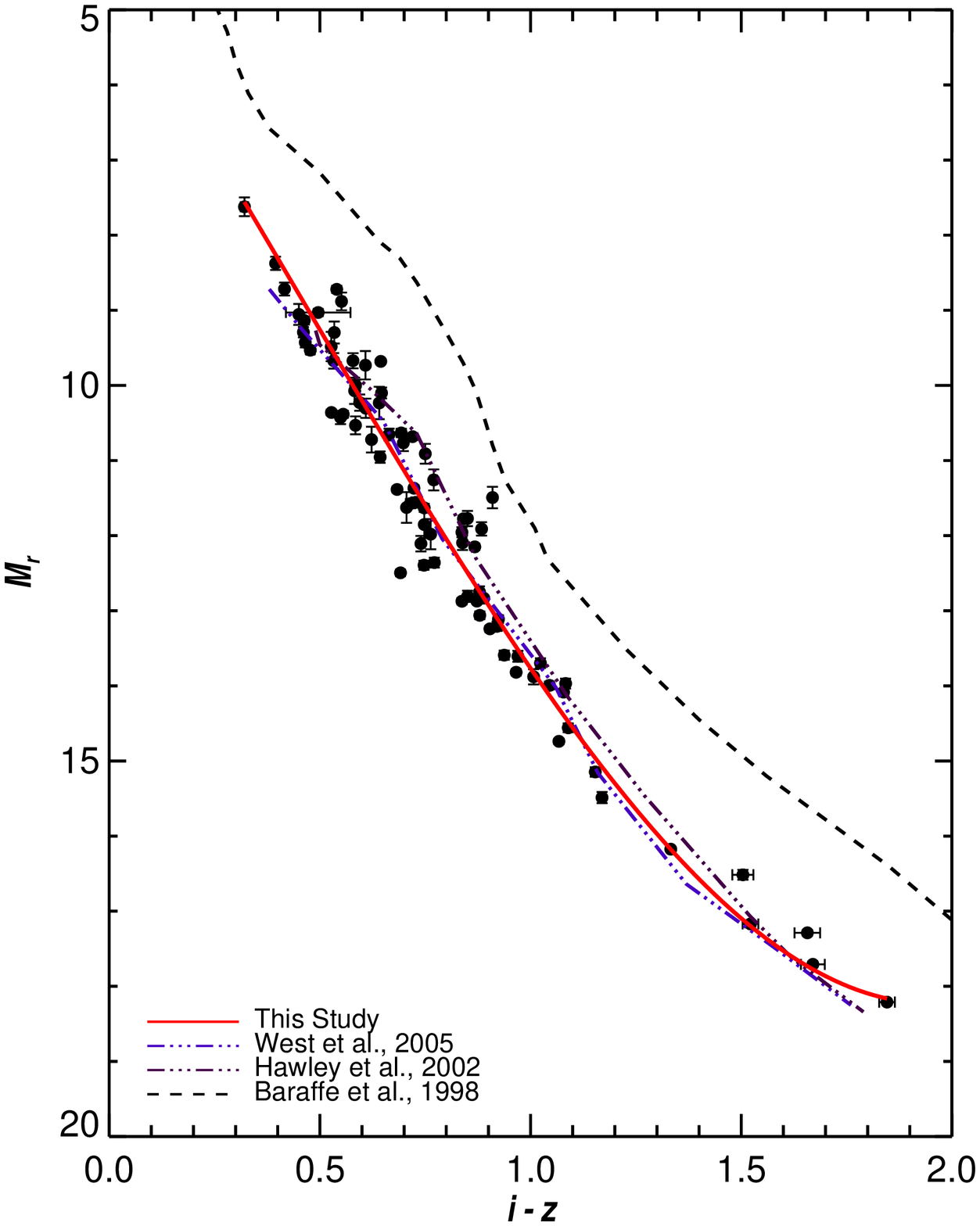}
    \caption{ $M_r$ vs. $i - z$ CMD. Same symbol definitions as Figures \ref{chap4:fig:photpis} \& \ref{chap4:fig:r_vs_rz}. Note the very poor agreement between the observations and model predictions.}
\label{chap4:fig:r_vs_iz}
\end{figure}

We present three different CMRs: ($M_r, r-i$), ($M_r, r-z$) and ($M_r, i-z$) in Table \ref{table:cmrs}.  $M_r$ was used for absolute magnitude, as it contains significant flux in all late-type stars.   The $r-z$ color has the longest wavelength baseline and small residual rms scatter ($\sigma \lesssim 0.40$ mag).  Other long baseline colors ($g-r$, $g-z$) are metallicity sensitive \citep[][]{2004AJ....128..426W, 2008ApJ...681L..33L} but most of our sample does not have reliable $g$--band photometry.  The adopted photometric parallax relations in these colors did not include any discontinuities, although we note a slight increase in the dispersion of the main sequence around $M_r \sim$ 12.  The final fits are shown in Figures \ref{chap4:fig:photpis},  \ref{chap4:fig:r_vs_rz} and \ref{chap4:fig:r_vs_iz}, along with other published photometric parallax relations in the $ugriz$ system.  

\begin{deluxetable*}{crrl}
\centering
\tabletypesize{\scriptsize}
\tablecaption{Color--Absolute Magnitude Relations in the $ugriz$ system} 
 \tablewidth{0pt}
 \tablehead{
 \colhead{Abs. Mag.} &
 \colhead{Color range} &
 \colhead{Best Fit} &
 \colhead{$\sigma_{M_r}$}
 }
 \startdata
 $M_r$ & $0.50 < r-z < 4.53$ & 5.190 + 2.474 ($r-z$) + 0.4340 ($r-z)^2$ - 0.08635 ($r-z)^3$ & 0.394 \\
 $M_r$ & $0.62 < r-i < 2.82$ & 5.025 - 4.548 ($r-i$) + 0.4175 ($r-i)^2$ - 0.18315 ($r-i)^3$ & 0.403 \\
 $M_r$ & $0.32 < i-z < 1.85$ & 4.748 + 8.275 ($i-z$) + 2.2789 ($i-z)^2$ - 1.5337 ($i-z)^3$  & 0.481 \\
 \enddata
 \label{table:cmrs}
\end{deluxetable*}

\section{Analysis}\label{sec:analysis}
Our photometric sample comprises a data--set three orders of magnitude larger (in number) than any previous LF study (see Table \ref{chap5:table:existing_studies}).  Furthermore, it is spread over 8,400 sq$.$ deg$.$, nearly 300 times larger than the sample analyzed by \cite{covey08}.  This large sky coverage represents the main challenge in measuring the LF from this sample.  Most of the previous studies in Table \ref{chap5:table:existing_studies} either assumed a uniform density distribution (for nearby stars) or calculated a Galactic density profile, $\rho(r)$ along one line of sight.  With millions of stars spread over nearly 1/4 of the sky, numerically integrating Galactic density profiles for each star is computationally prohibitive.   

To address this issue, we introduced the following technique for measuring the luminosity function.  First, absolute magnitudes were assigned and distances to each star were computed using the $r-z$ and $r-i$ CMRs from Table \ref{table:cmrs}.  Each CMR was processed separately.  Next, a small range in absolute magnitude (0.5 mag) was selected and the stellar density was measured \textit{in situ} as a function of Galactic radius (\textit{R}) and Galactic height (\textit{Z}).  This range in absolute magnitude was selected to provide high resolution in the LF, while maintaining a large number of stars ($\sim 10^6$) in each bin.   Finally, a Galactic profile was fit to the $R,Z$ density maps, solving for the shape of the thin and thick disks, as well as the local density.  The LF was then constructed by combining the local density of each absolute magnitude slice.

\subsection{Stellar Density Maps}\label{sec:density}
To assemble a $(R,Z)$ density map, an accurate count of the number of stars in a given $R,Z$ bin, as well as the volume spanned by each bin, was required.  A cylindrical ($R,Z,\phi$) coordinate system was taken as the natural coordinates of stellar density in the Milky Way.  In this frame, the Sun's position was set at $R_\odot$ = 8.5 kpc \citep{1986MNRAS.221.1023K} and $Z_\odot = 15$ pc above the Plane \citep{1995ApJ...444..874C,1997A&A...324...65N,1997MNRAS.288..365B}.  Azimuthal symmetry was assumed (and was recently verified by \cite{2008ApJ...673..864J} and found to be appropriate for the local Galaxy).   The following analysis was carried out in $R$ and $Z$.  We stress that we are not presenting any information on $\phi = 0$ plane.  Rather, the density maps are summed over $\phi$, collapsing the three-dimensional SDSS volume into a two--dimensional density map.

\begin{figure}[!htbp]
\centering
\includegraphics[scale=0.4]{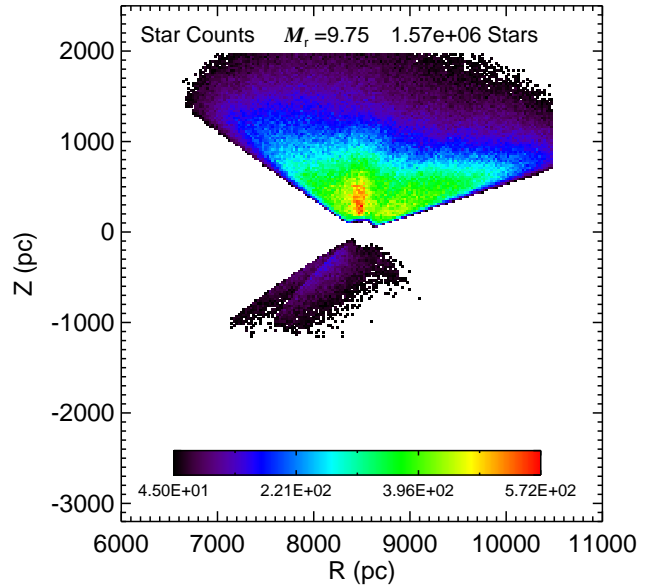} 
\caption{Star counts as a function of Galactic $R$ and $Z$ for a 0.5 magnitude slice in absolute magnitude centered on $M_r = 9.75$.  The color bar on the lower part of the plot displays the scale of the image, with redder colors corresponding to larger stellar counts.  The number of stars in this absolute magnitude slice is at the top of the plot.  The majority of the stars in the sample were found in the northern Galactic hemisphere, since SDSS was centered on the Northern Galactic Cap. }
\label{chap5:fig:counts_example}
\end{figure}

The coordinate transformation from a spherical coordinate system ($\ell ,b,$ and $d$) to a cylindrical ($R,Z$) system was performed with the following equations:
\begin{equation}
R = \sqrt{ (d \cos b)^2 +  R_\odot (R_\odot - 2d \cos b \cos \ell)} \\
\end{equation}

\begin{equation}
Z = Z_\odot + d \sin ( b - \arctan ( Z_\odot / R_\odot))
\end{equation} 
where $d$ was the distance (as determined by Equation \ref{chap4:eqn:distance_mod} and the $(M_r, r-z$) CMR), $\ell$ and $b$ are Galactic longitude and latitude, respectively, and $R_\odot$ and $Z_\odot$ are the position of the Sun, as explained above\footnote{Note that Eqn. (3) should contain a term with $\ell$.  We ignore this term due to its small size relative to the $d \sin b$ term.}.  The density maps were binned in $R$ and $Z$.  The bin width needed to be large enough to contain many stars (to minimize Poisson noise), but small enough to accurately resolve the structure of the thin and thick disks.  The $R,Z$ bin size was set at 25 pc.  An example of the star counts as a function of $R$ and $Z$ is shown in Figure \ref{chap5:fig:counts_example}.  

The volume sampled by each $R,Z$ bin was estimated with the following numerical method.  A $4 \times 4 \times 4$ kpc cube of ``test'' points was laid down, centered on the Sun, at uniform intervals 1/10$^{\rm th}$ the $R,Z$ bin size (every 2.5 pc).  This grid discretizes the volume, with each point corresponding to a fraction of the total volume.  Here, the volume associated with each grid point was $k =$ $2.5^3$ pc$^3$ point$^{-1}$ or 15.625 pc$^3$ point$^{-1}$.

\begin{figure}[!htbp]
\centering
\includegraphics[scale=0.4]{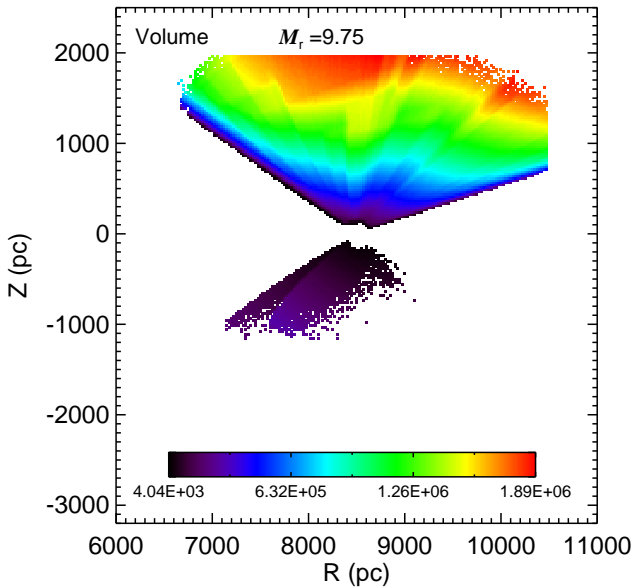} 
\caption{Volume probed by SDSS imaging as a function of Galactic $R$ and $Z$ for one 0.5 mag slice at $M_r = 9.75$.  The corresponding scale (in pc$^3$) is at the bottom of the plot, with redder colors corresponding to larger volumes.}
\label{chap5:fig:vol_example}
\end{figure}

The volume of an arbitrary shape is straightforward to calculate: simply count the points that fall within the shape, and multiply by $k$.   The $\alpha$, $\delta$ and distance of each point were calculated and compared to the SDSS volume.  The number of test points in each $R,Z$ bin was summed and multiplied by $k$ to obtain the final volume corresponding to that $R,Z$ bin.  This process was repeated for each absolute magnitude slice.  The maximum and minimum distances were calculated for each absolute magnitude slice (corresponding to the faint and bright apparent magnitude limits of the sample), and only test points within those bounds were counted. The same volume was used for all stars within the sample.  The bluer stars in our sample were found at distances beyond 4 kpc, but computing volumes at these distances would be computationally prohibitive. Furthermore, this method minimizes galaxy contamination, which is largest for bluer, faint objects (see \S \ref{sec:star-gal}).  Since the volumes are fully discretized, the error associated with $N_{\rm points}$ is Poisson--distributed.  A fiducial example of the volume calculations is shown in Figure \ref{chap5:fig:vol_example}.  

\begin{figure}[!htbp]
\centering
\includegraphics[scale=0.4]{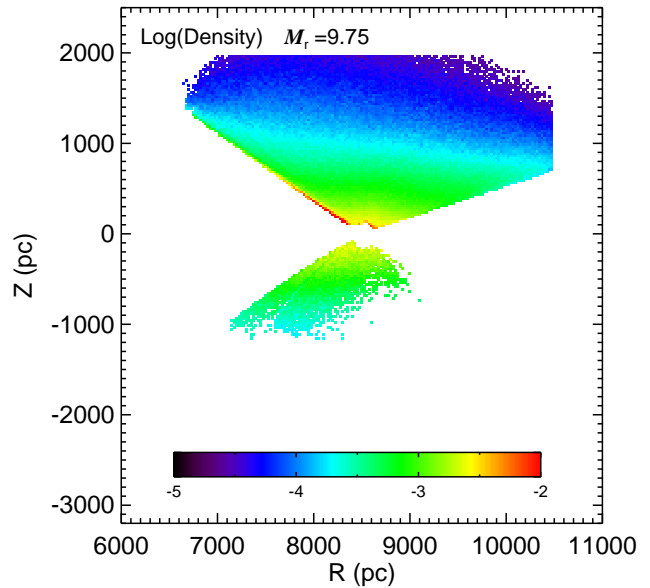} 
\caption{Density (in stars pc$^{-3}$) as a function of Galactic $R$ and $Z$.  The logarithmic scale is shown beneath the density map, with redder colors corresponding to larger densities.  The disk structure of the Milky Way is clearly evident, with a smooth decline towards larger $R$, and an increase in density approaching the Plane ($Z$ = 0).}
\label{chap5:fig:density_example}
\end{figure}

\begin{figure}[!htbp]
\centering
\includegraphics[scale=0.4]{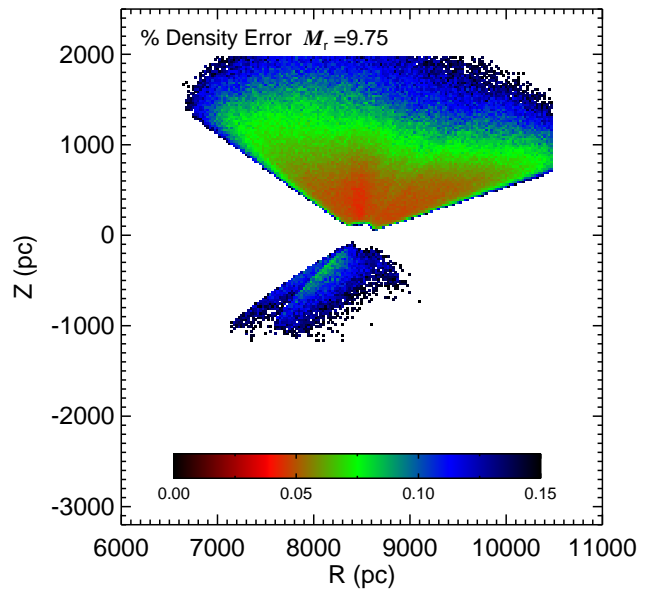} 
\caption{The fractional error in density as a function of $R$ and $Z$.  As in the previous figures, the scale is below the map, with bluer colors indicating larger errors.  The errors, calculated in Equation \ref{chap5:eqn:error} are $\lesssim 7\%$ for the majority of the sample.  }
\label{chap5:fig:density_error_example}
\end{figure}

After calculating the volume of each $R,Z$ bin, the density (in units of stars pc$^{-3}$) is simply:
\begin{equation}
\rho(R,Z) = \frac{N(R,Z)}{V(R,Z)}
\label{chap5:eqn:density}
\end{equation}      
with the error given by:
\begin{equation}
\sigma{_\rho} = \rho \sqrt{ \left( \frac{\sqrt{N(R,Z)}}{N(R,Z)} \right )^2 + \left ( \frac {k\sqrt{N_{points}(R,Z)}}{V(R,Z)} \right )^2}
\label{chap5:eqn:error}
\end{equation}
where $N(R,Z)$ is the star counts in each $R,Z$ bin and $V(R,Z)$ is the corresponding volume.  Fiducial density and error maps are shown in Figures \ref{chap5:fig:density_example} and \ref{chap5:fig:density_error_example}.  Note that the error in Eqn. \ref{chap5:eqn:error} is dominated by the first term on the right hand side.  While a smaller $k$ could make the second term less significant, it would be computationally prohibitive to include more test points.  We discuss systematic errors which may influence the measured density in \S \ref{sec:correction}.

\subsection{Galactic Model Fits}
Using the method described above, $(R,Z)$ stellar density maps were constructed for each 0.5 mag slice in $M_r$, from $M_r = 7.25$ to $M_r = 15.75$, roughly corresponding to spectral types M0-M8.  The bin size in each map was constant, at 25 pc in the $R$ and $Z$ directions.  For $R,Z$ bins with density errors (Equation \ref{chap5:eqn:error}) of $< 15\%$, the following disk density structure was fit:
\begin{equation}
 \rho_{\rm thin}(R,Z) = \rho_{\circ} f e^{\left( { - \frac{ R - R_\odot}{R_{\rm \circ,thin}}}\right)}  e^{ \left( - {\frac{\mid Z \mid - Z_\odot}{Z_{\rm \circ,thin}}}\right)} \label{chap5:eqn:rho_thin}
\end{equation}
\begin{equation}
 \rho_{\rm thick}(R,Z) = \rho_{\circ} (1-f) e^{\left( -{ \frac{ R - R_\odot}{R_{\rm \circ,thick}}}\right)}  e^{\left( - {\frac{\mid Z \mid - Z_\odot}{Z_{\rm \circ, thick}}}\right)} 
\label{chap5:eqn:rho_thick}
\end{equation}
\begin{equation}
 \rho(R,Z) = \rho_{\rm thin}(R,Z) + \rho_{\rm thick}(R,Z)
\label{chap5:eqn:rho_combined}
\end{equation}
where $\rho_{\circ}$ is the local density at the solar position ($R_\odot = 8500$ pc, $Z_\odot = 15$ pc), $f$ is the fraction of the local density contributed by the thin disk, $R_{\rm \circ,thin}$ and $R_{\rm \circ,thick}$ are the thin and thick disk scale lengths, and $Z_{\rm \circ,thin}$ and $Z_{\rm \circ, thick}$ are the thin and thick disk scale heights, respectively.  Since the density maps are dominated by nearby disk structure, the halo was neglected.  Furthermore, \cite{2008ApJ...673..864J} demonstrated that the halo structure is only important at $|Z| > 3$ kpc, well outside the volumes probed here.  Restricting the sample to  bins with density errors $< 15\%$ ensures that they are well-populated by stars and have precise volume measurements, and should accurately trace the underlying Milky Way stellar distribution.  Approximately $50\%$ of the $R,Z$ bins have errors $ \gtrsim 15\%$, while containing $ > 90\%$ of the stars in the sample.

\begin{figure}[htbp]
\centering
\includegraphics[scale=0.4]{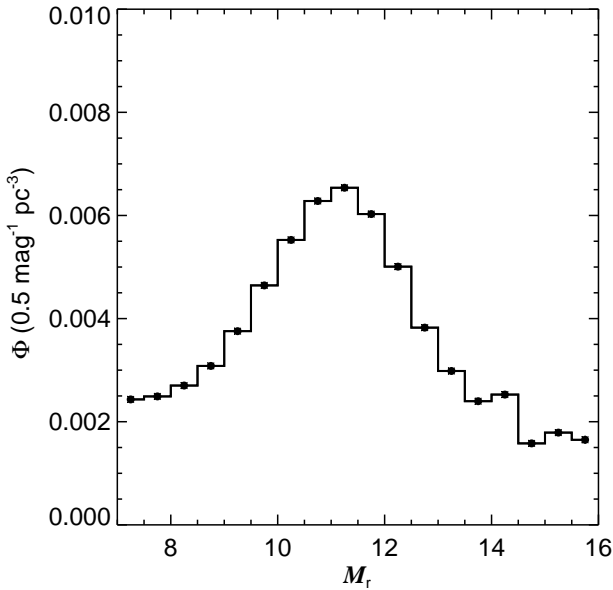}

\caption{The raw $r$-band LF for the stellar sample, using the $(M_r, r-z)$ CMR.  Note the smooth behavior, with a peak near $M_r \sim 11$, corresponding to a spectral type of $\sim$ M4.  The error bars (many of which are smaller than the points) are the formal uncertainties from fitting the local densities in each 0.5 mag absolute magnitude slice in stellar density.}
\label{chap5:fig:rawlf}
\end{figure}

The density maps were fit using Equation \ref{chap5:eqn:rho_combined} and a standard Levenberg--Marquardt algorithm \citep{1992nrfa.book.....P}, using the following approach.  First, the thin and thick disk scale heights and lengths, and their relative scaling, were measured using ten absolute magnitude slices, from $M_r = 7.25-11.75$.  These relatively more luminous stars yield the best estimates for Galactic structure parameters. Including lower-luminosity stars biases the fits, artificially shrinking the scale heights and lengths to compensate for density differences between a small number of adjacent $R,Z$ bins.  The scale lengths and heights and their relative normalization were fit for the entire $M_r = 7.25-11.75$ range simultaneously.  The resulting Galactic structure parameters ($Z_{\rm o,thin}, Z_{\rm o,thick}, R_{\rm o,thin}, Z_{\rm o,thick}, f$) are listed in Table \ref{chap5:table:raw_GS} as raw values, not yet corrected for systematic effects (see \S \ref{sec:correction} and Table \ref{table:corrected_GS}). After the relative thin/thick normalization ($f$) and the scale heights and lengths of each component are fixed, the local densities were fit for each absolute magnitude slice, using a progressive sigma clipping method similar to that of \cite{2008ApJ...673..864J}.  This clipping technique excludes obvious density anomalies from biasing the final best-fit.  First, a density model was computed, and the standard deviation ($\sigma$) of the residuals was calculated.  The $R,Z$ density maps were refit and bins with density residuals greater than 50$\sigma$ are excluded.  This process was repeated multiple times, with $\sigma$ smoothly decreasing by the following series: $\sigma = (40,30, 20, 10, 5)$.   An example luminosity function, constructed from the local densities of each absolute magnitude slice and derived from the $M_R, r-z$ CMR, is shown in Figure \ref{chap5:fig:rawlf}.  

\begin{deluxetable}{rll}
\tablewidth{0pt}
 
 \tablecaption{Measured Galactic Structure}
 \tabletypesize{\small}
 
\tablehead{
\colhead{Property} &
\colhead{Raw Value} &
\colhead{Uncertainty}
}
\startdata
$Z_{\rm o,thin}$   &   255 pc  & 12 pc  \\  
$R_{\rm o,thin}$   &   2200 pc & 65 pc  \\
$Z_{\rm o,thick}$  &   1360 pc  & 300 pc \\
$R_{\rm o,thick}$  &   4100 pc & 740 pc  \\
$f$                &   0.97    & 0.006  \\
\enddata
\label{chap5:table:raw_GS}
\end{deluxetable}

\section{Systematic Corrections}\label{sec:correction}
The observed LF is subject to systematics imposed by nature, such as unresolved binarity and  metallicity gradients, as well as those from the observations and analysis, e$.$g$.$ Malmquist bias.  The systematic differences manifested in different CMRs, which vary according to stellar metallicity, interstellar extinction, and color, are isolated and discussed in \S \ref{sec:sys:feh} and \S \ref{sec:sys:extinction} , and the results are used in \S \ref{sec:sys_uncert} to estimate the systematic uncertainties in the LF and Galactic structure.

Malmquist bias (\S \ref{sec:malm}) and unresolved binarity (\S \ref{sec:binaries}) were quantified using Monte Carlo (MC) models.  Each model was populated with synthetic stars that were consistent with the observed Galactic structure and LF.  The mock stellar catalog was analyzed with the same pipeline as the actual observations and the differences between the input and ``observed'' Galactic structure and LF were used to correct the observed values.  

\subsection{Systematic CMRs: Metallicity}\label{sec:sys:feh}
A star with low metallicity will have a higher luminosity and temperature compared to its solar--metallicity counterpart of the same mass, as first described by \cite{1959MNRAS.119..278S}.  However, at a fixed color, stars with lower metallicities have fainter absolute magnitudes.  Failing to account for this effect artificially brightens low-metallicity stars, increasing their estimated distance.  This inflates densities at large distances, increasing the observed scale heights \citep[e.g.,][]{1990mwg..conf.....K}.  

Quantifying the effects of metallicity on low-mass dwarfs is complicated by multiple factors.  First, direct metallicity measurements of these cool stars are difficult \citep[e.g.,][]{2006PASP..118..218W, 2009arXiv0904.3092J}, as current models do not accurately reproduce their complex spectral features.  Currently, measurements of metallicity-sensitive molecular bandheads (CaH and TiO) are used to estimate the metallicity of M dwarfs at the $\sim 1$ dex level \citep[see][]{1997AJ....113..806G,Lepine03, Burgasser06, 2008AJ....135..785W}, but detailed measurements are only available for a few stars.  
The effects of metallicity on the absolute magnitudes of low--mass stars are poorly constrained.  Accurate parallaxes for nearby subdwarfs do exist \citep{1992AJ....103..638M,1997AJ....114..161R, 2008ApJ...672.1159B}, but measurements of  their precise metal abundances are difficult given the extreme complexity of calculating the opacity of the molecular absorption bands that dominate the spectra of M dwarfs.  Observations of clusters with known metallicities could mitigate this problem \citep{2008AJ....135..682C, 2008ApJS..179..326A}, but there are no comprehensive observations in the $ugriz$ system that probe the lower main sequence.   

To test the systematic effects of metallicity on this study, the ($[Fe/H],\Delta M_r$) relation from \cite{Ivezic08} was adopted.  We note that this relation is appropriate for more luminous F and G stars, near the main-sequence turnoff, but should give us a rough estimate for the magnitude offset. The adopted Galactic metallicity gradient is:

\begin{equation}
[Fe/H] = -0.0958 - 2.77 \times 10^{-4} |Z| 
\label{chap5:eqn:metallicity}
\end{equation}

At small Galactic heights ($Z \lesssim 100$ pc), this linear gradient produces a metallicity of about [Fe/H] = -0.1, appropriate for nearby, local stars \citep{2004A&A...420..183A}. At a height of $\sim 2$ kpc (the maximum height probed by this study), the metallicity is [Fe/H] $\sim -0.65$, consistent with measured distributions \citep{Ivezic08}.  The actual metallicity distribution is probably more complex, but given the uncertainties associated with the effects of metallicity on M dwarfs, adopting a more complex description is not justified.  The correction to the absolute magnitude, $\Delta M_r$, measured from F and G stars in clusters of known metallicity and distance \citep{Ivezic08}, is given by:

\begin{equation}
\Delta M_r = -0.10920 - 1.11[Fe/H] - 0.18[Fe/H]^2
\label{chap5:eqn:deltamr}
\end{equation}

Substituting Equation \ref{chap5:eqn:metallicity} into Equation \ref{chap5:eqn:deltamr}, yields a quadratic equation for $\Delta M_r$ in Galactic height.  After initially assigning absolute magnitudes and distances with the CMRs appropriate for nearby stars, each star's estimated height above the Plane, $Z_{\rm ini}$, was computed.  This is related to the star's actual height, $Z_{\rm true}$, through the following equation:

\begin{equation}
Z_{\rm true} = Z_{\rm ini}10^{\frac{-\Delta M_r(Z_{\rm true})}{5}}
\label{chap5:eqn:z_relation}
\end{equation}

A star's true height above the Plane was calculated by finding the root of this non-linear equation.   Since $\Delta M_r$ is a positive value, the actual distance from the Galactic plane, $Z_{\rm true}$ is smaller than the initial estimate, $Z_{\rm ini}$.  As explained above, this effect becomes important at larger distances, moving stars inwards and decreasing the density gradient.  Thus, if metallicity effects are neglected, the scale heights and lengths are overestimated.

\begin{figure}[htbp]
\centering
\includegraphics[scale=0.35]{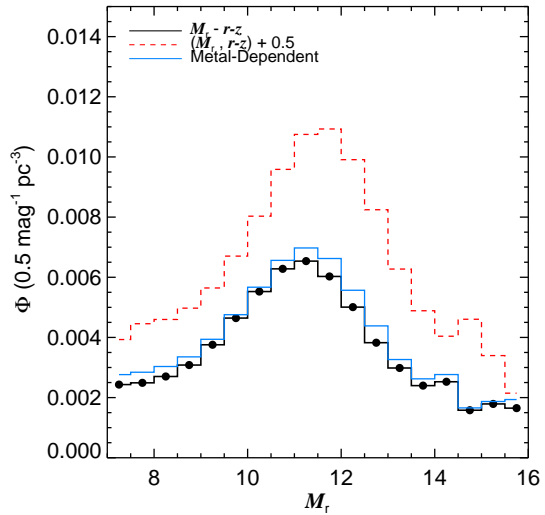}
\caption{The differences in the LF induced by metallicity gradients, along with the raw LF (black line).  The red histogram corresponds to the extreme limit, where all stars are metal-poor ($[Fe/H] \sim -0.65$).  The  blue histogram shows the effect of the metallicity gradient from Equation \ref{chap5:eqn:metallicity}.}
\label{chap5:fig:metal_LF}
\end{figure}

In Figure \ref{chap5:fig:metal_LF}, the systematic effects of metallicity dependent CMRs are shown.  The first is the extreme limit, shown as the red histogram, where all stars in the sample have an $[Fe/H] \sim -0.65$, corresponding to a $\Delta M_r$ of roughly 0.5 magnitudes.  All of the stars in the sample are shifted to smaller distances, greatly enhancing the local density.  This limit is probably not realistic, as prior luminosity function studies \citep[e.g.,][]{1997AJ....113.2246R, 2007AJ....133..439C} would have demonstrated similar behavior.  The effect of the metallicity gradient given in Equation \ref{chap5:eqn:metallicity} is shown with the solid blue line.  Note that local densities are increased, since more stars are shifted to smaller distances.

\subsection{Systematic CMRs: Extinction}\label{sec:sys:extinction}
\begin{figure}
\centering
\includegraphics[scale=0.35]{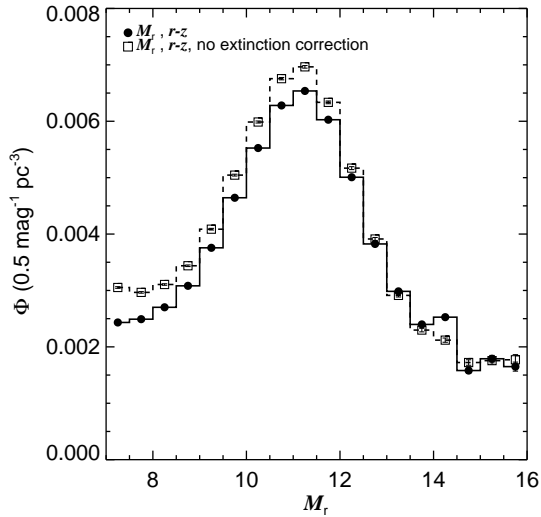}
\caption{The systematic effect of extinction on the raw LF.  When no extinction correction is applied (open squares), distant stars act to inflate the local densities of the brightest stars, compared to the fiducial case (filled circles).  At fainter luminosities, this effect becomes less important.}
\label{chap5:fig:extinction}
\end{figure}

The extinction and reddening corrections applied to SDSS photometry are derived from the \cite{1998ApJ...500..525S} dust maps and an assumed dust law of $R_V = 3.1$ \citep{1989ApJ...345..245C}.  The median extinction in the sample is $A_r = 0.09$, while 95\% of the sample has $A_r < 0.41$.  Typical absolute magnitude differences due to reddening range up to $\sim 1$ magnitude, producing distance corrections of $\sim 40$ pc, enough to move stars between adjacent $R,Z$ bins and absolute magnitude bins. This effect introduces strong covariances between adjacent luminosity bins, and implies that the final LF depends on the assumed extinction law.  Most of the stars in our sample lie beyond the local dust column, and the full correction is probably appropriate \citep{2006A&A...453..635M}.  To bracket the effects of extinction on our analysis, two LFs were computed.  The first is the ($M_r, r-z$) LF, which employs the entire extinction correction.  The second uses the same CMR, but without correcting for extinction.   The two LFs are compared in Figure \ref{chap5:fig:extinction}.  When the extinction correction is neglected, stellar distances are underestimated, which increases the local density.  This effect is most pronounced for larger luminosities.  The dominant effect in this case is not the attenuation of light due to extinction, but rather the reddening of stars, which causes the stellar absolute magnitudes to be underestimated. 

\subsection{Systematic Uncertainties}\label{sec:sys_uncert}
\begin{figure}[htbp]
\centering
\includegraphics[scale=0.4]{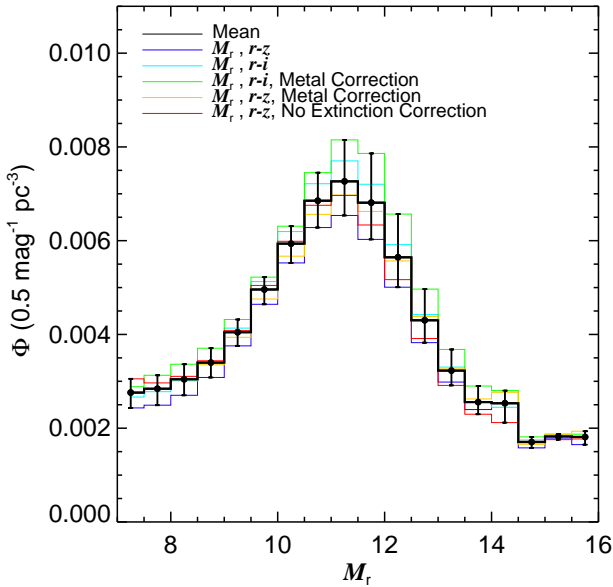}
\caption{The mean observed system LF, derived from five different CMRs.  Each CMR is described in the legend. The uncertainty in a given LF bin is set by the full range spanned by the five different LFs.}
\label{fig:raw_lf_spread}
\end{figure}

The statistical error in a given LF bin is quite small, typically $\lesssim 0.1\%$, and does not represent a major source of uncertainty in this analysis.  The assumed CMR dominates the systematic uncertainty, affecting the shape of the LF and resulting MF.  To quantify the systematic uncertainty in the LF and Galactic structure, the following procedure was employed.  The LF was computed five times using different CMRs:  The $(M_r, r-z)$ and $(M_r, r-i)$ CMRs with and without metallicity corrections, and the $(M_r, r-z)$ CMR without correcting for Galactic extinction.  The LFs measured by each CMR are plotted in Figure \ref{fig:raw_lf_spread}, along with the unweighted mean of the five LF determinations.   The uncertainty in a given LF mag bin was set by the maximum and minimum of the five test cases, often resulting in asymmetric error bars.  This uncertainty was propagated through the entire analysis pipeline using three LFs:  the mean, the ``maximum'' LF, corresponding to the maximum $\Phi$ in each magnitude bin, and the ``minimum'' LF, corresponding to the lowest $\Phi$ value.  We adopted the mean LF as the observed system LF  and proceeded to correct it for the effects of Malmquist bias and binarity, as described below.

\subsection{Monte Carlo Models:  Malmquist Bias}\label{sec:malm}
Malmquist bias \citep{m36} arises in flux-limited surveys (such as SDSS), when distant stars with brighter absolute magnitudes (either intrinsically, from the width of the main sequence, or artificially, due to measurement error) scatter into the survey volume.  These stars have their absolute magnitudes systematically overestimated (i.e., they are assigned fainter absolute magnitudes than they actually possess), which leads to underestimated intrinsic luminosities.  Thus, their distances will be systematically underestimated.  This effect artificially shrinks the observed scale heights, and inflates the measured LF densities.   Assuming a Gaussian distribution about a ``true'' mean absolute magnitude $M_{\circ}$, classical Malmquist bias is given by:
\begin{equation}
\bar{M}(m) = M_{\circ} - \frac{\sigma^2}{{\rm log}~e} \frac{dA(m)}{dm}
\end{equation}
where $\sigma$ is the spread in the main sequence (or CMR),  $ dA(m) / dm$ is the slope of the star counts as a function of apparent magnitude $m$, and $\bar{M}(m)$ is the observed mean absolute magnitude.  Qualitatively, $\bar{M}(m)$ is always less than $M_{\circ}$ (assuming  $ dA(m) / dm$ is positive), meaning that the observed absolute magnitude distribution is skewed towards more luminous objects.

Malmquist bias effects were quantified by including dispersions in absolute magnitude of $\sigma_{M_r} = 0.3$ and $\sigma_{M_r} = 0.5$ mag and a color dispersion of $\sigma_{r-z, r-i} = 0.05$ mag.  These values were chosen to bracket the the observed scatter in the color--magnitude diagrams (see Table \ref{table:cmrs}). The LF measured with the Malmquist bias model is shown in Figure \ref{fig:both_corr_vs_raw_lf}.  The correction is important for most of the stars in the sample, especially the brightest stars ($M_r < 10$). Stars at this magnitude and color ($r-i \sim 0.5$, $r-z \sim 1$, see Figure \ref{chap5:fig:color_hists}) are very common in the SDSS sample because they span a larger volume than lower luminosity stars.  Thus, they are more susceptible to having over-luminous stars scattered into their absolute magnitude bins.  However, the dominant factor that produced the differences between the raw and corrected LFs was the value of the thin disk scale height.

\subsection{Monte Carlo Models: Unresolved Binarity}\label{sec:binaries}
For all but the widest pairs, binaries in our sample will masquerade as a single star.  The unresolved duo will be over-luminous at a given color, leading to an underestimate of its distance.  This compresses the density maps, leading to decreased scale heights and lengths, as binary systems are assigned smaller distances appropriate to single stars.  
  
Currently, the parameter space that describes M dwarf binaries: binary fraction, mass ratio, and average separation, is not well constrained.  However, there are general trends that are useful for modeling their gross properties.  First, the binary fraction ($f_b$) seems to steadily decline from $\sim 50\%$ at F and G stars \citep{1991A&A...248..485D} to about 30\% for M dwarfs \citep{1992ApJ...396..178F,2004ASPC..318..166D,2006ApJ...640L..63L, 2007prpl.conf..427B}.  Next, the mass ratio distribution becomes increasingly peaked towards unity at lower masses.  That is, F and G stars are more likely to have a companion from a wide range of masses, while M dwarfs are commonly found with a companion of nearly the same mass, when the M dwarf is the primary (warmer) star \citep{2007prpl.conf..427B}.  The average separation distribution is not well known, but many companions are found with separations of $\sim 10-30$ AU \citep{1992ApJ...396..178F}, while very--low mass stars have smaller average separations \citep{2007prpl.conf..427B}.   At the typical distances probed by the SDSS sample (100s of pc) these binary systems would be unresolved by SDSS imaging with an average PSF width of 1.4\arcsec~in $r$.

We introduced binaries into our simulations with four different binary fraction prescriptions.  The first three ($f_b = 30\%, 40\%$ and 50\%) are independent of primary star mass.  The fourth binary fraction follows the methodology of \cite{covey08} and is a primary--star mass--dependent binary fraction, given by:
\begin{equation}
f_b({\cal M}_p) = 0.45 - \frac{0.7 - {\cal M}_p}{4}
\end{equation}
where ${\cal M}_p$ is the mass of the primary star, estimated using the \cite{2000A&A...364..217D} mass--luminosity relations.  This linear equation reflects the crude observational properties described above for stars with ${\cal M}_p < 0.7 {\cal M_{\odot}}$.  Near 1 ${\cal M_{\odot}}$, the binary fraction is $\sim 50\%$, while at smaller masses, the binary fraction falls to $\sim 30\%$.  Secondary stars are forced to be less massive than their primaries.  This is the only constraint on the mass--ratio distribution. 

\begin{figure}[htbp]
\centering
\includegraphics[scale=0.4]{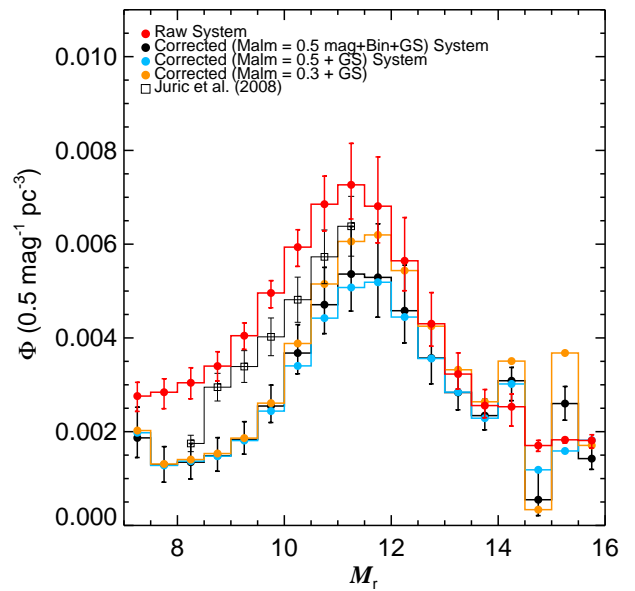}
\caption{The effects of Malmquist bias (orange and blue filled circles), unresolved binarity (black) and Galactic structure (GS) on the survey.  The mean observed LF is larger than the corrected LF at most bins.  The largest effects are seen for the brightest stars, which are subject to the largest shifts due to a change in the thin disk scale height.  The difference between the orange and blue LFs demonstrate the sensitivity of the Malmquist correction to the assumed scatter in the main sequence.  The binary correction becomes relatively more important at fainter absolute magnitudes.  The LF from \citet[open squares]{2008ApJ...673..864J} is shown for comparison to our raw system LF.  They did not probe faint absolute magnitudes, employed a different CMR, and did not correct their densities for Malmquist bias, which accounts for the offsets between their LF and ours.}
\label{fig:both_corr_vs_raw_lf}
\end{figure}

 An iterative process, similar to that described in \cite{covey08}, is employed to estimate the binary--star population.  First, the mean observed LF from Figure \ref{fig:raw_lf_spread} is input as a primary--star LF (PSLF).  A mock stellar catalog is drawn from the PSLF, and binary stars are generated with the prescriptions described above.  Next, the flux from each pair is merged, and new colors and brightnesses are calculated for each system.   Scatter is introduced in color and absolute magnitude, as described in \S \ref{sec:malm}.  The stellar catalog is analyzed with the same pipeline as the data, and the output model LF is compared to the observed LF.  The input PSLF is then tweaked according to the differences between the observed system LF and the model system LF.  This loop is repeated until the artificial system LF matches the observed system LF.  Note that the Galactic structure parameters are also adjusted during this process, and the bias--corrected values are given in Table \ref{table:corrected_GS}.   The thin disk scale height, which has a strong effect on the derived LF, is in very good agreement with previous values.   As the measured thin disk scale height increases, the density gradients decrease, and a smaller local density is needed to explain distant structures.   This change is most pronounced at the bright end, where the majority of the stars are many thin disk scale heights away from the Sun (see Figure \ref{fig:both_corr_vs_raw_lf}).  The preferred model thin disk and thick disk scale lengths were found to be similar.  This is most likely due to the limited radial extent of the survey compared to their typical scale lengths.  Upcoming IR surveys of disk stars, such as APOGEE \citep{2008AN....329.1018A}, should provide more accurate estimates of these parameters.  

SDSS observations form a sensitive probe of the thin disk and thick disk scale heights, since the survey focused mainly on the Northern Galactic Cap.  Our estimates suggest a larger thick disk scale height and smaller thick disk fraction than recent studies \citep[e.g.,][]{2002ApJ...578..151S, 2008ApJ...673..864J}.  However, these two parameters are highly degenerate \citep[see Figure 1 of ][]{2002ApJ...578..151S}.  In particular, the differences between our investigation and the \cite{2008ApJ...673..864J} study highlight the sensitivity of these parameters to the assumed CMR and density profiles, as they included a halo in their study and we did not.  The \cite{2008ApJ...673..864J} study sampled larger distances than our work, which may affect the resulting Galactic parameters.   However, the smaller normalization found in our study is in agreement with recent results from a kinematic analysis of nearby M dwarfs with SDSS spectroscopy \citep{pineda}.   They find a relative normalization of $\sim 5\%$, similar to the present investigation.  The discrepancy in scale height highlights the need for additional investigations into the thick disk and suggests that future investigations should be presented in terms of stellar mass contained in the thick disk, not scale height and normalization.

\begin{figure}[htbp]
\centering
\includegraphics[scale=0.4]{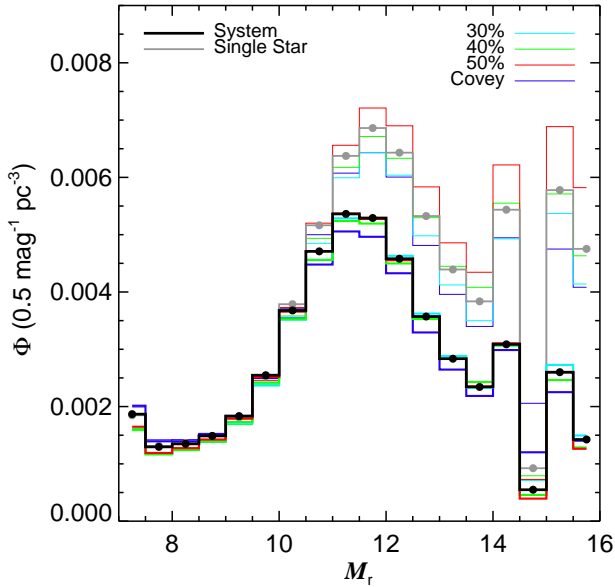}
\caption{The system and single star $M_r$ LFs for our four different binary prescriptions.  The spread between prescriptions in each bin is used to calculate the final uncertainty in the system and single--star LFs.}
\label{fig:single_vs_system}
\end{figure}

 The iterative process described above accounts for binary stars in the sample, and allows us to compare the system LF and single--star LF in Figure \ref{fig:single_vs_system}.  Most observed LFs are $system$ LFs, except for the local volume--limited surveys.  However, most theoretical investigations into the IMF predict the form of the single star MF.  Note that for all binary prescriptions, the largest differences between the two LFs are seen at the faintest $M_r$, since the lowest luminosity stars are most easily hidden in binary systems.

\begin{deluxetable}{rll}
\tablewidth{0pt}

 \tablecaption{Bias--Corrected Galactic Structure}
 \tabletypesize{\small}
 
\tablehead{
\colhead{Property} &
\colhead{Corrected Value} &
\colhead{Uncertainty}
}
\startdata
$Z_{\rm o,thin}$   &   300 pc  & 15 pc  \\  
$R_{\rm o,thin}$   &   3100 pc & 100 pc  \\
$Z_{\rm o,thick}$  &   2100 pc  & 700 pc \\
$R_{\rm o,thick}$  &   3700 pc & 800 pc  \\
$f$                &   0.96    & 0.02  \\
\enddata
\label{table:corrected_GS}
\end{deluxetable}

\section{Results: Luminosity Function}\label{sec:LF} 
The final adopted system and single star $M_r$ LFs are presented in Figure \ref{fig:lf_sing_sys_w_juric}.  The LFs were corrected for unresolved binarity and Malmquist bias.  The uncertainty in each bin is computed from the spread due to CMR differences, binary prescriptions and Malmquist corrections.  The mean LFs and uncertainties are listed in Tables \ref{table:system_r_band_lf} and \ref{table:single_r_band_lf}.    The differences between the single and system LFs are discussed below and compared to previous studies in both $M_r$ and $M_J$.

\subsection{Single--Star vs. System Luminosity Function}
\begin{figure}[htbp]
\centering
\includegraphics[scale=0.4]{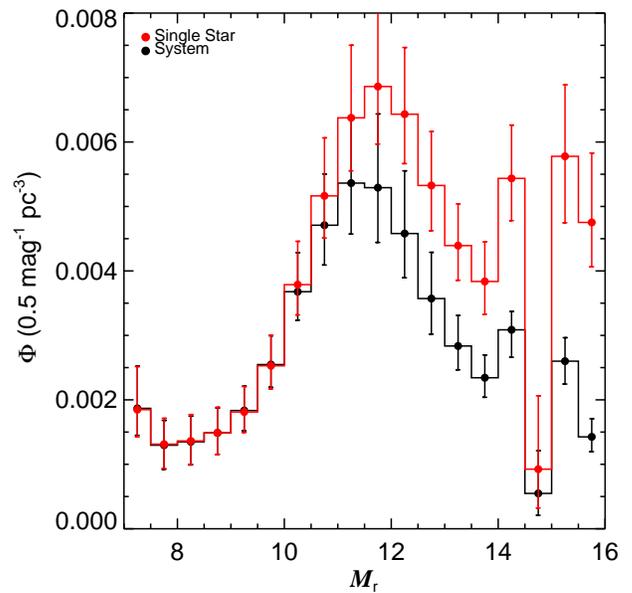}
\caption{Single star (red filled circles) and system (black filled circles) luminosity functions.  Note that the major differences between our system and single--star LFs occur at low luminosities, since low mass stars can be companions to stars of any higher mass, including masses above those sampled here.} 
\label{fig:lf_sing_sys_w_juric}
\end{figure}

Figure \ref{fig:lf_sing_sys_w_juric} demonstrates a clear difference between the single--star LF and the system LF.  The single--star LF rises above the system LF near the peak at $M_r \sim 11$ (or a spectral type $\sim$ M4), and maintains a density about twice that of the system LF\footnote{We note that the differences between our system and single--star LFs disagree considerably with those reported by \cite{covey08}.  These differences were investigated, and the \cite{covey08} binary corrections were found to be erroneous, with companion stars sampled from the MF convolved with the full sample volume, which is inappropriate for companion stars.  The authors regret the error.}.  This implies that lower luminosity stars are easily hidden in binary systems, but isolated low--luminosity systems are intrinsically rare.  The agreement between the system and single--star LFs at high luminosities is a byproduct of our binary prescription, which enforced that a secondary be less massive than its primary--star counterpart.  Since our LF does not extend to higher masses (G and K stars), we may be missing some secondary companions to these stars, which would inflate the single--star LF at high luminosities.  However, only $\sim$ 700,000 G dwarfs are present in the volume probed by this study.  Even if all of these stars harbored an M--dwarf binary companion, the resulting differences in a given bin would be only a fraction of a percent.

\subsection{$M_r$ LF}
Since many traditional LF studies have not employed the $r$--band, our ability to compare to previous results is hampered. The most extensive study of the $M_r$ LF was conducted by \cite{2008ApJ...673..864J}, using 48 million photometric SDSS observations, over different color ranges.  Figure \ref{fig:both_corr_vs_raw_lf} compares the $M_r$ system LF determined here to the ``joint fit, bright parallax'' results of \citet[their Table 3]{2008ApJ...673..864J}, assuming $10\%$ error bars.  The two raw system LFs broadly agree statistically, although the \cite{2008ApJ...673..864J} work only probes to $M_r \sim$ 11, due to their red limit of $r-i \sim$ 1.4.  We compare system LFs, since the \cite{2008ApJ...673..864J} study did not explicitly compute a SSLF and their reported LF was not corrected for Malmquist bias.

\begin{deluxetable}{llll}

 \tablewidth{3in}
 \tablecaption{Final System $M_r$ Luminosity Function}
 
 \tablehead{
 \colhead{$M_r$ bin} &
 \colhead{$\Phi_{\rm Mean}$}&
 \colhead{$\Phi_{\rm Max}$} &
 \colhead{$\Phi_{\rm Min}$ } 
  }
 \startdata
 7.25 &  1.87 &  2.52 &  1.45 \\
 7.75 &  1.30 &  1.68 &  0.92 \\
 8.25 &  1.35 &  1.75 &  0.99 \\
 8.75 &  1.49 &  1.87 &  1.16 \\
 9.25 &  1.83 &  2.21 &  1.52 \\
 9.75 &  2.55 &  2.99 &  2.20 \\
10.25 &  3.68 &  4.28 &  3.24 \\
10.75 &  4.71 &  5.50 &  4.09 \\
11.25 &  5.36 &  6.38 &  4.57 \\
11.75 &  5.29 &  6.44 &  4.44 \\
12.25 &  4.58 &  5.55 &  3.89 \\
12.75 &  3.57 &  4.29 &  3.02 \\
13.25 &  2.84 &  3.31 &  2.47 \\
13.75 &  2.34 &  2.70 &  2.04 \\
14.25 &  3.09 &  3.37 &  2.66 \\
14.75 &  0.55 &  1.21 &  0.21 \\
15.25 &  2.60 &  2.96 &  2.24 \\
15.75 &  1.43 &  1.71 &  1.20 \\
\enddata
\tablecomments{Densities are reported in units of (stars pc$^{-3}$ 0.5 mag$^{-1}$) $\times 10^{-3}$.  }
\label{table:system_r_band_lf}
\end{deluxetable}

\begin{deluxetable}{llll}

 \tablewidth{3in}
 \tablecaption{Final Single $M_r$ Luminosity Function}
 
 \tablehead{
 \colhead{$M_r$ bin} &
 \colhead{$\Phi_{\rm Mean}$}&
 \colhead{$\Phi_{\rm Max}$} &
 \colhead{$\Phi_{\rm Min}$ } 
  }
 \startdata
 7.25 &  1.85 &  2.51 &  1.43 \\
 7.75 &  1.31 &  1.71 &  0.93 \\
 8.25 &  1.36 &  1.77 &  1.00 \\
 8.75 &  1.49 &  1.89 &  1.15 \\
 9.25 &  1.81 &  2.21 &  1.49 \\
 9.75 &  2.53 &  3.00 &  2.17 \\
10.25 &  3.79 &  4.46 &  3.32 \\
10.75 &  5.16 &  6.06 &  4.51 \\
11.25 &  6.38 &  7.50 &  5.55 \\
11.75 &  6.86 &  8.10 &  5.97 \\
12.25 &  6.43 &  7.47 &  5.66 \\
12.75 &  5.33 &  6.16 &  4.62 \\
13.25 &  4.39 &  5.04 &  3.86 \\
13.75 &  3.84 &  4.45 &  3.33 \\
14.25 &  5.44 &  6.26 &  4.78 \\
14.75 &  0.92 &  2.06 &  0.32 \\
15.25 &  5.78 &  6.89 &  4.75 \\
15.75 &  4.75 &  5.83 &  4.06 \\
\enddata
\tablecomments{Densities are reported in units of (stars pc$^{-3}$ 0.5 mag$^{-1}$) $\times 10^{-3}$.  }
\label{table:single_r_band_lf}
\end{deluxetable}

\subsection{$M_J$ LF}
\begin{figure*}[htbp]
\begin{center}$
\begin{array}{cc}
\includegraphics[scale=0.3]{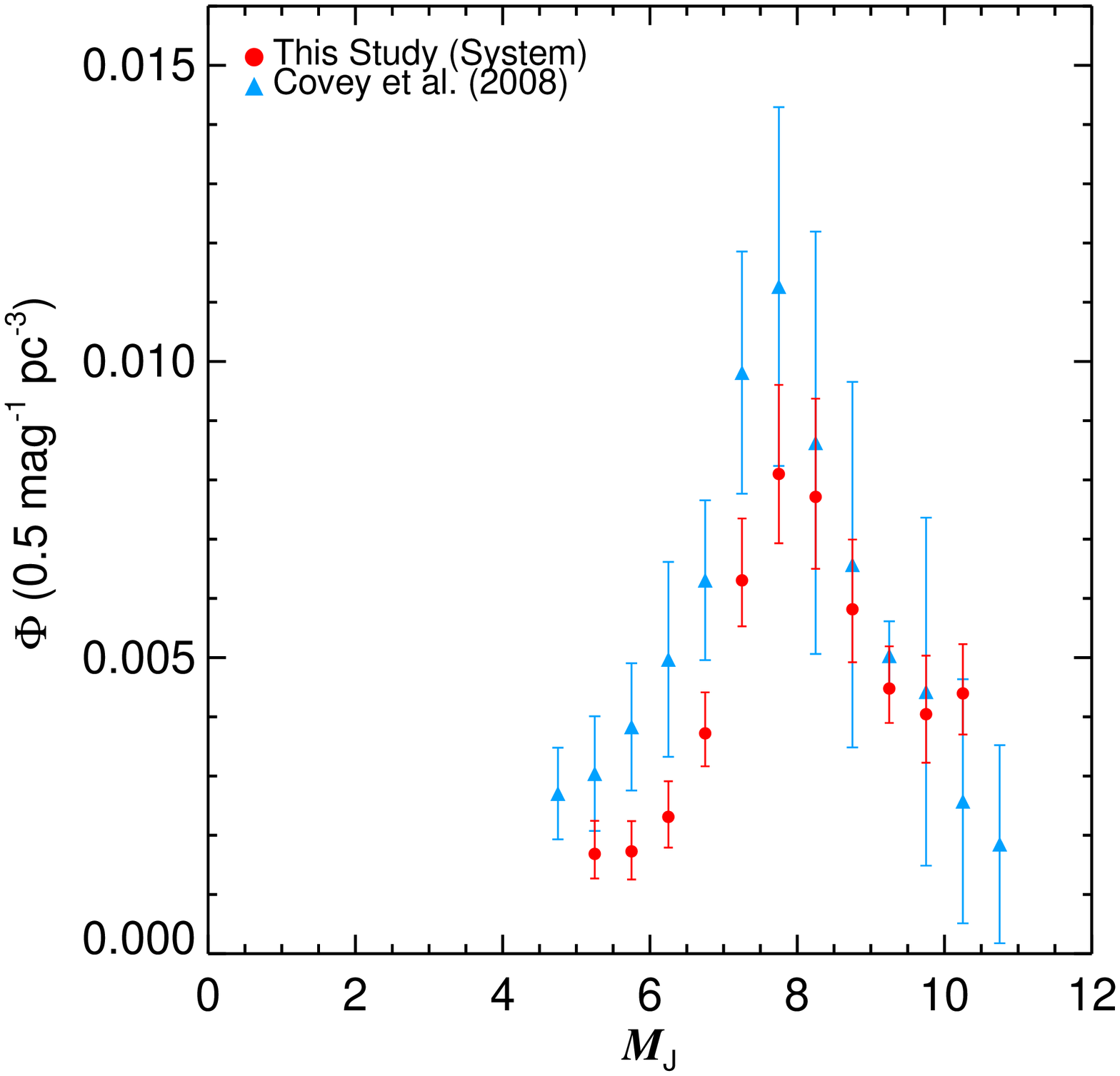} 
\includegraphics[scale=0.3]{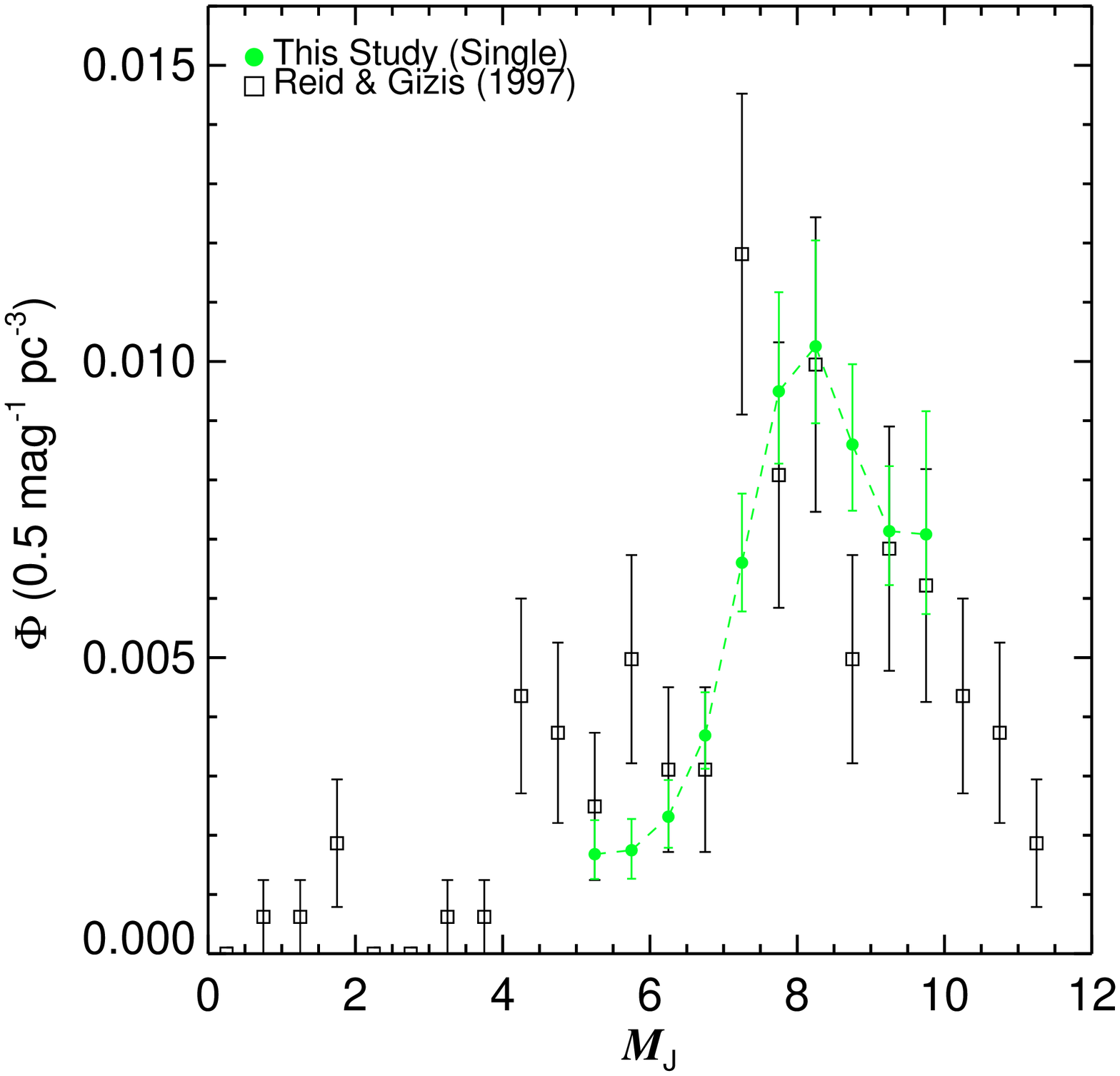}  
\end{array}$
\end{center}
 \caption{Left Panel: The $M_J$ system LF.  We compare our system LF (red filled circles) to the system LF measured by \citet[blue triangles]{covey08}. Our system LF and the \cite{covey08} results agree reasonably well.  Right Panel:   The $M_J$ single--star LF.  We compare our LF for single stars (green filled circles \& dashed line) compared to the single--star LF measured by \citet[open squares]{2002AJ....124.2721R}.  The single--star LFS also agree within the uncertainties, except for a few bins, resolving previous discrepancies between photometric and volume--complete samples.} 
 \label{fig:mj_all_lf}
\end{figure*}

We next converted $M_r$ to $M_J$ using relations derived from the calibration sample described in \cite{bochanskithesis}.  The $J$ filter has traditionally been used as a tracer of mass \citep{2000A&A...364..217D} and bolometric luminosity \citep{2004AJ....127.3516G} in low--mass stars, since it samples the SED near its peak.  The largest field LF investigation to date, \cite{covey08}, determined the $J$ band LF from $M_J = 4$ to  $M_J = 12$.   In Figure \ref{fig:mj_all_lf}, our transformed system $M_J$ LF (given in Table \ref{table:system_j_band_lf}) is plotted with the $M_J$ LF from \cite{covey08}.  The shape of these two LFs agree quite well, both peaking near $M_J = 8$, although there appears to be a systematic offset, in that our $M_J$ LF is consistently lower than the one from the \cite{covey08} study.  This is most likely due to the different CMRs employed by the two studies.  \cite{covey08} used an $(M_i,i-J)$ CMR, as opposed to the various CMRs employed in the current study.

Figure \ref{fig:mj_all_lf} also compares our single star $M_J$ LF to the LF of primaries and secondaries first measured by \cite{1997AJ....113.2246R} and updated by \cite{2007AJ....133..439C}.  These stars are drawn from a volume--complete sample with $d < 8$ pc.   A total of 146 stars in 103 systems are found within this limit.  The distances for the stars in this volume--complete sample are primarily found from trigonometric parallaxes, with $ < 5\%$ of the stellar distances estimated by spectral type.  There is reasonable agreement between our $M_J$ LF and the LF from the volume--complete sample, within the estimated uncertainties, indicating that the photometric and volume--complete methods now give similar results.   Furthermore, it indicates that our assumed CMRs are valid for local, low--mass stars.  Finally, the agreement between the single--star LFs validates our assumed corrections for unresolved binarity.

\begin{deluxetable}{llll}
  \tablewidth{3in}
 
 \tablecaption{System $M_J$ Luminosity Function}
 
 \tablehead{
 \colhead{$M_J$ bin} &
 \colhead{$\Phi_{\rm Mean}$} &
 \colhead{$\Phi_{\rm Max}$} &
 \colhead{$\Phi_{\rm Min}$}  
  }
 \startdata
 5.25 &   1.69 &   2.24 &   1.27 \\
 5.75 &   1.73 &   2.24 &   1.25 \\
 6.25 &   2.31 &   2.91 &   1.79 \\
 6.75 &   3.72 &   4.41 &   3.16 \\
 7.25 &   6.30 &   7.35 &   5.53 \\
 7.75 &   8.10 &   9.60 &   6.93 \\
 8.25 &   7.71 &   9.37 &   6.50 \\
 8.75 &   5.82 &   6.99 &   4.92 \\
 9.25 &   4.48 &   5.19 &   3.90 \\
 9.75 &   4.04 &   5.03 &   3.22 \\
\enddata
\tablecomments{Densities are reported in units of (stars pc$^{-3}$ 0.5 mag$^{-1}$) $\times 10^{-3}$.  }
\label{table:system_j_band_lf}
\end{deluxetable}

\begin{deluxetable}{llll}
  \tablewidth{3in}
 
 \tablecaption{Single Star $M_J$ Luminosity Function}
 
 \tablehead{
 \colhead{$M_J$ bin} &
 \colhead{$\Phi_{\rm Mean}$} &
 \colhead{$\Phi_{\rm Max}$} &
 \colhead{$\Phi_{\rm Min}$}  
  }
 \startdata
 5.25 &   1.68 &   2.25 &   1.26 \\
 5.75 &   1.75 &   2.28 &   1.27 \\
 6.25 &   2.31 &   2.93 &   1.79 \\
 6.75 &   3.69 &   4.41 &   3.12 \\
 7.25 &   6.60 &   7.77 &   5.78 \\
 7.75 &   9.50 &  11.17 &   8.27 \\
 8.25 &  10.26 &  12.04 &   8.96 \\
 8.75 &   8.60 &   9.95 &   7.48 \\
 9.25 &   7.14 &   8.23 &   6.22 \\
 9.75 &   7.08 &   9.16 &   5.73 \\

 \enddata
\tablecomments{Densities are reported in units of (stars pc$^{-3}$ 0.5 mag$^{-1}$) $\times 10^{-3}$.  }
\label{table:single_j_band_lf}
\end{deluxetable}

\section{Results: Mass Function}\label{sec:MF}
The mass function (MF) was calculated from the $M_J$ LFs and the mass--luminosity ($M_J$) relation from \cite{2000A&A...364..217D}.  We computed both a single star MF and system MF.  As discussed in \cite{covey08}, some past discrepancies between MFs are probably due to comparing analytic fits and the actual MF data.  This effect is discussed below, where we compare our results to available MF data from nearby \citep[e.g.,][]{1997AJ....113.2246R} and distant \citep[e.g.,][]{2001ApJ...555..393Z} samples.  We also compare our analytic fits to seminal IMF studies.  

\begin{figure}[htbp]
 \centering
 \includegraphics[scale=0.35]{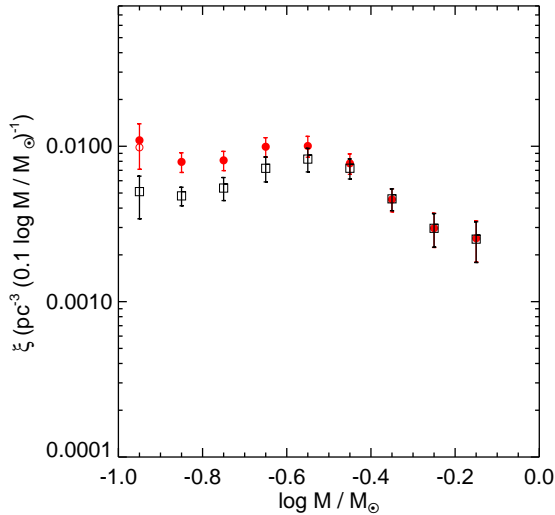}
 \caption{The single star MF (red filled circles) and system MF (open squares) for this study.  Note that the largest differences between the two MFs occur at smaller masses, since low--mass stars are easily obscured in binary systems with a higher mass primary.  The possible correction for young brown dwarfs discussed in \S \ref{sec:MF} is shown as an open circle in the single--star MF.}
 \label{fig:single_vs_system_mf}
\end{figure}

The single star and system MFs are shown in Figure \ref{fig:single_vs_system_mf}.  As seen in the LFs, there is agreement between the two relations at higher masses.  At masses less than 0.5 $\msun$, the shapes of the MFs are roughly equivalent,  but the single star density is roughly twice that of the systems.  
We note a small possible correction to the lowest mass bin (log $\cal{M} / \msun =$ -0.95).  At these masses, young brown dwarfs, with ages less than 1 Gyr and masses near $\cal{M} \sim$ 0.075 $\msun$ will have luminosities similar to late--type M dwarfs.  Since these objects are not stellar, they should be removed from our mass function.  Assuming a constant star formation rate, \cite{2003PASP..115..763C} estimated that brown dwarfs contribute $\sim 10\%$ of the observed densities at the faintest absolute magnitudes (or lowest masses).  Recent studies of nearby, young M dwarfs \cite[e.g.][]{2009arXiv0904.3323S} show that $\sim 10\%$ have ages less than 300 Myr, further supporting the presence of young brown dwarfs in our sample.  Thus, a correction of $10\%$ to the lowest mass bin would account for young brown dwarfs (see Figure \ref{fig:single_vs_system_mf}).  We don't apply the correction, but include its impact on the uncertainty of the last MF bin.

\begin{figure}[htbp]
 \centering
 \includegraphics[scale=0.35]{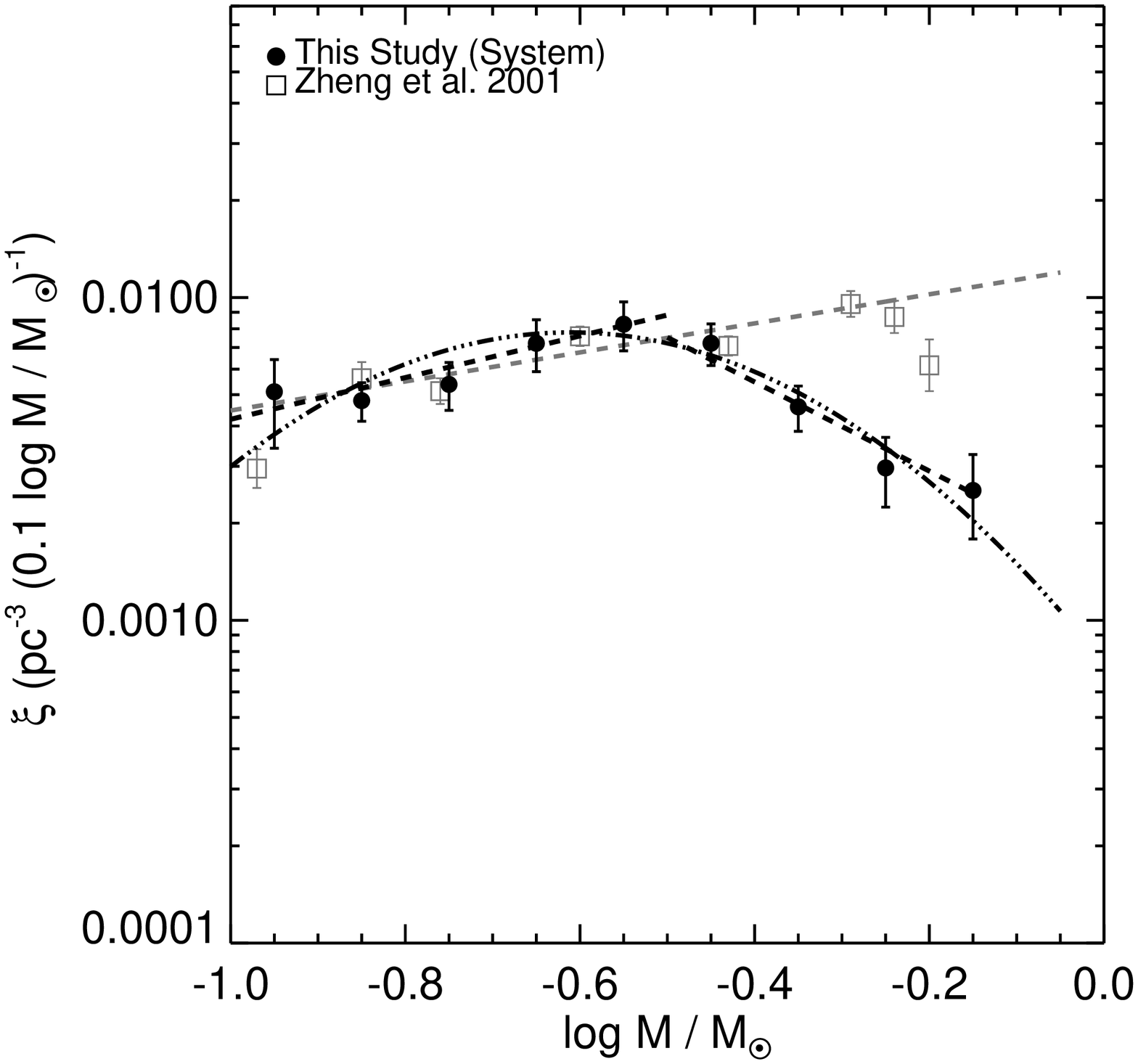}
 \caption{The system MF (filled circles), along with broken power law (dashed line) and log--normal (dot--dashed line) fits.  The power law break occurs near 0.3 $\msun$.  The parameters of both fits are found in Table \ref{table:sys_mf_fits}.  The MF data (open squares) and power law fit from \citet[light dashed line]{2001ApJ...555..393Z} are also shown.  The agreement between our data and the \cite{2001ApJ...555..393Z} data is good at lower masses, but the data diverge at higher masses. }
 \label{fig:sys_mf_fit_w_zheng}
\end{figure}

\begin{figure}[htbp]
 \centering
 \includegraphics[scale=0.35]{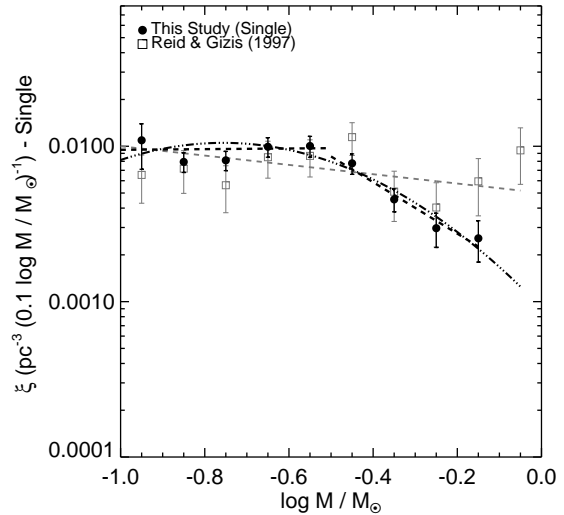}
 \caption{The single star MF (filled circles).  We fit this distribution with a broken power law (dashed line) and a log--normal distribution (dot--dashed line).  The parameters of these fits are found in Table \ref{table:sing_mf_fits}. The MF data (open squares) and single power law fit from \citet[light dashed line]{1997AJ....113.2246R} are also shown.  The data are in reasonable agreement, with discrepancies larger than the error bars in only two bins. }
 \label{fig:sing_mf_fit_w_reid}
\end{figure}

In Figures \ref{fig:sys_mf_fit_w_zheng} and \ref{fig:sing_mf_fit_w_reid} we display the log--normal and broken power law fits to the system and single star MFs.  While the broken  power law is preferred by many observers \citep{covey08, 2002Sci...295...82K}, the log--normal formalism has been popularized by some theorists \citep[e.g.,][]{2008ApJ...684..395H}.  Our MF data are best fit by a log--normal distribution, as confirmed by an $f$ test.  We suggest using the log--normal form when comparing to previous MF fits, but stress that comparisons using the actual MF data (Tables \ref{table:sys_mf} and \ref{table:sing_mf}) are preferred.

The system MF is compared to the pencil beam survey of \cite{2001ApJ...555..393Z} MF in Figure \ref{fig:sys_mf_fit_w_zheng}.  Their data were acquired with the Hubble Space Telescope, and the stars in their sample are at distances similar to this study.   The \cite{2001ApJ...555..393Z} study sparked some discussion in the literature \citep[e.g.,][]{2003PASP..115..763C, 2003ApJ...586L.133C}, as their results differed dramatically from the nearby star sample \citep[e.g.,][]{1997AJ....113.2246R}.   The proposed solution was unresolved binarity.  Binary systems would be resolved easily at small distances, but not at the larger distances probed by the HST sample.  We compare our system MF to the \cite{2001ApJ...555..393Z} sample,  and find agreement over a large range of masses (${\cal M} < 0.4 \msun$).    At larger masses ($\cal{M} \sim$ 0.4 $\msun$), the MFs diverge.   This can most likely be attributed to differing CMRs, in particular differences in the corrections for stellar metallicity gradients.

Our single star MF is compared to the nearby star sample \citep{1997AJ....113.2246R} in Figure \ref{fig:sing_mf_fit_w_reid}. The MFs agree remarkably well, with discrepancies in only two bins (likely the result of small numbers in the nearby star sample).    This indicates that our CMRs and methodology are valid, since the output densities are in agreement.  It suggests that our binary corrections are reasonable, as both our system and single star MFs agree with previous results.  Moreover, the discrepancy between the \cite{1997AJ....113.2246R} and \cite{2001ApJ...555..393Z} MFs can be attributed to unresolved binarity.  The  \cite{1997AJ....113.2246R} MF results present a single--star MF, while the \cite{2001ApJ...555..393Z} MF is a system MF.

\begin{figure}[htbp]
 \centering
 \includegraphics[scale=0.35]{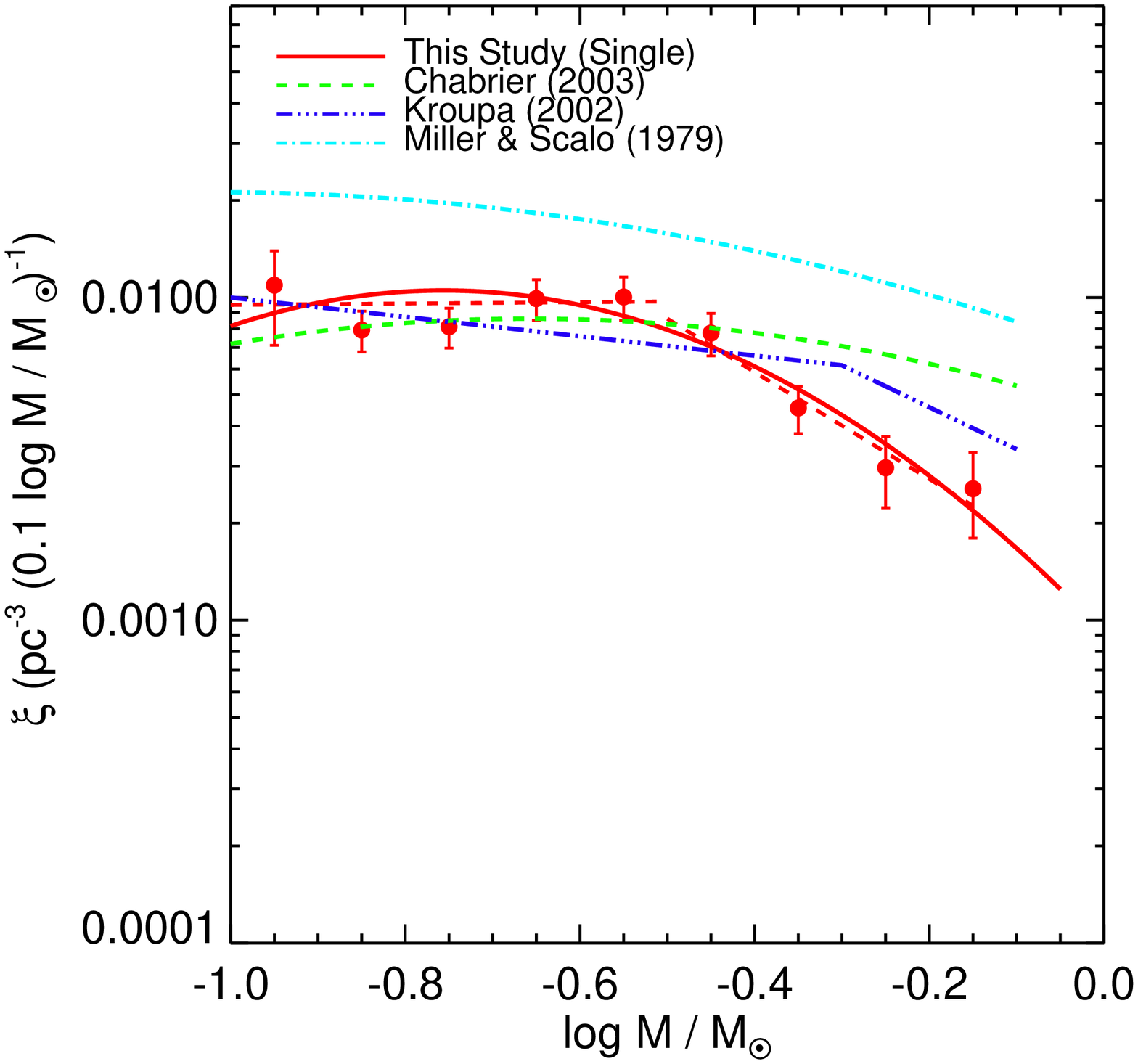}
 \caption{Shown is the MF data and best log-normal fit from this study (solid circles and red line), along with the analytic MF fits of \citet[green dashed line]{2003PASP..115..763C}, \citet[dark blue dash--dot--dot--dot line]{2002Sci...295...82K} and \citet[light blue dash-dot line]{1979ApJS...41..513M}.  We stress that comparing actual MF data is more valid than comparing analytic fits. }
 \label{fig:single_vs_bigshots_mf}
\end{figure}

We compare our single--star LF to seminal IMF analytic fits in Figure \ref{fig:single_vs_bigshots_mf} (see Table \ref{chap6:table:field_mfs}).   While we advocate the comparison of MF data whenever possible (as discussed in \citealp{covey08}), it is informative to compare our results to these studies.   The \cite{2002Sci...295...82K} and \cite{2003PASP..115..763C} studies demonstrate the best agreement with our data at masses ${\cal M} < 0.4 \msun$, but diverge
at higher masses, predicting larger space densities than we infer here. The disagreement of the \cite{1979ApJS...41..513M} MF with the other three MFs suggest an  issue with their normalization.

\begin{deluxetable}{llll}
\tablewidth{3in}
 \tablecaption{System Mass Function}
\tablehead{
 \colhead{log (${\cal M} / \msun$) } &
 \colhead{$\xi_{\rm Mean}$} &
\colhead{$\xi_{\rm Max}$} &
\colhead{$\xi_{\rm Min}$} 
 }
 \startdata
-0.15 &   2.53 &   3.26 &   1.79 \\
-0.25 &   2.96 &   3.69 &   2.24 \\
-0.35 &   4.59 &   5.32 &   3.85 \\
-0.45 &   7.22 &   8.28 &   6.15 \\
-0.55 &   8.27 &   9.70 &   6.84 \\
-0.65 &   7.21 &   8.53 &   5.89 \\
-0.75 &   5.38 &   6.29 &   4.47 \\
-0.85 &   4.79 &   5.45 &   4.14 \\
-0.95 &   5.11 &   6.42 &   3.41 \\

\enddata
 \tablecomments{Densities are reported in units of (stars pc$^{-3}$ 0.1 log~${\rm M}^{-1}$) $\times 10^{-3}$.  }
 \label{table:sys_mf}
\end{deluxetable}

\begin{deluxetable}{llll}
\tablewidth{3in}
 \tablecaption{Single Star Mass Function}
\tablehead{
 \colhead{log (${\cal M} / \msun$) } &
 \colhead{$\xi_{\rm Mean}$} &
\colhead{$\xi_{\rm Max}$} &
\colhead{$\xi_{\rm Min}$} 
 }
 \startdata
-0.15 &   2.56 &   3.31 &   1.80 \\
-0.25 &   2.97 &   3.71 &   2.23 \\
-0.35 &   4.55 &   5.31 &   3.79 \\
-0.45 &   7.76 &   8.93 &   6.59 \\
-0.55 &  10.04 &  11.58 &   8.50 \\
-0.65 &   9.93 &  11.36 &   8.50 \\
-0.75 &   8.12 &   9.27 &   6.97 \\
-0.85 &   7.93 &   9.08 &   6.78 \\
-0.95 &  10.92 &  13.95 &   7.11 \\
\enddata
 \tablecomments{Densities are reported in units of (stars pc$^{-3}$ 0.1 log~${\rm M}^{-1}$) $\times 10^{-3}$.  }
 \label{table:sing_mf}
\end{deluxetable}

\begin{deluxetable*}{lll}

\tablecaption{System Mass Function Analytic Fits}
 
\tablehead{
\colhead{Form} &
\colhead{Mass Range} & 
\colhead{Parameter} 
} 
\startdata
Log--Normal &  $-1.0 <$ log ${\cal M} / \msun$ $< -0.1   $ & $C_{\circ} =$ 0.008 $\pm$ 0.001 \\ 
                   &      & ${\cal M}_{\circ}$ = 0.25 $\pm$ 0.01 \\
                  &                  &$\sigma =$ 0.28 $\pm$ 0.02  \\
Broken Power Law (Low--Mass) &  0.1 $\msun < {\cal M} <$ 0.32 $\msun$  & $\alpha$ = 0.35$\pm$0.07   \\
Broken Power Law (High--Mass) &  0.32 $\msun < {\cal M} <$ 0.8 $\msun$  & $\alpha$ = 2.38$\pm$0.05  \\
\enddata

\tablecomments{We use the form $\psi({\cal M}) = C_{\circ} e^\frac{(log {\cal M} - log {\cal M}_{\circ})^2}{2\sigma^2}$ for the log--normal fit and  $\psi({\cal M}) \propto {\cal M}^{-\alpha}$ for the power law fit.}

\label{table:sys_mf_fits}
\end{deluxetable*}

\begin{deluxetable*}{lll}

\tablecaption{Single Mass Function Analytic Fits}
 
\tablehead{
\colhead{Form} &
\colhead{Mass Range} & 
\colhead{Parameter} 
} 
\startdata
Log--Normal &  $-1.0 <$ log ${\cal M} / \msun$ $< -0.1   $ & $C_{\circ} =$ 0.011 $\pm$ 0.002 \\ 
                   &      & ${\cal M}_{\circ}$ = 0.18 $\pm$ 0.02 \\
                  &                  &$\sigma =$ 0.34 $\pm$ 0.05  \\
Broken Power Law (Low--Mass) &  0.1 $\msun < {\cal M} <$ 0.32 $\msun$  & $\alpha$ = 0.98$\pm$0.15   \\
Broken Power Law (High--Mass) &  0.32 $\msun < {\cal M} <$ 0.8 $\msun$  & $\alpha$ = 2.66$\pm$0.10  \\
\enddata

\label{table:sing_mf_fits}
\end{deluxetable*}

\subsection{The IMF in other Mass Regimes}
\begin{figure}[htbp]
 \centering
 \includegraphics[scale=0.35]{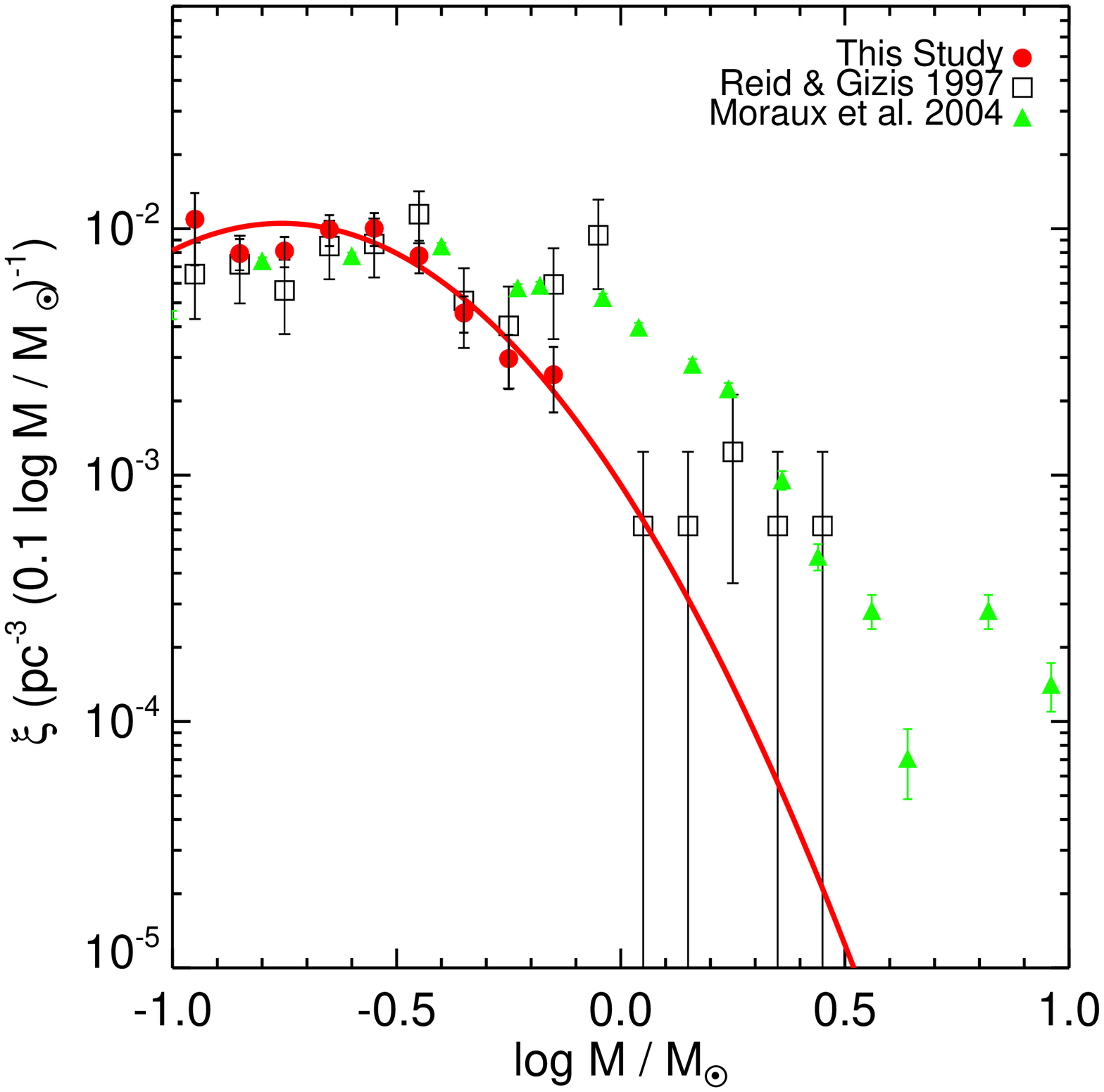}
 \caption{Shown is the single star MF and best log normal fit from this study (red filled circles and solid line), the \citet[open squares]{1997AJ....113.2246R} MF (open squares) and the Pleiades MF \citet[green triangles, ]{2004A&A...426...75M}.  The best fit extrapolated from our study systematically under--predicts the density at masses outside the bounds of our data.}
 \label{chap6:fig:high_mass_mf}
\end{figure}

A single analytic description of the IMF over a wide range in mass may not be appropriate.  Figure \ref{chap6:fig:high_mass_mf} shows the derived mass functions from this study, and those from the \cite{1997AJ....113.2246R} sample and the Pleiades \citep{2004A&A...426...75M}.  The log--normal fit from this study is extended to higher masses and it clearly fails to match the Pleiades MF.  Therefore, it is very important to only use the analytic fits over the mass ranges where they are appropriate.  Extending analytic fits beyond their quoted bounds can result in significant inaccuracies in the predicted number of stars.

\subsection{Theoretical Implications of the IMF}
Any successful model of star formation must accurately predict the IMF.  The measured field MF traces the IMF of low--mass stars averaged over the star formation history of the Milky Way.  Thus, the field MF is not a useful tool for investigating changes in the IMF due to physical conditions in the star forming regions, such as density or metallicity.  However, it does lend insight into the dominant physical processes that shape the IMF.  Recent theoretical investigations \citep[see][and references therein]{2007ASPC..362..269E} have mainly focused on three major mechanisms that would shape the low--mass IMF:  turbulent fragmentation, competitive accretion \& ejection, and thermal cooling arguments.

Turbulent fragmentation occurs when supersonic shocks compress the molecular gas \citep{1981MNRAS.194..809L, 2001ApJ...553..227P}.  Multiple shocks produce filaments within the gas, with properties tied to the shock properties.  Clumps then form along these filaments and collapse ensues.  In general, the shape of the IMF depends on the Mach number and power spectrum of shock velocities \citep{2006ApJ...637..384B, 2006A&A...452..487G} and the molecular cloud density \citep{2002ApJ...576..870P}.  Turbulence readily produces a clump distribution similar to the ubiquitous Salpeter IMF at high masses ($>$ 1 $\msun$). However, the flattening at lower masses is reproduced if only a fraction of clumps are dense enough to form stars \citep{2007A&A...471..499M}.  

An alternative model to turbulent fragmentation is accretion and ejection \citep{2005MNRAS.356.1201B}.  Briefly, small cores form near the opacity limit ($\sim$ 0.003 $\msun$), which is set by cloud composition, density and temperature.  These clumps proceed to accrete nearby gas.  Massive stars form near the center of the cloud's gravitational potential, thus having access to a larger gas reservoir.  Accretion ends when the nascent gas is consumed or the accreting object is ejected via dynamical interactions.  The characteristic mass is set by the accretion rate and the typical time scale for ejection, with more dense star forming environments producing more low--mass stars.  This method has fallen out of favor recently, as brown dwarfs have been identified in weakly bound binaries \citep{2004ApJ...614..398L,2009ApJ...691.1265L}, which should be destroyed if ejection is a dominant mechanism.  Furthermore, if ejection is important, the spatial density of brown dwarfs should be higher near the outskirts of a cluster compared to stars, and this is not observed in Taurus \citep{2004ApJ...617.1216L,2006ApJ...645..676L} or Chamaeleon \citep{2006A&A...448..655J}.

\cite{2005MNRAS.359..211L} suggested that thermal cooling arguments are also important in star formation.  This argument has gained some popularity, as it predicts a relative insensitivity of the IMF to initial conditions, which is supported by many observations \citep[e.g.,][and references therein]{2002Sci...295...82K,2007A&A...471..499M, coveyreview}.  This insensitivity is due to changes in the cooling rate with density.  At low densities, cooling is controlled by atomic and molecular transitions, while at higher densities, the gas is coupled with dust grains, and these dust grains dominate the cooling.   The result is an equation of state with cooling at low densities and a slight heating term at high densities.  This equation of state serves as a funneling mechanism and imprints a characteristic mass on the star formation process, with little sensitivity to the initial conditions.  

The general shape of the IMF has been predicted by star formation theories that account for all of these effects \citep{2003PASP..115..763C, 2005ASSL..327...41C}.  In particular, the high mass IMF is regulated by the power spectrum of the turbulent flows \citep{2009ApJ...702.1428H} and is probably affected by the coagulation of less massive cores, while the flatter, low--mass distribution can be linked to the dispersions in gas density and temperatures \citep{2007A&A...471..499M, 2006MNRAS.368.1296B}.  As the IMF reported in this study is an average over the star formation history of the Milky Way, changes in the characteristic shape of the IMF cannot be recovered.  However, our observational IMF can rule out star formation theories that do not show a flattening at low masses, with a characteristic mass $\sim 0.2 \msun$.  Recent numerical simulations have shown favorable agreement with our results \citep{2009MNRAS.392.1363B}, however most numerical simulations of star formation are restricted in sample size and suffer significant Poisson uncertainties.  Analytical investigations of the IMF \citep{2008ApJ...684..395H} are also showing promising results, reproducing characteristic masses $\sim 0.3 \msun$ and log--normal distributions in the low--mass regime.

\section{Conclusions}\label{sec:conclusions}
We have assembled the largest set of photometric observations of M dwarfs to date and used it to study the low--mass stellar luminosity and mass functions.  Previous studies were limited by sample size, mostly due to the intrinsic faintness of M dwarfs.  The precise photometry of the SDSS allowed us to produce a clean, complete sample of M dwarfs, nearly two orders of magnitude larger than other studies.

To accurately estimate the brightness and distances to these stars, we constructed new photometric parallax relations from data kindly provided to us prior to publication \citep{Golimowski08}.  These relations were derived from $ugrizJHK_s$ photometry of nearby stars with known trigonometric parallax measurements.  We compared our new relations to those previously published for SDSS observations.

We also introduced a method for measuring the LF within large surveys.  Previous LF investigations either assumed a Galactic profile (for pencil beam surveys, such as \citealp{2001ApJ...555..393Z}) or a constant density (for nearby stars, \citealp[i.e.][]{1997AJ....113.2246R}).  However, none of these samples have approached the solid angle or number of the stars observed in this study.  We solved for the LF and Galactic structure simultaneously, using a technique similar to \cite{2008ApJ...673..864J}.  Our LF is measured in the $r$--band.  Using multiple CMRs, we investigated systematic errors in the LF, and computed the effects of Malmquist bias, unresolved binarity and Galactic structure changes using Monte Carlo models.  This allowed us to compare our results both to distant LF studies (which sampled mostly the system LF) and nearby star samples (which can resolve single stars in binary systems).  

Finally, we computed mass functions for single stars and systems.  Low luminosity stars are more common in the single star MF, since they can be companions to any higher mass star.  We fitted both MFs with a broken power law, a form preferred by \cite{2002Sci...295...82K}, and a log--normal distribution, which is favored by \cite{2003PASP..115..763C}.  The log--normal distribution at low masses seems to be ubiquitous: it is evident both in the field \citep[this study; ][]{2003PASP..115..763C} and open clusters, such as Blanco 1 \citep{2007A&A...471..499M}, the Pleiades \citep{2004A&A...426...75M}, and NGC 6611 \citep{2009MNRAS.392.1034O}. The best fits for this study are reported in Tables \ref{table:sys_mf_fits} and \ref{table:sing_mf_fits} .  We stress the point first made in \cite{covey08}, that comparing MF $data$ is preferable to comparing analytic fits, since the latter are often heavily swayed by slight discrepancies amongst the data.  We also caution the reader against extrapolating our reported mass function beyond  0.1 $\msun < {\cal M} <$ 0.8 $\msun$, the masses that bound our sample. In the future, we plan to investigate the luminosity function in other SDSS bandpasses, such as $i$ and $z$.  Our system and single-star MFs represent the best current values for this important quantity for low--mass stars.

JJB would like to acknowledge the support of the UW Astronomy department over the course of his thesis study.  In particular, he would like to acknowledge Ivan King, Andrew Becker, Nick Cowan, Lucianne Walkowicz, Nathan Kaib and Richard Plotkin for fruitful and helpful discussions.  He also acknowledges the support and guidance of Adam Burgasser.  We also thank the anonymous referee, whose insightful comments, especially on star--galaxy separation, greatly strengthened this paper.
We also gratefully acknowledge the support of NSF grants AST 02-05875 and AST 06-07644 and NASA ADP grant NAG5-13111.   We appreciate the serene and productive atmosphere fostered by Friday Harbor Labs, where this project first started.

 Funding for the SDSS and SDSS-II has been provided by the Alfred P. Sloan Foundation, the Participating Institutions, the National Science Foundation, the U.S. Department of Energy, the National Aeronautics and Space Administration, the Japanese Monbukagakusho, the Max Planck Society, and the Higher Education Funding Council for England. The SDSS Web Site is http://www.sdss.org/.

    The SDSS is managed by the Astrophysical Research Consortium for the Participating Institutions. The Participating Institutions are the American Museum of Natural History, Astrophysical Institute Potsdam, University of Basel, University of Cambridge, Case Western Reserve University, University of Chicago, Drexel University, Fermilab, the Institute for Advanced Study, the Japan Participation Group, Johns Hopkins University, the Joint Institute for Nuclear Astrophysics, the Kavli Institute for Particle Astrophysics and Cosmology, the Korean Scientist Group, the Chinese Academy of Sciences (LAMOST), Los Alamos National Laboratory, the Max-Planck-Institute for Astronomy (MPIA), the Max-Planck-Institute for Astrophysics (MPA), New Mexico State University, Ohio State University, University of Pittsburgh, University of Portsmouth, Princeton University, the United States Naval Observatory, and the University of Washington.


\begin{thebibliography}{0}
\expandafter\ifx\csname natexlab\endcsname\relax\def\natexlab#1{#1}\fi

\end{thebibliography}


\begin{thebibliography}{152}
\expandafter\ifx\csname natexlab\endcsname\relax\def\natexlab#1{#1}\fi

\bibitem[{{Abazajian} {et~al.}(2009)}]{2009ApJS..182..543A}
{Abazajian}, K.~N., {et~al.} 2009, \apjs, 182, 543

\bibitem[{{Adelman-McCarthy} {et~al.}(2008)}]{2008ApJS..175..297A}
{Adelman-McCarthy}, J.~K., {et~al.} 2008, \apjs, 175, 297

\bibitem[{{Allende Prieto} {et~al.}(2004){Allende Prieto}, {Barklem},
  {Lambert}, \& {Cunha}}]{2004A&A...420..183A}
{Allende Prieto}, C., {Barklem}, P.~S., {Lambert}, D.~L., \& {Cunha}, K. 2004,
  \aap, 420, 183

\bibitem[{{Allende Prieto} {et~al.}(2008)}]{2008AN....329.1018A}
{Allende Prieto}, C., {et~al.} 2008, Astronomische Nachrichten, 329, 1018

\bibitem[{{An} {et~al.}(2008)}]{2008ApJS..179..326A}
{An}, D., {et~al.} 2008, \apjs, 179, 326

\bibitem[{{Ballesteros-Paredes} {et~al.}(2006){Ballesteros-Paredes}, {Gazol},
  {Kim}, {Klessen}, {Jappsen}, \& {Tejero}}]{2006ApJ...637..384B}
{Ballesteros-Paredes}, J., {Gazol}, A., {Kim}, J., {Klessen}, R.~S., {Jappsen},
  A., \& {Tejero}, E. 2006, \apj, 637, 384

\bibitem[{{Baraffe} {et~al.}(1998){Baraffe}, {Chabrier}, {Allard}, \&
  {Hauschildt}}]{1998A&A...337..403B}
{Baraffe}, I., {Chabrier}, G., {Allard}, F., \& {Hauschildt}, P.~H. 1998, \aap,
  337, 403

\bibitem[{{Bastian} {et~al.}(2010){Bastian}, {Meyer}, \& {Covey}}]{coveyreview}
{Bastian}, N., {Meyer}, M.~R., \& {Covey}, K.~R. 2010, {in preparation}

\bibitem[{{Bate}(2009)}]{2009MNRAS.392.1363B}
{Bate}, M.~R. 2009, \mnras, 392, 1363

\bibitem[{{Bate} \& {Bonnell}(2005)}]{2005MNRAS.356.1201B}
{Bate}, M.~R., \& {Bonnell}, I.~A. 2005, \mnras, 356, 1201

\bibitem[{{Belokurov} {et~al.}(2006)}]{2006ApJ...642L.137B}
{Belokurov}, V., {et~al.} 2006, \apjl, 642, L137

\bibitem[{{Belokurov} {et~al.}(2007)}]{2007ApJ...654..897B}
---. 2007, \apj, 654, 897

\bibitem[{{Binney} {et~al.}(1997){Binney}, {Gerhard}, \&
  {Spergel}}]{1997MNRAS.288..365B}
{Binney}, J., {Gerhard}, O., \& {Spergel}, D. 1997, \mnras, 288, 365

\bibitem[{{Bochanski} {et~al.}(2007{\natexlab{a}}){Bochanski}, {Munn},
  {Hawley}, {West}, {Covey}, \& {Schneider}}]{2007AJ....134.2418B}
{Bochanski}, J.~J., {Munn}, J.~A., {Hawley}, S.~L., {West}, A.~A., {Covey},
  K.~R., \& {Schneider}, D.~P. 2007{\natexlab{a}}, \aj, 134, 2418

\bibitem[{{Bochanski} {et~al.}(2007{\natexlab{b}}){Bochanski}, {West},
  {Hawley}, \& {Covey}}]{2007AJ....133..531B}
{Bochanski}, J.~J., {West}, A.~A., {Hawley}, S.~L., \& {Covey}, K.~R.
  2007{\natexlab{b}}, \aj, 133, 531

\bibitem[{{Bochanski}(2008)}]{bochanskithesis}
{Bochanski}, Jr., J.~J. 2008, PhD thesis, University of Washington

\bibitem[{{Bonnell} {et~al.}(2006){Bonnell}, {Clarke}, \&
  {Bate}}]{2006MNRAS.368.1296B}
{Bonnell}, I.~A., {Clarke}, C.~J., \& {Bate}, M.~R. 2006, \mnras, 368, 1296

\bibitem[{{Burgasser} \& {Kirkpatrick}(2006)}]{Burgasser06}
{Burgasser}, A.~J., \& {Kirkpatrick}, J.~D. 2006, \apj, 645, 1485

\bibitem[{{Burgasser} {et~al.}(2002){Burgasser}, {Kirkpatrick}, {Brown},
  {Reid}, {Burrows}, {Liebert}, {Matthews}, {Gizis}, {Dahn}, {Monet}, {Cutri},
  \& {Skrutskie}}]{2002ApJ...564..421B}
{Burgasser}, A.~J., {Kirkpatrick}, J.~D., {Brown}, M.~E., {Reid}, I.~N.,
  {Burrows}, A., {Liebert}, J., {Matthews}, K., {Gizis}, J.~E., {Dahn}, C.~C.,
  {Monet}, D.~G., {Cutri}, R.~M., \& {Skrutskie}, M.~F. 2002, \apj, 564, 421

\bibitem[{{Burgasser} {et~al.}(2007){Burgasser}, {Reid}, {Siegler}, {Close},
  {Allen}, {Lowrance}, \& {Gizis}}]{2007prpl.conf..427B}
{Burgasser}, A.~J., {Reid}, I.~N., {Siegler}, N., {Close}, L., {Allen}, P.,
  {Lowrance}, P., \& {Gizis}, J. 2007, in Protostars and Planets V, ed.
  B.~{Reipurth}, D.~{Jewitt}, \& K.~{Keil}, 427--441

\bibitem[{{Burgasser} {et~al.}(2008){Burgasser}, {Vrba}, {L{\'e}pine}, {Munn},
  {Luginbuhl}, {Henden}, {Guetter}, \& {Canzian}}]{2008ApJ...672.1159B}
{Burgasser}, A.~J., {Vrba}, F.~J., {L{\'e}pine}, S., {Munn}, J.~A.,
  {Luginbuhl}, C.~B., {Henden}, A.~A., {Guetter}, H.~H., \& {Canzian}, B.~C.
  2008, \apj, 672, 1159

\bibitem[{{Cardelli} {et~al.}(1989){Cardelli}, {Clayton}, \&
  {Mathis}}]{1989ApJ...345..245C}
{Cardelli}, J.~A., {Clayton}, G.~C., \& {Mathis}, J.~S. 1989, \apj, 345, 245

\bibitem[{{Carollo} {et~al.}(2007)}]{2007Natur.450.1020C}
{Carollo}, D., {et~al.} 2007, \nat, 450, 1020

\bibitem[{{Chabrier}(2001)}]{2001ApJ...554.1274C}
{Chabrier}, G. 2001, \apj, 554, 1274

\bibitem[{{Chabrier}(2003{\natexlab{a}})}]{2003PASP..115..763C}
---. 2003{\natexlab{a}}, \pasp, 115, 763

\bibitem[{{Chabrier}(2003{\natexlab{b}})}]{2003ApJ...586L.133C}
---. 2003{\natexlab{b}}, \apjl, 586, L133

\bibitem[{{Chabrier}(2005)}]{2005ASSL..327...41C}
{Chabrier}, G. 2005, in Astrophysics and Space Science Library, Vol. 327, The
  Initial Mass Function 50 Years Later, ed. {E.~Corbelli, F.~Palla, \&
  H.~Zinnecker}, 41--+

\bibitem[{{Clem} {et~al.}(2008){Clem}, {Vanden Berg}, \&
  {Stetson}}]{2008AJ....135..682C}
{Clem}, J.~L., {Vanden Berg}, D.~A., \& {Stetson}, P.~B. 2008, \aj, 135, 682

\bibitem[{{Cohen}(1995)}]{1995ApJ...444..874C}
{Cohen}, M. 1995, \apj, 444, 874

\bibitem[{{Covey} {et~al.}(2008){Covey}, {Hawley}, {Bochanski}, {West}, {Reid},
  {Golimowski}, {Davenport}, {Henry}, {Uomoto}, \& {Holtzman}}]{covey08}
{Covey}, K.~R., {Hawley}, S.~L., {Bochanski}, J.~J., {West}, A.~A., {Reid},
  I.~N., {Golimowski}, D.~A., {Davenport}, J.~R.~A., {Henry}, T., {Uomoto}, A.,
  \& {Holtzman}, J.~A. 2008, \aj, 136, 1778

\bibitem[{{Covey} {et~al.}(2007)}]{2007AJ....134.2398C}
{Covey}, K.~R., {et~al.} 2007, \aj, 134, 2398

\bibitem[{{Cruz} {et~al.}(2007)}]{2007AJ....133..439C}
{Cruz}, K.~L., {et~al.} 2007, \aj, 133, 439

\bibitem[{{Cutri} {et~al.}(2003)}]{2003yCat.2246....0C}
{Cutri}, R.~M., {et~al.} 2003, VizieR Online Data Catalog, 2246, 0

\bibitem[{{Dahn} {et~al.}(1986){Dahn}, {Liebert}, \&
  {Harrington}}]{1986AJ.....91..621D}
{Dahn}, C.~C., {Liebert}, J., \& {Harrington}, R.~S. 1986, \aj, 91, 621

\bibitem[{{Dahn} {et~al.}(2002)}]{2002AJ....124.1170D}
{Dahn}, C.~C., {et~al.} 2002, \aj, 124, 1170

\bibitem[{{Davenport} {et~al.}(2007){Davenport}, {Bochanski}, {Covey},
  {Hawley}, {West}, \& {Schneider}}]{2007AJ....134.2430D}
{Davenport}, J.~R.~A., {Bochanski}, J.~J., {Covey}, K.~R., {Hawley}, S.~L.,
  {West}, A.~A., \& {Schneider}, D.~P. 2007, \aj, 134, 2430

\bibitem[{{Delfosse} {et~al.}(2004){Delfosse}, {Beuzit}, {Marchal}, {Bonfils},
  {Perrier}, {S{\'e}gransan}, {Udry}, {Mayor}, \&
  {Forveille}}]{2004ASPC..318..166D}
{Delfosse}, X., {Beuzit}, J.-L., {Marchal}, L., {Bonfils}, X., {Perrier}, C.,
  {S{\'e}gransan}, D., {Udry}, S., {Mayor}, M., \& {Forveille}, T. 2004, in
  Astronomical Society of the Pacific Conference Series, Vol. 318,
  Spectroscopically and Spatially Resolving the Components of the Close Binary
  Stars, ed. R.~W. {Hilditch}, H.~{Hensberge}, \& K.~{Pavlovski}, 166--174

\bibitem[{{Delfosse} {et~al.}(2000){Delfosse}, {Forveille}, {S{\'e}gransan},
  {Beuzit}, {Udry}, {Perrier}, \& {Mayor}}]{2000A&A...364..217D}
{Delfosse}, X., {Forveille}, T., {S{\'e}gransan}, D., {Beuzit}, J.-L., {Udry},
  S., {Perrier}, C., \& {Mayor}, M. 2000, \aap, 364, 217

\bibitem[{{Duquennoy} \& {Mayor}(1991)}]{1991A&A...248..485D}
{Duquennoy}, A., \& {Mayor}, M. 1991, \aap, 248, 485

\bibitem[{{Elmegreen}(2007)}]{2007ASPC..362..269E}
{Elmegreen}, B.~G. 2007, in Astronomical Society of the Pacific Conference
  Series, Vol. 362, The Seventh Pacific Rim Conference on Stellar Astrophysics,
  ed. {Y.~W.~Kang, H.-W.~Lee, K.-C.~Leung, \& K.-S.~Cheng}, 269--+

\bibitem[{{ESA}(1997)}]{1997yCat.1239....0E}
{ESA}. 1997, VizieR Online Data Catalog, 1239, 0

\bibitem[{{Fischer} \& {Marcy}(1992)}]{1992ApJ...396..178F}
{Fischer}, D.~A., \& {Marcy}, G.~W. 1992, \apj, 396, 178

\bibitem[{{Fuchs} {et~al.}(2009)}]{2009AJ....137.4149F}
{Fuchs}, B., {et~al.} 2009, \aj, 137, 4149

\bibitem[{{Fukugita} {et~al.}(1996){Fukugita}, {Ichikawa}, {Gunn}, {Doi},
  {Shimasaku}, \& {Schneider}}]{1996AJ....111.1748F}
{Fukugita}, M., {Ichikawa}, T., {Gunn}, J.~E., {Doi}, M., {Shimasaku}, K., \&
  {Schneider}, D.~P. 1996, \aj, 111, 1748

\bibitem[{{Giclas} {et~al.}(1971){Giclas}, {Burnham}, \&
  {Thomas}}]{1971lpms.book.....G}
{Giclas}, H.~L., {Burnham}, R., \& {Thomas}, N.~G. 1971, {Lowell proper motion
  survey Northern Hemisphere. The G numbered stars. 8991 stars fainter than
  magnitude 8 with motions $<$ 0''.26/year} (Flagstaff, Arizona: Lowell
  Observatory, 1971)

\bibitem[{{Girardi} {et~al.}(2004){Girardi}, {Grebel}, {Odenkirchen}, \&
  {Chiosi}}]{2004A&A...422..205G}
{Girardi}, L., {Grebel}, E.~K., {Odenkirchen}, M., \& {Chiosi}, C. 2004, \aap,
  422, 205

\bibitem[{{Gizis}(1997)}]{1997AJ....113..806G}
{Gizis}, J.~E. 1997, \aj, 113, 806

\bibitem[{{Gliese} \& {Jahreiss}(1991)}]{1991adc..rept.....G}
{Gliese}, W., \& {Jahreiss}, H. 1991, {Preliminary Version of the Third
  Catalogue of Nearby Stars}, Tech. rep.

\bibitem[{{Golimowski} {et~al.}(2004)}]{2004AJ....127.3516G}
{Golimowski}, D.~A., {et~al.} 2004, \aj, 127, 3516

\bibitem[{{Golimowski} {et~al.}(2010)}]{Golimowski08}
---. 2010, {AJ, in preparation}

\bibitem[{{Goodwin} {et~al.}(2006){Goodwin}, {Whitworth}, \&
  {Ward-Thompson}}]{2006A&A...452..487G}
{Goodwin}, S.~P., {Whitworth}, A.~P., \& {Ward-Thompson}, D. 2006, \aap, 452,
  487

\bibitem[{{Gould} {et~al.}(1996){Gould}, {Bahcall}, \&
  {Flynn}}]{1996ApJ...465..759G}
{Gould}, A., {Bahcall}, J.~N., \& {Flynn}, C. 1996, \apj, 465, 759

\bibitem[{{Gould} {et~al.}(1997){Gould}, {Bahcall}, \&
  {Flynn}}]{1997ApJ...482..913G}
---. 1997, \apj, 482, 913

\bibitem[{{Gunn} {et~al.}(1998)}]{1998AJ....116.3040G}
{Gunn}, J.~E., {et~al.} 1998, \aj, 116, 3040

\bibitem[{{Gunn} {et~al.}(2006)}]{2006AJ....131.2332G}
---. 2006, \aj, 131, 2332

\bibitem[{{Hanson}(1979)}]{1979MNRAS.186..875H}
{Hanson}, R.~B. 1979, \mnras, 186, 875

\bibitem[{{Hawley} {et~al.}(1996){Hawley}, {Gizis}, \&
  {Reid}}]{1996AJ....112.2799H}
{Hawley}, S.~L., {Gizis}, J.~E., \& {Reid}, I.~N. 1996, \aj, 112, 2799

\bibitem[{{Hawley} {et~al.}(2002)}]{2002AJ....123.3409H}
{Hawley}, S.~L., {et~al.} 2002, \aj, 123, 3409

\bibitem[{{Hennebelle} \& {Chabrier}(2008)}]{2008ApJ...684..395H}
{Hennebelle}, P., \& {Chabrier}, G. 2008, \apj, 684, 395

\bibitem[{{Hennebelle} \& {Chabrier}(2009)}]{2009ApJ...702.1428H}
---. 2009, \apj, 702, 1428

\bibitem[{{Henry} {et~al.}(1994){Henry}, {Kirkpatrick}, \&
  {Simons}}]{1994AJ....108.1437H}
{Henry}, T.~J., {Kirkpatrick}, J.~D., \& {Simons}, D.~A. 1994, \aj, 108, 1437

\bibitem[{{Henry} \& {McCarthy}(1990)}]{1990ApJ...350..334H}
{Henry}, T.~J., \& {McCarthy}, Jr., D.~W. 1990, \apj, 350, 334

\bibitem[{{Henry} {et~al.}(2004){Henry}, {Subasavage}, {Brown}, {Beaulieu},
  {Jao}, \& {Hambly}}]{2004AJ....128.2460H}
{Henry}, T.~J., {Subasavage}, J.~P., {Brown}, M.~A., {Beaulieu}, T.~D., {Jao},
  W.-C., \& {Hambly}, N.~C. 2004, \aj, 128, 2460

\bibitem[{{Herbst} {et~al.}(1999){Herbst}, {Thompson}, {Fockenbrock}, {Rix}, \&
  {Beckwith}}]{1999ApJ...526L..17H}
{Herbst}, T.~M., {Thompson}, D., {Fockenbrock}, R., {Rix}, H.-W., \&
  {Beckwith}, S.~V.~W. 1999, \apjl, 526, L17

\bibitem[{{Hogg} {et~al.}(2001){Hogg}, {Finkbeiner}, {Schlegel}, \&
  {Gunn}}]{2001AJ....122.2129H}
{Hogg}, D.~W., {Finkbeiner}, D.~P., {Schlegel}, D.~J., \& {Gunn}, J.~E. 2001,
  \aj, 122, 2129

\bibitem[{{Ivezi{\'c}} {et~al.}(2004)}]{2004AN....325..583I}
{Ivezi{\'c}}, {\v Z}., {et~al.} 2004, Astronomische Nachrichten, 325, 583

\bibitem[{{Ivezi{\'c}} {et~al.}(2007)}]{2007AJ....134..973I}
---. 2007, \aj, 134, 973

\bibitem[{{Ivezi{\'c}} {et~al.}(2008)}]{Ivezic08}
{Ivezi{\'c}}, Z., {et~al.} 2008, \apj, 684, 287

\bibitem[{{Joergens}(2006)}]{2006A&A...448..655J}
{Joergens}, V. 2006, \aap, 448, 655

\bibitem[{{Johnson} \& {Apps}(2009)}]{2009arXiv0904.3092J}
{Johnson}, J.~A., \& {Apps}, K. 2009, ArXiv e-prints

\bibitem[{{Juri{\'c}} {et~al.}(2008)}]{2008ApJ...673..864J}
{Juri{\'c}}, M., {et~al.} 2008, \apj, 673, 864

\bibitem[{{Kerr} \& {Lynden-Bell}(1986)}]{1986MNRAS.221.1023K}
{Kerr}, F.~J., \& {Lynden-Bell}, D. 1986, \mnras, 221, 1023

\bibitem[{{King} {et~al.}(1990){King}, {Gilmore}, \& {van der
  Kruit}}]{1990mwg..conf.....K}
{King}, I., {Gilmore}, G., \& {van der Kruit}, P.~C., eds. 1990, {The Milky Way
  As Galaxy}

\bibitem[{{Kirkpatrick} {et~al.}(1995){Kirkpatrick}, {Henry}, \&
  {Simons}}]{1995AJ....109..797K}
{Kirkpatrick}, J.~D., {Henry}, T.~J., \& {Simons}, D.~A. 1995, \aj, 109, 797

\bibitem[{{Kirkpatrick} {et~al.}(1999)}]{1999ApJ...519..802K}
{Kirkpatrick}, J.~D., {et~al.} 1999, \apj, 519, 802

\bibitem[{{Kowalski} {et~al.}(2009){Kowalski}, {Hawley}, {Hilton}, {Becker},
  {West}, {Bochanski}, \& {Sesar}}]{2009arXiv0906.2030K}
{Kowalski}, A.~F., {Hawley}, S.~L., {Hilton}, E.~J., {Becker}, A.~C., {West},
  A.~A., {Bochanski}, J.~J., \& {Sesar}, B. 2009, ArXiv e-prints

\bibitem[{{Kroupa}(2002)}]{2002Sci...295...82K}
{Kroupa}, P. 2002, Science, 295, 82

\bibitem[{{Kroupa} {et~al.}(1990){Kroupa}, {Tout}, \&
  {Gilmore}}]{1990MNRAS.244...76K}
{Kroupa}, P., {Tout}, C.~A., \& {Gilmore}, G. 1990, \mnras, 244, 76

\bibitem[{{Kroupa} {et~al.}(1993){Kroupa}, {Tout}, \&
  {Gilmore}}]{1993MNRAS.262..545K}
---. 1993, \mnras, 262, 545

\bibitem[{{Lada}(2006)}]{2006ApJ...640L..63L}
{Lada}, C.~J. 2006, \apjl, 640, L63

\bibitem[{{Larson}(1981)}]{1981MNRAS.194..809L}
{Larson}, R.~B. 1981, \mnras, 194, 809

\bibitem[{{Larson}(2005)}]{2005MNRAS.359..211L}
---. 2005, \mnras, 359, 211

\bibitem[{{Laughlin} {et~al.}(1997){Laughlin}, {Bodenheimer}, \&
  {Adams}}]{1997ApJ...482..420L}
{Laughlin}, G., {Bodenheimer}, P., \& {Adams}, F.~C. 1997, \apj, 482, 420

\bibitem[{{L{\'e}pine} {et~al.}(2003){L{\'e}pine}, {Rich}, \&
  {Shara}}]{Lepine03}
{L{\'e}pine}, S., {Rich}, R.~M., \& {Shara}, M.~M. 2003, \aj, 125, 1598

\bibitem[{{L{\'e}pine} \& {Scholz}(2008)}]{2008ApJ...681L..33L}
{L{\'e}pine}, S., \& {Scholz}, R.-D. 2008, \apjl, 681, L33

\bibitem[{{Luhman}(2004{\natexlab{a}})}]{2004ApJ...617.1216L}
{Luhman}, K.~L. 2004{\natexlab{a}}, \apj, 617, 1216

\bibitem[{{Luhman}(2004{\natexlab{b}})}]{2004ApJ...614..398L}
---. 2004{\natexlab{b}}, \apj, 614, 398

\bibitem[{{Luhman}(2006)}]{2006ApJ...645..676L}
---. 2006, \apj, 645, 676

\bibitem[{{Luhman} {et~al.}(2009){Luhman}, {Mamajek}, {Allen}, {Muench}, \&
  {Finkbeiner}}]{2009ApJ...691.1265L}
{Luhman}, K.~L., {Mamajek}, E.~E., {Allen}, P.~R., {Muench}, A.~A., \&
  {Finkbeiner}, D.~P. 2009, \apj, 691, 1265

\bibitem[{{Lupton} {et~al.}(2001){Lupton}, {Gunn}, {Ivezi{\'c}}, {Knapp}, \&
  {Kent}}]{2001ASPC..238..269L}
{Lupton}, R., {Gunn}, J.~E., {Ivezi{\'c}}, Z., {Knapp}, G.~R., \& {Kent}, S.
  2001, in Astronomical Society of the Pacific Conference Series, Vol. 238,
  Astronomical Data Analysis Software and Systems X, ed. F.~R. {Harnden}, Jr.,
  F.~A. {Primini}, \& H.~E. {Payne}, 269--+

\bibitem[{{Lutz} \& {Kelker}(1973)}]{1973PASP...85..573L}
{Lutz}, T.~E., \& {Kelker}, D.~H. 1973, \pasp, 85, 573

\bibitem[{{Luyten}(1939)}]{1939POMin...2..121L}
{Luyten}, W.~J. 1939, Publications of the Astronomical Observatory University
  of Minnesota, 2, 121

\bibitem[{{Luyten}(1941)}]{1941NYASA..42..201L}
---. 1941, New York Academy Sciences Annals, 42, 201

\bibitem[{{Luyten}(1968)}]{1968MNRAS.139..221L}
---. 1968, \mnras, 139, 221

\bibitem[{{Luyten}(1979)}]{1979lccs.book.....L}
---. 1979, {LHS catalogue. A catalogue of stars with proper motions exceeding
  0''5 annually} (Minneapolis: University of Minnesota, 1979, 2nd ed.)

\bibitem[{{Majewski} {et~al.}(2003){Majewski}, {Skrutskie}, {Weinberg}, \&
  {Ostheimer}}]{2003ApJ...599.1082M}
{Majewski}, S.~R., {Skrutskie}, M.~F., {Weinberg}, M.~D., \& {Ostheimer}, J.~C.
  2003, \apj, 599, 1082

\bibitem[{{Malmquist}(1936)}]{m36}
{Malmquist}, K.~G. 1936, Stockholm Obs. Medd., 26

\bibitem[{{Marshall} {et~al.}(2006){Marshall}, {Robin}, {Reyl{\'e}},
  {Schultheis}, \& {Picaud}}]{2006A&A...453..635M}
{Marshall}, D.~J., {Robin}, A.~C., {Reyl{\'e}}, C., {Schultheis}, M., \&
  {Picaud}, S. 2006, \aap, 453, 635

\bibitem[{{Martini} \& {Osmer}(1998)}]{1998AJ....116.2513M}
{Martini}, P., \& {Osmer}, P.~S. 1998, \aj, 116, 2513

\bibitem[{{McCuskey}(1966)}]{1966VA......7..141M}
{McCuskey}, S.~W. 1966, Vistas in Astronomy, 7, 141

\bibitem[{{Miller} \& {Scalo}(1979)}]{1979ApJS...41..513M}
{Miller}, G.~E., \& {Scalo}, J.~M. 1979, \apjs, 41, 513

\bibitem[{{Monet} {et~al.}(1992){Monet}, {Dahn}, {Vrba}, {Harris}, {Pier},
  {Luginbuhl}, \& {Ables}}]{1992AJ....103..638M}
{Monet}, D.~G., {Dahn}, C.~C., {Vrba}, F.~J., {Harris}, H.~C., {Pier}, J.~R.,
  {Luginbuhl}, C.~B., \& {Ables}, H.~D. 1992, \aj, 103, 638

\bibitem[{{Moraux} {et~al.}(2007){Moraux}, {Bouvier}, {Stauffer}, {Barrado y
  Navascu{\'e}s}, \& {Cuillandre}}]{2007A&A...471..499M}
{Moraux}, E., {Bouvier}, J., {Stauffer}, J.~R., {Barrado y Navascu{\'e}s}, D.,
  \& {Cuillandre}, J. 2007, \aap, 471, 499

\bibitem[{{Moraux} {et~al.}(2004){Moraux}, {Kroupa}, \&
  {Bouvier}}]{2004A&A...426...75M}
{Moraux}, E., {Kroupa}, P., \& {Bouvier}, J. 2004, \aap, 426, 75

\bibitem[{{Ng} {et~al.}(1997){Ng}, {Bertelli}, {Chiosi}, \&
  {Bressan}}]{1997A&A...324...65N}
{Ng}, Y.~K., {Bertelli}, G., {Chiosi}, C., \& {Bressan}, A. 1997, \aap, 324, 65

\bibitem[{{Oliveira} {et~al.}(2009){Oliveira}, {Jeffries}, \& {van
  Loon}}]{2009MNRAS.392.1034O}
{Oliveira}, J.~M., {Jeffries}, R.~D., \& {van Loon}, J.~T. 2009, \mnras, 392,
  1034

\bibitem[{{O'Mullane} {et~al.}(2005){O'Mullane}, {Li}, {Nieto-Santisteban},
  {Szalay}, {Thakar}, \& {Gray}}]{2005cs........2072O}
{O'Mullane}, W., {Li}, N., {Nieto-Santisteban}, M., {Szalay}, A., {Thakar}, A.,
  \& {Gray}, J. 2005, ArXiv Computer Science e-prints

\bibitem[{{Padoan} {et~al.}(2001){Padoan}, {Juvela}, {Goodman}, \&
  {Nordlund}}]{2001ApJ...553..227P}
{Padoan}, P., {Juvela}, M., {Goodman}, A.~A., \& {Nordlund}, {\AA}. 2001, \apj,
  553, 227

\bibitem[{{Padoan} \& {Nordlund}(2002)}]{2002ApJ...576..870P}
{Padoan}, P., \& {Nordlund}, {\AA}. 2002, \apj, 576, 870

\bibitem[{{Pier} {et~al.}(2003){Pier}, {Munn}, {Hindsley}, {Hennessy}, {Kent},
  {Lupton}, \& {Ivezi{\'c}}}]{2003AJ....125.1559P}
{Pier}, J.~R., {Munn}, J.~A., {Hindsley}, R.~B., {Hennessy}, G.~S., {Kent},
  S.~M., {Lupton}, R.~H., \& {Ivezi{\'c}}, {\v Z}. 2003, \aj, 125, 1559

\bibitem[{{Pineda} {et~al.}(2010)}]{pineda}
{Pineda}, J.~S., {et~al.} 2010, {AJ, in preparation}

\bibitem[{{Press} {et~al.}(1992){Press}, {Teukolsky}, {Vetterling}, \&
  {Flannery}}]{1992nrfa.book.....P}
{Press}, W.~H., {Teukolsky}, S.~A., {Vetterling}, W.~T., \& {Flannery}, B.~P.
  1992, {Numerical recipes in FORTRAN. The art of scientific computing}
  (Cambridge: University Press, |c1992, 2nd ed.)

\bibitem[{{Reid}(1997)}]{1997AJ....114..161R}
{Reid}, I.~N. 1997, \aj, 114, 161

\bibitem[{{Reid} \& {Cruz}(2002)}]{2002AJ....123.2806R}
{Reid}, I.~N., \& {Cruz}, K.~L. 2002, \aj, 123, 2806

\bibitem[{{Reid} {et~al.}(2003){Reid}, {Cruz}, {Laurie}, {Liebert}, {Dahn},
  {Harris}, {Guetter}, {Stone}, {Canzian}, {Luginbuhl}, {Levine}, {Monet}, \&
  {Monet}}]{2003AJ....125..354R}
{Reid}, I.~N., {Cruz}, K.~L., {Laurie}, S.~P., {Liebert}, J., {Dahn}, C.~C.,
  {Harris}, H.~C., {Guetter}, H.~H., {Stone}, R.~C., {Canzian}, B.,
  {Luginbuhl}, C.~B., {Levine}, S.~E., {Monet}, A.~K.~B., \& {Monet}, D.~G.
  2003, \aj, 125, 354

\bibitem[{{Reid} \& {Gizis}(1997)}]{1997AJ....113.2246R}
{Reid}, I.~N., \& {Gizis}, J.~E. 1997, \aj, 113, 2246

\bibitem[{{Reid} {et~al.}(2002){Reid}, {Gizis}, \&
  {Hawley}}]{2002AJ....124.2721R}
{Reid}, I.~N., {Gizis}, J.~E., \& {Hawley}, S.~L. 2002, \aj, 124, 2721

\bibitem[{{Reid} \& {Hawley}(2005)}]{2005nlds.book.....R}
{Reid}, I.~N., \& {Hawley}, S.~L. 2005, {New light on dark stars : red dwarfs,
  low-mass stars, brown dwarfs} (New Light on Dark Stars Red Dwarfs, Low-Mass
  Stars, Brown Stars, by I.N.~Reid and S.L.~Hawley.~ Springer-Praxis books in
  astrophysics and astronomy.~Praxis Publishing Ltd, 2005.~ ISBN 3-540-25124-3)

\bibitem[{{Reid} {et~al.}(1995){Reid}, {Hawley}, \&
  {Gizis}}]{1995AJ....110.1838R}
{Reid}, I.~N., {Hawley}, S.~L., \& {Gizis}, J.~E. 1995, \aj, 110, 1838

\bibitem[{{Reid} {et~al.}(1999){Reid}, {Kirkpatrick}, {Liebert}, {Burrows},
  {Gizis}, {Burgasser}, {Dahn}, {Monet}, {Cutri}, {Beichman}, \&
  {Skrutskie}}]{1999ApJ...521..613R}
{Reid}, I.~N., {Kirkpatrick}, J.~D., {Liebert}, J., {Burrows}, A., {Gizis},
  J.~E., {Burgasser}, A., {Dahn}, C.~C., {Monet}, D., {Cutri}, R., {Beichman},
  C.~A., \& {Skrutskie}, M. 1999, \apj, 521, 613

\bibitem[{{Salpeter}(1955)}]{1955ApJ...121..161S}
{Salpeter}, E.~E. 1955, \apj, 121, 161

\bibitem[{{Sandage} \& {Eggen}(1959)}]{1959MNRAS.119..278S}
{Sandage}, A.~R., \& {Eggen}, O.~J. 1959, \mnras, 119, 278

\bibitem[{{Scalo}(1986)}]{1986FCPh...11....1S}
{Scalo}, J.~M. 1986, Fundamentals of Cosmic Physics, 11, 1

\bibitem[{{Schlegel} {et~al.}(1998){Schlegel}, {Finkbeiner}, \&
  {Davis}}]{1998ApJ...500..525S}
{Schlegel}, D.~J., {Finkbeiner}, D.~P., \& {Davis}, M. 1998, \apj, 500, 525

\bibitem[{{Schultheis} {et~al.}(2006){Schultheis}, {Robin}, {Reyl{\'e}},
  {McCracken}, {Bertin}, {Mellier}, \& {Le F{\`e}vre}}]{2006AA...447..185S}
{Schultheis}, M., {Robin}, A.~C., {Reyl{\'e}}, C., {McCracken}, H.~J.,
  {Bertin}, E., {Mellier}, Y., \& {Le F{\`e}vre}, O. 2006, \aap, 447, 185

\bibitem[{{Scoville} {et~al.}(2007)}]{2007ApJS..172....1S}
{Scoville}, N., {et~al.} 2007, \apjs, 172, 1

\bibitem[{{Sesar} {et~al.}(2008){Sesar}, {Ivezi{\'c}}, \&
  {Juri{\'c}}}]{Sesar08}
{Sesar}, B., {Ivezi{\'c}}, {\v Z}., \& {Juri{\'c}}, M. 2008, \apj, 689, 1244

\bibitem[{{Shkolnik} {et~al.}(2009){Shkolnik}, {Liu}, \&
  {Reid}}]{2009arXiv0904.3323S}
{Shkolnik}, E., {Liu}, M.~C., \& {Reid}, I.~N. 2009, ArXiv e-prints

\bibitem[{{Siegel} {et~al.}(2002){Siegel}, {Majewski}, {Reid}, \&
  {Thompson}}]{2002ApJ...578..151S}
{Siegel}, M.~H., {Majewski}, S.~R., {Reid}, I.~N., \& {Thompson}, I.~B. 2002,
  \apj, 578, 151

\bibitem[{{Skrutskie} {et~al.}(2006)}]{2006AJ....131.1163S}
{Skrutskie}, M.~F., {et~al.} 2006, \aj, 131, 1163

\bibitem[{{Smith} {et~al.}(2002)}]{2002AJ....123.2121S}
{Smith}, J.~A., {et~al.} 2002, \aj, 123, 2121

\bibitem[{{Stobie} {et~al.}(1989){Stobie}, {Ishida}, \&
  {Peacock}}]{1989MNRAS.238..709S}
{Stobie}, R.~S., {Ishida}, K., \& {Peacock}, J.~A. 1989, \mnras, 238, 709

\bibitem[{{Stoughton} {et~al.}(2002)}]{2002AJ....123..485S}
{Stoughton}, C., {et~al.} 2002, \aj, 123, 485

\bibitem[{{Strauss} {et~al.}(1999)}]{1999ApJ...522L..61S}
{Strauss}, M.~A., {et~al.} 1999, \apjl, 522, L61

\bibitem[{{Tinney}(1993)}]{1993ApJ...414..279T}
{Tinney}, C.~G. 1993, \apj, 414, 279

\bibitem[{{Tucker} {et~al.}(2006)}]{2006AN....327..821T}
{Tucker}, D.~L., {et~al.} 2006, Astronomische Nachrichten, 327, 821

\bibitem[{{van Altena} {et~al.}(1995){van Altena}, {Lee}, \&
  {Hoffleit}}]{1995gcts.book.....V}
{van Altena}, W.~F., {Lee}, J.~T., \& {Hoffleit}, E.~D. 1995, {The general
  catalogue of trigonometric [stellar] paralaxes} (New Haven, CT: Yale
  University Observatory, |c1995, 4th ed., completely revised and enlarged)

\bibitem[{{van Leeuwen}(2007)}]{2007A&A...474..653V}
{van Leeuwen}, F. 2007, \aap, 474, 653

\bibitem[{{van Rhijn}(1925)}]{1925PGro...38D...1V}
{van Rhijn}, P.~J. 1925, Publications of the Kapteyn Astronomical Laboratory
  Groningen, 38, D1+

\bibitem[{{van Rhijn}(1936)}]{1936PGro...47....1V}
---. 1936, Publications of the Kapteyn Astronomical Laboratory Groningen, 47, 1

\bibitem[{{Vrba} {et~al.}(2004)}]{2004AJ....127.2948V}
{Vrba}, F.~J., {et~al.} 2004, \aj, 127, 2948

\bibitem[{{West} {et~al.}(2008){West}, {Hawley}, {Bochanski}, {Covey}, {Reid},
  {Dhital}, {Hilton}, \& {Masuda}}]{2008AJ....135..785W}
{West}, A.~A., {Hawley}, S.~L., {Bochanski}, J.~J., {Covey}, K.~R., {Reid},
  I.~N., {Dhital}, S., {Hilton}, E.~J., \& {Masuda}, M. 2008, \aj, 135, 785

\bibitem[{{West} {et~al.}(2005){West}, {Walkowicz}, \&
  {Hawley}}]{2005PASP..117..706W}
{West}, A.~A., {Walkowicz}, L.~M., \& {Hawley}, S.~L. 2005, \pasp, 117, 706

\bibitem[{{West} {et~al.}(2004)}]{2004AJ....128..426W}
{West}, A.~A., {et~al.} 2004, \aj, 128, 426

\bibitem[{{Wielen}(1974)}]{1974HiA.....3..395W}
{Wielen}, R. 1974, Highlights of Astronomy, 3, 395

\bibitem[{{Wielen} {et~al.}(1983){Wielen}, {Jahreiss}, \&
  {Kr{\"u}ger}}]{1983nssl.conf..163W}
{Wielen}, R., {Jahreiss}, H., \& {Kr{\"u}ger}, R. 1983, in IAU Colloq. 76:
  Nearby Stars and the Stellar Luminosity Function, ed. A.~G.~D. {Philip} \&
  A.~R. {Upgren}, 163--170

\bibitem[{{Williams} {et~al.}(2002)}]{2002AAS...20111511W}
{Williams}, C.~C., {et~al.} 2002, in Bulletin of the American Astronomical
  Society, Vol.~34, Bulletin of the American Astronomical Society, 1292--+

\bibitem[{{Willman} {et~al.}(2005){Willman}, {Blanton}, {West}, {Dalcanton},
  {Hogg}, {Schneider}, {Wherry}, {Yanny}, \& {Brinkmann}}]{2005AJ....129.2692W}
{Willman}, B., {Blanton}, M.~R., {West}, A.~A., {Dalcanton}, J.~J., {Hogg},
  D.~W., {Schneider}, D.~P., {Wherry}, N., {Yanny}, B., \& {Brinkmann}, J.
  2005, \aj, 129, 2692

\bibitem[{{Woolf} \& {Wallerstein}(2006)}]{2006PASP..118..218W}
{Woolf}, V.~M., \& {Wallerstein}, G. 2006, \pasp, 118, 218

\bibitem[{{Yanny} {et~al.}(2003){Yanny}, {Newberg}, {Grebel}, {Kent},
  {Odenkirchen}, {Rockosi}, {Schlegel}, {Subbarao}, {Brinkmann}, {Fukugita},
  {Ivezic}, {Lamb}, {Schneider}, \& {York}}]{2003ApJ...588..824Y}
{Yanny}, B., {Newberg}, H.~J., {Grebel}, E.~K., {Kent}, S., {Odenkirchen}, M.,
  {Rockosi}, C.~M., {Schlegel}, D., {Subbarao}, M., {Brinkmann}, J.,
  {Fukugita}, M., {Ivezic}, {\v Z}., {Lamb}, D.~Q., {Schneider}, D.~P., \&
  {York}, D.~G. 2003, \apj, 588, 824

\bibitem[{{York} {et~al.}(2000)}]{2000AJ....120.1579Y}
{York}, D.~G., {et~al.} 2000, \aj, 120, 1579

\bibitem[{{Zheng} {et~al.}(2001){Zheng}, {Flynn}, {Gould}, {Bahcall}, \&
  {Salim}}]{2001ApJ...555..393Z}
{Zheng}, Z., {Flynn}, C., {Gould}, A., {Bahcall}, J.~N., \& {Salim}, S. 2001,
  \apj, 555, 393

\end{thebibliography}
\end{document}